\documentclass[letterpaper,11pt]{article}

\usepackage{jheppub}
\usepackage{graphicx}
\usepackage{amsmath}
\usepackage{amssymb}
\usepackage{xspace}

\newcommand{\eq}[1]{eq.~\eqref{eq:#1}}
\newcommand{\eqs}[2]{eqs.~\eqref{eq:#1} and \eqref{eq:#2}}
\renewcommand{\sec}[1]{sec.~\ref{sec:#1}}

\newcommand{\subsec}[1]{sec.~\ref{subsec:#1}}
\newcommand{\app}[1]{app.~\ref{app:#1}} 
\newcommand{\fig}[1]{fig.~\ref{fig:#1}}

\newcommand{\abs}[1]{\lvert#1\rvert}
\newcommand{\Abs}[1]{\bigl\lvert#1\bigr\rvert}
\newcommand{\ord}[1]{{\mathcal O}(#1)}
\newcommand{\ORd}[1]{{\mathcal O}\Bigl(#1\Bigr)}

\newcommand{\Mae}[3]{\bigl\langle#1\bigr\rvert#2\bigr\rvert#3\bigr\rangle}

\newcommand{\nn}{\nonumber}

\newcommand{\df}{\mathrm{d}}
\newcommand{\img}{\mathrm{i}}
\newcommand{\sdt}{\!\cdot\!}
\newcommand{\tr}{\mathrm{tr}}
\newcommand{\lra}{\leftrightarrow}

\def\dblone{\hbox{$1\hskip -1.2pt\vrule depth 0pt height 1.6ex width 0.7pt
\vrule depth 0pt height 0.3pt width 0.12em$}}

\newcommand{\al}{\alpha}
\newcommand{\bt}{\beta}
\newcommand{\ga}{\gamma}
\newcommand{\Ga}{\Gamma}
\newcommand{\de}{\delta}

\newcommand{\eps}{\epsilon}
\newcommand{\ve}{\varepsilon}
\newcommand{\la}{\lambda}
\newcommand{\si}{\sigma}
\newcommand{\w}{\omega}

\newcommand{\cB}{{\mathcal B}}
\newcommand{\cG}{{\mathcal G}}
\newcommand{\cJ}{{\mathcal J}}
\newcommand{\cL}{{\mathcal L}}
\newcommand{\cP}{{\mathcal P}}

\newcommand{\nslash}{n\!\!\!\slash}
\newcommand{\bnslash}{\bar{n}\!\!\!\slash}
\newcommand{\pslash}{p\!\!\!\slash}
\newcommand{\ellslash}{\ell\!\!\!\slash}

\newcommand{\tG}{\tilde \cG}

\newcommand{\lp}{p_\ell}         

\newcommand{\bn}{\bar{n}}
\newcommand{\bnP}{\overline {\mathcal P}}

\newcommand{\GeV}{\,\mathrm{GeV}}

\newcommand{\lqcd}{\Lambda_\mathrm{QCD}}

\newcommand{\cut}{\mathrm{cut}}
\newcommand{\bare}{\mathrm{bare}}
\newcommand{\cusp}{\mathrm{cusp}}
\newcommand{\jet}{\mathrm{jet}}
\newcommand{\pol}{\mathrm{pol}}

\newcommand{\zero}{{(0)}}
\newcommand{\one}{{(1)}}

\newcommand{\SCETa}{\ensuremath{{\rm SCET}_{\rm I}}\xspace}
\newcommand{\SCETb}{\ensuremath{{\rm SCET}_{\rm II}}\xspace}

\newcommand{\MSbar}{\overline{\rm MS}}

\allowdisplaybreaks[2]

\title{Parton Fragmentation within an Identified Jet at NNLL}

\author[a]{Ambar Jain,}
\author[b,c]{Massimiliano Procura,}
\author[d]{Wouter J.~Waalewijn}

\affiliation[a]{Department of Physics, Carnegie Mellon University, Pittsburgh, PA~15213, U.S.A.}
\affiliation[b]{Physik-Department, Technische Universit\"at M\"unchen, 
D-85748 Garching, Germany}
\affiliation[c]{Albert Einstein Center for Fundamental Physics, Institute for Theoretical Physics, \\ University of Bern, CH-3012 Bern, Switzerland}
\affiliation[d]{Department of Physics, University of California at San Diego, 
La Jolla, CA 92093, U.S.A. \vspace{2ex}}

\emailAdd{ambar@andrew.cmu.edu}
\emailAdd{mprocura@itp.unibe.ch}
\emailAdd{wouterw@physics.ucsd.edu}

\abstract{
The fragmentation of a light parton $i$ to a jet containing a light energetic hadron $h$, where the momentum fraction of this hadron as well as the invariant mass of the jet is measured, is described by ``fragmenting jet functions". We calculate the one-loop matching coefficients $\cJ_{ij}$ that relate the fragmenting jet functions $\cG_i^h$ to the standard, unpolarized fragmentation functions $D_j^h$ for quark and gluon jets. We perform this calculation using various IR regulators and show explicitly how the IR divergences cancel in the matching. We derive the relationship between the coefficients $\cJ_{ij}$ and the quark and gluon jet functions. This provides a cross-check of our results. As an application we study the process $e^+ e^- \to X \pi^+$ on the $\Upsilon(4S)$ resonance where we measure the momentum fraction of the $\pi^+$ and restrict to the dijet limit by imposing a cut on thrust $T$. In our analysis we sum the logarithms of $\tau=1-T$ in the cross section to next-to-next-to-leading-logarithmic accuracy (NNLL). We find that including contributions up to NNLL (or NLO) can have a large impact on extracting fragmentation functions from $e^+ e^- \to {\rm dijet} + h$.
}
\begin{document}
{\flushright TUM--EFT 18/10 \\[-6ex]}

\maketitle

%%%%%%%%%%%%%%%%%%%%%%%%%%%%%%%%%%%%%%%%%%%%%%%%%%%%%%%%%%%%%%%%%%%%%%%%%%%%%%%%
\section{Introduction}
\label{sec:intro}
%%%%%%%%%%%%%%%%%%%%%%%%%%%%%%%%%%%%%%%%%%%%%%%%%%%%%%%%%%%%%%%%%%%%%%%%%%%%%%%%

In single-inclusive hadron production, an energetic parton $i=\{g,u, \bar u, d, \ldots\}$ produces an observed energetic hadron $h$ and accompanying hadrons $X$. Factorization theorems allow one to identify perturbative (calculable) and non-perturbative (universal) contributions to these processes. For example, in $e^+e^-\to X h$ at a high center-of-mass (c.m.) energy $Q$ it has been proven that, to all orders in $\alpha_s$ and at leading power of $\lqcd/Q$, the cross-section has the following factorized form (see e.g.~ref.~\cite{Collins:1989gx})
%%%
\begin{equation} \label{eq:eeXh}
\frac{\df \sigma}{\df z} =  \si_0 \sum_{i=g,u, \bar u, d, \ldots}
\int_z^1 \frac{\df x}{x}\,C_i\Big(Q, \frac{z}{x}, \mu\Big) D_i^h(x, \mu) \,.
\end{equation}
%%%
Here, $z=2 E_h/Q$ is the energy fraction of the hadron $h$ in the c.m.~frame, $\si_0$ is the Born cross section, and $\mu$ is the $\MSbar$ renormalization scale. The coefficient functions $C_i$ incorporate the short-distance partonic process producing the fragmenting parton $i$: they are calculable in perturbation theory and independent of the observed hadron $h$. The long-distance physics of the hadronization resides in the non-perturbative fragmentation function $D_i^h(x,\mu)$, which is the number density of hadrons of type $h$ in the ``decay" products of the parton $i$, for a specific value of $x$~\cite{Georgi:1977mg,Ellis:1978ty,Curci:1980uw,Collins:1981uw}. The convolution variable $x$ in \eq{eeXh} is the fraction of the energy of the parent parton $i$ carried by the observed hadron $h$. At leading order (LO) the hard partonic process is $e^+ e^- \to q \bar q$ and $x = z$. Beyond LO, radiation will be emitted before the parton $i$ produces $h$, and thus $x \geq z$.

In experimental studies of fragmentation, additional measurements on the hadronic final state $X$ may be needed. For example, the Belle collaboration studies light-quark fragmentation by restricting to dijet final-state configurations, which removes
$B$-meson events from the data sample on the $\Upsilon(4S)$ resonance~\cite{Seidl:2008xc}. This is achieved by imposing a cut on thrust, which is an event shape variable defined as~\cite{Farhi:1977sg}
%%%
\begin{equation}
  T = \max{}_{\hat t}\; \frac{\sum_i |\hat t \sdt {\vec p}_i|}{\sum_i |{\vec p}_i|} 
  \,,
\end{equation}
%%%
where the sum is over all final-state particles. 
In terms of $\tau=1-T$, a more convenient quantity to describe dijet events, $\tau$ close to 0 corresponds to configurations with two narrow, pencil-like, back-to-back jets; while the other extreme $\tau=1/2$ corresponds to a spherically symmetric event. Since at $\sqrt{s}=10.58 \GeV$ the $B$ mesons decay nearly at rest in the c.m., a thrust cut of $\tau < 0.2$ removes $98\%$ of the $B$ data leaving the thrust distribution dominated by the fragmentation of light ($u d s$) and charmed quark pairs~\cite{Seidl:2008xc}. This cut on $\tau$ constrains the (squared) invariant masses $s_i$ of the final-state jets. Indeed, in the dijet limit, 
%%%
\begin{equation} \label{eq:tau}
 \tau=\frac{s_a+s_b}{Q^2} + \frac{k}{Q}
\end{equation}
%%%
where $k$ is the contribution from soft radiation between jets.

We focus on such restrictions on the hadronic final state, by studying the (spin-averaged) fragmentation of a light hadron $h$ inside a collimated jet originating from a light parton $i$, when the jet invariant mass 
is constrained. These features cannot be solely described by $D_i^h(x,\mu)$, which only depends on the momentum fraction $x$: in our case fragmentation is probed at a more differential level. In ref.~\cite{Procura:2009vm} a novel ``fragmenting jet function" $\cG_i^h(s,z,\mu)$ was introduced which depends both on the fragmentation variable $z$ and on the invariant mass $s$ of the collinear radiation that forms the jet. 
The relevant hierarchy of scales is given by $m_h  \ll \sqrt{s} \ll E_\jet$, where $E_\jet$ is the jet energy. This hierarchy allows us to employ Soft-Collinear Effective Theory (SCET)~\cite{Bauer:2000ew, Bauer:2000yr, Bauer:2001ct, Bauer:2001yt}, which is an effective field theory of QCD suitable for processes with well-separated energetic jets. Collinear and soft degrees of freedom describe, respectively, the energetic radiation inside jets and the soft emissions between them. A different collinear sector is associated with each jet. The collinear sectors and the soft sector decouple at leading power~\cite{Bauer:2001yt}. This leads to factorization formulae for inclusive observables at high energies which are characterized by convolutions of jet functions $J(s_i,\mu)$ (describing the invariant mass distribution of each jet) with a soft function $S$ encoding the contribution of the soft degrees of freedom, see e.g. \eq{factth} below.

The fragmenting jet function $\cG_i^h(s,z,\mu)$ has features of the standard fragmentation function $D_i^h(x,\mu)$ and the leading inclusive jet function $J_i(s,\mu)$, which is calculable in perturbation theory. 
As was shown in ref.~\cite{Procura:2009vm}, the following simple replacement rule holds
%%%
\begin{equation} \label{eq:repl}
J_i(s,\mu) \to \frac{1}{2 (2 \pi)^3}\, \cG_i^h(s,z,\mu)\, \df z 
\,,
\end{equation}
%%%
which allows us to obtain factorization formulae for semi-inclusive processes with fragmentation within a jet, from the corresponding inclusive ones. The factor $2(2 \pi)^3$ is related to the normalization of $\cG_i^h$  and to the phase space factor for the hadron $h$.

Here we will focus on the relation between ${\cG}_i^h(s,z,\mu)$ and $D_i^h(z,\mu)$. At leading order in $\lqcd^2/s \ll 1$, the fragmenting jet function can be expressed as a convolution between short-distance coefficients and the standard fragmentation functions at the scale $\mu_J \simeq \sqrt{s}$~\cite{Procura:2009vm},
%%%
\begin{align} \label{eq:GisD}
 {\cal G}_i^h(s,z,\mu_J) &= \sum_{j=g,\, u,\, \bar{u},\,d, \dots} \int_z^1 \frac{\df x}{x} \,
   {\cal J}_{ij}\Big(s,\frac{z}{x},\mu_J\Big)\,
   D_{j}^h(x,\mu_J) \,.
\end{align}
%%%
The ${\cal J}_{ij}$ describe the emission of collinear radiation, forming a jet with invariant mass $s$, within which the non-perturbative, long-distance fragmentation process takes place. 

In this paper we present the one-loop calculation of the matching coefficients ${\cal J}_{ij}$, where the initiating parton $i$ can be either an (anti)quark or a gluon. 
This completes the picture detailed in ref.~\cite{Procura:2009vm}
with the information necessary to relate factorization theorems for semi-inclusive processes, where the
jet invariant mass is probed, with the standard $D_{j}^h(x,\mu)$.

In the presentation of our results particular attention will be devoted to show how the infrared (IR) divergences of the partonic ${\cal G}_i$ and $D_i$ cancel in the matching. Cross-checks of our results are provided by the known anomalous dimensions of the fragmentation functions and the (fragmenting) jet functions, as well as by a relationship between $\cG_i(s,z,\mu)$ and $J_i(s,\mu)$, which we work out in \subsec{jet}.
As we will see explicitly, the $\cJ_{ij}$ contain double logarithms, \emph{i.e.}~contributions of the form $\al_s^n\ln^m(s/\mu_J^2)$ with $m \leq 2n$,  so that \eq{GisD} should be evaluated at $\mu_J \simeq \sqrt{s}$ to avoid the breakdown of the standard perturbative expansion.

\begin{figure}[t]
\centering
\includegraphics[width=0.9\textwidth]{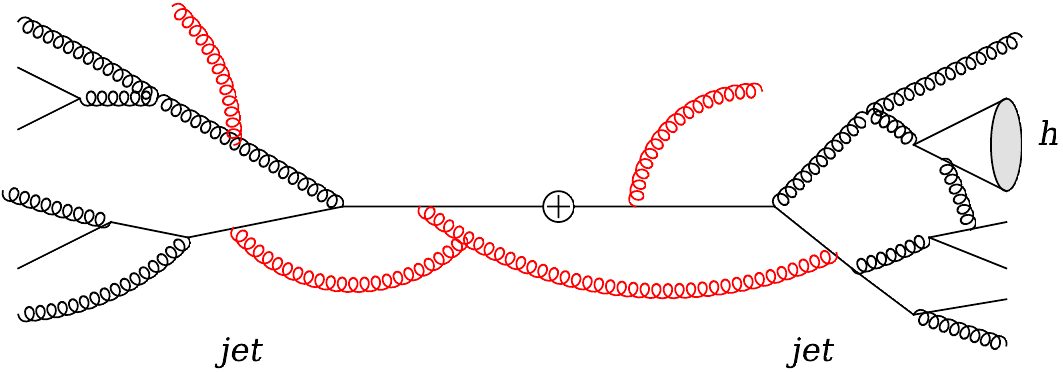}
\caption{Schematic display of the factorization in \eq{factth} for the fragmentation process $e^+ e^- \to {\rm dijet} + h$. The cross denotes the short-distance process $e^+ e^- \to q \bar q$ producing the back-to-back jets. In one of these jets a hadron of type $h$ is observed and its momentum fraction $z$ is measured. Each event may contribute more than once to the cross-section if the final state contains several of these hadrons $h(z_1), \dots, h(z_n)$. The dijet limit restricts the radiation to be either collinear or soft, drawn respectively in black and red color. At leading power, the collinear radiation is described by a (fragmenting) jet function, and the two jets only interact through soft radiation, described by the soft function.
}
\label{fig:eplemin}
\end{figure}

We will illustrate our results through a numerical analysis of the process $e^+ e^- \to X h$ where we restrict to the dijet limit by a cut on $\tau$, as in the Belle study of light quark fragmentation mentioned above. 
Using \eq{repl} in the leading-order factorization theorem for the cross-section where an inclusive measurement of thrust is performed for $\tau \ll 1$ \cite{Catani:1992ua,Korchemsky:1999kt,Fleming:2007qr,Schwartz:2007ib}, we obtain:
%%%
\begin{align} \label{eq:factth}
  \frac{\df^2 \si}{\df \tau\, \df z} &= 
\sum_q \frac{\sigma_0^q}{2 (2 \pi)^3}\,  H(Q^2, \mu) \int\! \df s_a\, \df s_b\,\df k\, \Big[\cG_q^h(s_a, z, \mu)\, J_{\bar{q}}(s_b,\mu) + J_q(s_a, \mu)\, \cG_{\bar{q}}^h(s_b,z,\mu)\Big]
 \nn \\
 & \quad \times S_\tau(k,\mu) \,\de\Big(\tau-\frac{s_a+s_b}{Q^2}-\frac{k}{Q} \Big) 
 \Big[1  + \ord{\tau}\Big]
 \nn \\
 &= \sum_{q,j} \frac{\sigma_0^q}{2 (2 \pi)^3}\,  H(Q^2, \mu) \int\! \df s_a\, \df s_b\,\frac{\df x}{x} \, \Big[\cJ_{q j} \Big(s_a, \frac{z}{x}, \mu \Big)\, J_{\bar{q}} (s_b,\mu) + J_q(s_a, \mu)\, \cJ_{\bar{q} j} \Big(s_b,\frac{z}{x},\mu \Big)\Big]
  \nn \\
 & \quad \times D_j^h(x,\mu)\, Q\, S_\tau \Big(Q \tau - \frac{s_a + s_b}{Q},\mu\Big) \bigg[1 + {\cal O} \Big(\tau, \frac{\lqcd^2}{\tau Q^2}\Big)\bigg]
  \,,
\end{align}
%%%
see \fig{eplemin}.
Here \eq{tau} is incorporated through the $\delta$-function.
The first line receives power corrections of $\ord{\tau}$ and in the second line we also have ${\cal O}[\lqcd^2/(\tau Q^2)]$ corrections from using \eq{GisD}.
We will only consider the contribution from the light quark flavors $q=u,\bar u, d, \bar d, s, \bar s$. The gluon fragmenting jet function does not appear in \eq{factth}, but the gluon fragmentation function does contribute because the sum over $j$ includes $j=g$.
The normalization factor $\sigma_0^q$ is the tree-level cross-section for the electroweak process $e^+ e^- \to (\gamma\,, Z) \to q \bar q$ given in \eq{si0}, which depends on the quark flavor. Since we assume that it is not known whether the observed hadron $h$ fragmented from the quark or the antiquark initiated jet, we have a sum over both possibilities in the factorization theorem. 

%%%
%%%

In \eq{factth}, the hard function $H(Q^2,\mu)$ encodes virtual effects arising from the production of the $q \bar q$ pair at the hard scale $\mu_H \simeq Q$, and is given by the square of Wilson coefficients in the matching of the relevant QCD onto SCET currents. 
The (real and virtual) collinear radiation of the jet from which the hadron fragments is described by $\cG_i^h$ or $\cJ_{ij}$ whereas the jet in the opposite hemisphere is represented through an inclusive jet function. The characteristic scale of these functions is the jet scale  $\mu_J \simeq \sqrt{\tau}Q$.
Finally, the soft function $S_\tau(k,\mu)$ describes the contribution to the hemisphere masses (and therefore to thrust) due to soft parton emissions.
$S_\tau$ is defined through the vacuum matrix element of eikonal Wilson lines and the corresponding soft scale is $\mu_S \simeq \tau Q$. 

In the two-jet limit $\tau \ll 1$, the cross section in \eq{factth} contains large double logarithms $\al_s^n \ln^m \tau$ $(m\leq 2n)$, which need to be resummed to make reliable predictions and uncertainty estimates. In our effective field theory approach this is achieved by evaluating the hard, (fragmenting) jet and soft functions at their natural scales $\mu_H$, $\mu_J$ and $\mu_S$ respectively, where they contain no large logarithms, and by running them to a common scale $\mu$ using their respective renormalization group equations (RGEs). In \eq{factth} all functions except the matching coefficients $\cJ_{ij}$ have already been studied in the literature. Our calculation therefore provides the missing ingredient necessary to sum these logarithms up to next-to-next-to-leading-logarithmic (NNLL) accuracy. We will discuss resummation effects in \eq{factth} for the case of unpolarized single charged pion production when the soft scale $\mu_S$ is perturbative. 

We stress that our results have a broad range of applicability. For example, \eq{repl} can be directly utilized when the hadronic final state is characterized via the event shape $N$-jettiness $\tau_N$~\cite{Stewart:2010tn}, since in this case the jets are described by the standard jet functions. This variable measures how $N$-jet-like an event is, and can be used to veto unwanted additional jets by requiring $\tau_N \leq \tau_N^\cut \ll 1$, which is the region of validity of the $N$-jettiness factorization theorem. In a more exclusive approach where jet algorithms are used, jet-algorithm-dependent jet functions arise~\cite{Ellis:2009wj,Ellis:2010rwa}. However, even for $N$-jettiness one may use a jet algorithm to determine the jet energies and directions, since all reasonable jet algorithms agree on these quantities in the exclusive $N$-jet regime $\tau_N \ll 1$, up to power corrections of $\ord{\tau_N}$~\cite{Stewart:2010tn}.

The paper is organized as follows. In section~\ref{sec:frag} we set up the theoretical framework, review the SCET definitions of fragmenting jet functions and standard fragmentation functions, and discuss their relationship and renormalization properties. We also discuss the relationship between the fragmenting jet function and jet function in detail. Our results for the matching coefficients $\cJ_{ij}$ are given in section \ref{subsec:OPE}. In section~\ref{sec:gluondelta} we present our calculation of the quark matching coefficients $\cJ_{q i}$ at one loop, where we 
show how the IR divergences cancel in the matching procedure. Section~\ref{sec:gluondimreg} is devoted to the gluon case at next-to-leading order (NLO). Numerical results for the fragmenting jet functions using \eq{GisD}, as well as, a numerical analysis of pion fragmentation in dijet-like $e^+ e^- \to X \pi^+$ with a cut on thrust are contained in section~\ref{sec:appl}. Conclusions and outlook are given in section~\ref{sec:concl}. Useful mathematical identities are given in appendix~\ref{app:plusdist}. An alternative calculation of $\cJ_{qi}$ using the optical theorem and a different IR regulator can be found in appendix~\ref{app:offshell}. All the ingredients necessary for our numerical analysis are collected in appendix~\ref{app:pert}.

%%%%%%%%%%%%%%%%%%%%%%%%%%%%%%%%%%%%%%%%%%%%%%%%%%%%%%%%%%%%%%%%%%%%%%%%%%%%%%%%
\section{Fragmentation within an Identified Jet}
\label{sec:frag}
%%%%%%%%%%%%%%%%%%%%%%%%%%%%%%%%%%%%%%%%%%%%%%%%%%%%%%%%%%%%%%%%%%%%%%%%%%%%%%%%

In this section we start by setting up the theoretical framework of our analysis, and introduce the SCET ingredients relevant for this paper.
We give the definitions of quark and gluon fragmentation functions in SCET and discuss their renormalization. We then focus on quark and gluon fragmenting jet functions, discuss their renormalization and their relationship with the standard fragmentation functions. At the end of this section, we consider their relationship with the jet function, which provides a powerful cross-check on our calculations.

%===============================================================================
\subsection{SCET Ingredients}
\label{subsec:SCET}
%===============================================================================

Light quark or gluon fragmentation within an identified jet is governed by three different scales: the (perturbative) \emph{hard scale} set by the jet energy $E_{X h}$, the intermediate (perturbative) \emph{jet scale} given by the jet invariant mass $m_{X h}$, and the \emph{soft scale} of order $\sim m_{X h}^2/E_{X h}$, with the hierarchy $m_h \ll m_{X h} \ll E_{X h}$; here we will always consider the hadron mass $m_h$ to be negligible as in ref.~\cite{Procura:2009vm}. 
After integrating out the hard dynamics that initiates the jet, we are left with collinear and soft modes in the $X h$ system. Therefore, SCET~\cite{Bauer:2000ew, Bauer:2000yr, Bauer:2001ct, Bauer:2001yt} -- which is an effective field theory of QCD that describes physics of collinear and soft degrees of freedom -- is well suited for this analysis.

Since the invariant mass of a jet is much smaller than its energy, the jet constituents are collimated and conveniently described using light-cone coordinates. To this end we introduce a light-cone vector $n^\mu$ whose spatial part is along the jet axis, and another light-cone vector $\bn^\mu$ such that $n^2 = \bn^2 = 0$ and $n\sdt\bn = 2$. Any four-vector $p^\mu$ can then be decomposed as $p^\mu = (p^+, p^-, p^\mu_\perp)$ with $p^+ = n\sdt p$, $p^- = \bn\sdt p$ and $p_\perp^\mu$, which contains the components of $p^\mu$ perpendicular to $n^\mu$ and $\bn^\mu$. The momentum $p^\mu$ of a particle within the jet scales collinearly, {\it i.e.} $p^\mu = (p^+, p^-, p_\perp^\mu) \sim p^- (\la^2, 1, \la)$, where $\la \sim m_{X h}/E_{{X h}} \ll 1$ is the SCET expansion parameter. For the soft degrees of freedom, the momentum scales like $q^\mu = (q^+, q^-, q_\perp^\mu) \sim p^- (\la^2, \la^2, \la^2)$. 

The collinear momentum $p^\mu$ is separated into a large part and a small residual part
%%%
\begin{equation}
p^\mu = \lp^\mu + p_r^\mu = \bn\cdot\lp\, \frac{n^\mu}{2} + p_{\ell\perp}^\mu + p_r^\mu
\,,\end{equation}
%%%
with $\lp^\mu = (0, \lp^-, p_{\ell\perp}) \sim \lp^- (0, 1, \la)$ and $p_r^\mu = (p_r^+, p_r^-, p_{r\perp}^\mu) \sim \lp^-(\la^2, \la^2, \la^2)$. The SCET fields for $n$-collinear quarks and gluons, $\xi_{n,\lp}(y)$ and $A_{n,\lp}(y)$ respectively, are labeled by $n$ and the label momentum $\lp$. Their argument $y$ is conjugate to the small residual momenta.
A derivative acting on these fields picks out the residual momentum dependence, $\img \partial^\mu \sim p_r^\mu \sim \la^2 \lp^-$, while label momentum operators $\bnP_n = \bn \sdt \cP_n$ ($\cP_{n\perp}^\mu$) return the sum of the minus (perpendicular) label components of all $n$-collinear fields on which they act.

Interactions between collinear fields cannot change the direction $n$ but change the momentum labels. 
It is therefore convenient to use the short-hand notation
%%%
\begin{equation} \label{eq:xi}
\xi_n(y) = \sum_{\lp \neq 0} \xi_{n,\lp}(y)
\,,\qquad
A_n^\mu(y) = \sum_{\lp \neq 0} A^\mu_{n,\lp}(y)
\,.\end{equation}
%%%
In the sum we explicitly exclude the case $\lp^\mu = 0$ to avoid double-counting of the soft degrees of freedom (which are described by separate soft quark and gluon fields). In practice, when calculating matrix elements, this is implemented using zero-bin subtractions~\cite{Manohar:2006nz} or alternatively by dividing out matrix elements of Wilson lines~\cite{Collins:1999dz, Lee:2006nr, Idilbi:2007ff}. We will study the zero-bin subtractions in detail since they play an important role in our calculation. This is most explicitly seen in section \ref{sec:gluondelta}, where we use a gluon mass and a $\de$-regulator \cite{Chiu:2009yz} to regulate the IR divergences in the one-loop quark fragmenting jet function.

Collinear operators are built out of products of fields and Wilson lines that are invariant under collinear gauge transformations~\cite{Bauer:2000yr,Bauer:2001ct}. The basic building blocks are the collinearly gauge-invariant quark and gluon fields, defined as
%%%
\begin{equation} \label{eq:chiB}
\chi_n(y) = W_n^\dagger(y)\, \xi_n(y)
\,,\qquad
\cB_{n\perp}^\mu(y) = \frac{1}{g} \bigl[W_n^\dagger(y)\, \img D_{n\perp}^\mu W_n(y) \bigr]
\,,\end{equation}
%%%
where $\img D_{n\perp}^\mu = \cP^\mu_{n\perp} + g A^\mu_{n\perp}$ is the $\perp$-collinear covariant derivative. The collinear Wilson line
%%%
\begin{equation} \label{eq:Wn}
W_n(y) = \biggl[\sum_\text{perms} \exp\Bigl(-\frac{g}{\bnP_n}\,\bn\sdt A_n(y)\Bigr)\biggr]
\end{equation}
%%%
sums up arbitrary emissions of $n$-collinear gluons from an $n$-collinear quark
or gluon, which are $\ord{1}$ in the power counting.

At leading order in the SCET power expansion, the interactions of soft gluons with collinear fields exponentiate to form eikonal Wilson lines. The soft gluons can thus be decoupled via the BPS field redefinition~\cite{Bauer:2001yt}
%%%
\begin{align} \label{eq:BPS}
\chi^\zero_{n,\w}(y) &= Y_n^\dagger(y)\,\chi_{n,\w}(y)
\,, \nn \\
\cB^{\mu\zero}_{n,\w\perp}(y) &= Y_n^\dagger(y)\,\cB^\mu_{n,\w\perp}(y)\,Y_n(y)
\,.\end{align}
%%%
The collinear fields we consider in this paper are those after this decoupling, and we drop the superscript $(0)$ for notational convenience. Here $Y_n(y)$ is a soft Wilson line in the fundamental representation 
%%%
\begin{align} \label{eq:Yn}
Y_n(y) &= \bar P\exp\biggl[-\img g\int_0^\infty \df u\, n\sdt A_{us}(y + u\,n) \biggr]  
\nn\\
&= \biggl[\sum_\text{perms} \exp\Bigl(-\frac{g}{\img n \sdt \partial}\, n\sdt A_{us}(y)\Bigr)\biggr]
\,.\end{align}
%%%
The symbol $\bar P$ in \eq{Yn} denotes anti-path ordering of the color generators along the integration path.  On the second line we write the Wilson line in momentum space akin to \eq{Wn}.

%===============================================================================
\subsection{Fragmentation Functions}
\label{subsec:frag}
%===============================================================================

The fragmentation functions $D_i^h(x)$ characterize the factorization theorems that describe high-energy single-inclusive hadron production processes at leading power~\cite{Collins:1989gx}, where no properties of the jet are probed, see e.g.~\eq{eeXh}. These functions encode the non-perturbative information on how the energetic parton $i$ (either a gluon or an (anti)quark of a certain flavor) produces the observed hadron $h$ which carries a fraction $x$ of the initial parton's large light-cone momentum component.

Let $k^\mu$ and $p^\mu_h$ denote the parton and hadron momenta, respectively. In a frame where $\vec k_{\perp}=0$, the hadron has $p_h^- \equiv x\, k^-$ and $p_h^+=(\vec{p}_{h\perp}^{\, 2} +m_h^2)/p_h^-$. With the gauge choice $\bar{n} \cdot A=0$, the bare unpolarized quark fragmentation function has the following operator definition in QCD~\cite{Collins:1981uw}
%%%
\begin{align} \label{eq:defDqQCD} 
 D_{q,\bare}^h(x)&=\frac{1}{x}\int\! \df^2 p_h^\perp 
\int\! \frac{\df y^+\,\df^2 y_\perp}{ 2(2 \pi)^3}\;e^{\,i k^-  y^+/2}
\sum_X  \frac{1}{2 N_c}\, \tr\Big[  \frac{\bnslash}{2}\,
  \langle 0 |  \psi (y^+, 0, y_\perp )| X h
  \rangle \langle X h | \bar{\psi}(0) |0 \rangle \Big] \,,
\end{align}
%%%
where $\psi$ is the quark field quantized on $y^-=0$ and the trace is taken over color and Dirac indices. The factor $1/(2N_c)$, where $N_c=3$ is the number of colors, comes from averaging over the color and spin of the parent parton. The state $|Xh\rangle = | X h(p_h)\rangle$ contains a hadron $h$ with momentum $p_h$, and a sum over the polarizations of $h$ is assumed. Boost invariance along the non-$\perp$ direction implies that $D$ can only be a function of $x=p_h^-/k^-$ and not $p_h^-$ or $k^-$ individually. According to factorization at leading power, the sum over the accompanying hadrons $X$ is dominated by jet-like configurations for the $|X h \rangle$ states~\cite{Collins:1989gx}.
 
In SCET notation, the fragmentation function takes on the following form~\cite{Procura:2009vm}
%%%
\begin{align} \label{eq:Dqdef}
  D_{q,\bare}^h(x) & = 
  \frac{1}{ x} \int\! \df^2 p_h^\perp  \sum_X \frac{1}{2 N_c} \tr
  \Big [\frac{\bnslash}{2}  \de(p_{X h, r}^-)\de^2(p_{X h, r}^\perp) \Mae{0}{[\de_{\w, \bnP} \,\de_{0, \cP_\perp} \chi_n(0)]} {X h}
  \Mae{X h}{\bar \chi_n(0)}{0} \Big ]\,.
\end{align}
%%%
Here, $\chi_n$ is the $n$-collinear quark field in \eq{chiB} that contains a Wilson line, making this definition (collinearly) gauge invariant. The $\bnP$ and $ {\cal{P}}_\perp$ operators pick out the ${\cal O}(\lambda^0)$ and ${\cal O}(\lambda)$ label momentum of the field, while the continuous ${\cal O}(\lambda^2)$ residual components of the jet momentum are denoted by $p_{X h, r}^\mu$. We use the notation $p_{X h}^\mu = p_X^\mu +p_h^\mu$.

The QCD definition for the bare gluon fragmentation function, in the $\bar{n} \cdot A=0$ gauge and in a frame where $p_{X h}^\perp=0$, for $d$ space-time dimensions, reads~\cite{Collins:1981uw}
%%%
\begin{align} \label{eq:defDgQCD} 
D_{g,\bare}^h(x)& = -\frac{1}{(d-2)(N_c^2-1) p_h^-} \int\! \df^2 p_h^\perp\! \int\! \frac{\df y^+\,\df^2 y_\perp}{ 2(2 \pi)^3}\;e^{\,i k^-  y^+/2}
\nn \\ & \quad \times
 \sum_X \bar{n}^\mu \bar{n}^\nu \langle 0 | G^a_{\mu \lambda}(y^+, 0, y_\perp ) | X h
  \rangle \langle X h | G^{\lambda,a}_\nu(0) |0 \rangle 
\end{align}
%%%
where $G_{\mu \nu} = \sum_a G_{\mu \nu}^a\, T^a$ is the QCD field-strength tensor and an average over colors and the $(d-2)$ polarizations of the gluon is performed. The corresponding expression in SCET is given in gauge invariant form by
%%%
\begin{align} \label{eq:Dgdef}
  D_{g,\bare}^h(x) & =   -\frac{\w}{(d-2)(N_c^2-1)\, x}  \int\! \df^2 p_h^\perp  \sum_X
 \de(p_{X h, r}^-)\, \de^2(p_{X h, r}^\perp)\,
\nn \\ & \quad \times 
    \Mae{0}{[\de_{\w, \bnP} \,\de_{0, \cP_\perp} \cB_{n\perp}^{\mu,a}(0)]} {X h}
  \Mae{X h}{\cB_{n\perp, \mu}^a(0)}{0} \,.
\end{align}
%%%

The operator products in the definitions of $D_q^h(x)$ and $D_g^h(x)$ are singular and require renormalization. The renormalized fragmentation functions are defined through~\cite{Collins:1981uw}
%%%
\begin{align} \label{eq:D_ren}
  D_{i,\text{bare}}^h(x) 
  & = \sum_{j=g, u, \bar{u}, d, \dots} \int_x^1 \frac{\df x'}{x'}\, Z^D_{ij}\Big(\frac{x}{x'},\mu\Big) D^h_j(x',\mu)
  \,.
\end{align}
%%%
Throughout this paper $\mu$ denotes the scale of dimensional regularization in the $\overline{\rm MS}$ scheme. The renormalization-group equation (RGE) for $D^h_i(x,\mu)$ follows from \eq{D_ren}
%%%
\begin{align}
  \mu \frac{\df} {\df\mu} D_i^h(x,\mu) 
  = \sum_j \int_x^1 \frac{\df x'}{x'}\, \ga^D_{ij}\Big(\frac{x}{x'}, \mu\Big) D_j^h(x', \mu)
  \,,
\end{align}
%%%
with anomalous dimension
%%%
\begin{align} \label{eq:gammaD}
  \gamma^D_{ij}(x,\mu) = -\! \int_x^1 \frac{\df x'}{x'}\, \big(Z^D\big)^{-1}_{ik}\Big(\frac{x}{x'},\mu\Big) \,
  \mu\frac{\df}{\df\mu} Z^D_{kj}(x',\mu) 
  \,.
\end{align}
%%%
Here, the inverse of the renormalization factor $(Z^D)^{-1}_{ik}$ is defined through
%%%
\begin{equation}
  \sum_k \int_x^1\! \frac{\df x'}{x'}\, \big(Z^D\big)^{-1}_{ik}\Big(\frac{x}{x'},\mu\Big) \,Z^D_{kj}(x',\mu) 
  = \de_{ij}\, \de(1-x)
  \,.
\end{equation}
%%%

In our perturbative calculations we will replace the hadron $h$ by either a quark or a gluon. The other ingredients of a factorization theorem are not affected by this, e.g.~the hard, jet and soft function in \eq{factth} are the same in both cases. Since the cross section is an observable, the $\mu$ dependence must cancel between all the factors of a factorization theorem. From this it follows that the renormalization and anomalous dimension of $D$ are the same if $h$ is a hadron or a parton.

We will use the variable $p$ for the momentum of the parton that replaces the hadron $h$. 
Denoting the discrete label parts of the momentum $p^\mu$ by $\lp^-$ and $p_{\ell\perp}$ and the continuous residual parts by $p_r^\mu$, the partonic fragmentation functions are at tree-level given by
%%%
\begin{align} \label{eq:Dtree}
  D_q^{q\zero}(x) & = 
  \frac{1}{x}\, \sum_{p_{\ell\perp}} \int\! \df^2p_r^\perp
 \frac{1}{2N_c} \, \tr\Big[\frac{\bnslash}{2} \de_{\w, p_\ell^-} \de_{0,p_\ell^\perp} \de(p_r^-) \de^2(p_r^\perp)\, \Mae{0}{ \xi_n(0)} {q_n(p)}
  \Mae{q_n(p)}{\bar \xi_n(0)}{0} \Big]
  \nn \\
  & = \frac{1}{2 x} \, \de(\w-p^-)\,  \tr\Big[\frac{\bnslash}{2} \sum_s u_n^s(p) \bar u_n^s(p)\Big] = \de(1-x)
  \,, \nn \\  
  D_g^{g\zero}(x) & = 
  -\frac{\w}{(d-2)(N_c^2-1)\, x} \, \sum_{p_{\ell\perp}} \int\! \df^2p_r^\perp
  \,\de_{\w, p_\ell^-} \de_{0,p_\ell^\perp} \de(p_r^-) \de^2(p_r^\perp)\,
  \nn \\ & \quad \times
  \Mae{0}{ A_{n\perp}^{\mu, a}(0)} {g_n(p)}
  \Mae{g_n(p)}{A_{n\perp, \mu}^a(0)}{0}
  \nn \\
  & = -\frac{\w}{(d-2)x} \de(\w-p^-)\, \sum_\text{pol} \ve_{n\perp}^*(p) \sdt \ve_{n\perp}(p) = \de(1-x)
  \,.
\end{align}
%%%
Here we recombined residuals and labels into the
continuous $p^-$, via 
%%%
\begin{equation}
\delta_{\omega,p_\ell^-} \,\delta(p_r^-) = \delta(\omega -p^-)~.
\end{equation}
%%%
We will use this relation in the rest of the paper. The partonic $D_q^g$ and $D_g^q$ vanish at tree level. At one-loop their UV divergences lead to the mixing of quark- and gluon fragmentation functions. The details of this calculation are given in section \ref{sec:gluondelta}.

The one-loop renormalization factors $Z^{D\one}_{ij}(x,\mu)$ can be easily extracted once the one-loop partonic result is known. Expanding the partonic version of \eq{D_ren} to one-loop,
%%%
\begin{align} \label{eq:DrenNLO}
  D_{i,\text{bare}}^{j\one}(x) & = \sum_k \int_{x}^1\! \frac{\df x'}{x'}\, 
  \Big[Z^{D\zero}_{ik}\Big(\frac{x}{x'},\mu\Big) \,D_k^{j \one}(x',\mu) + 
  Z^{D\one}_{ik}\Big(\frac{x}{x'},\mu\Big) \,D_k^{j\zero}(x',\mu)\Big]
  \nn \\
  & = \sum_k \int_x^1\! \frac{\df x'}{x'}\, 
  \Big[ \de_{ik}\, \de\Big(1-\frac{x}{x'}\Big)\, D_k^{j\one}(x',\mu) + 
  Z^{D\one}_{ik}\Big(\frac{x}{x'},\mu\Big)\, \de_{kj}\, \de(1-x') \Big]
  \nn \\
  & = D_i^{j \one}(x,\mu) + Z^{D\one}_{ij}(x,\mu)
  \, , 
\end{align}
%%%
where the superscripts $(0)$ and $(1)$ denote the tree-level and one-loop expressions, respectively.
From \eqs{gammaD}{DrenNLO} the one-loop $\ga^D_{ij}(x,\mu)$ can then be obtained straightforwardly.
At ${\cal O}(\alpha_s)$ the space-like and time-like Altarelli-Parisi evolution kernels are related to each other via a simple analytic continuation rule \cite{Drell:1969jm,Drell:1969wb} based on symmetries of the relevant diagrams under crossing, and via the so-called Gribov-Lipatov reciprocity relation which connects space-like and time-like structure functions in their respective physical regions \cite{Gribov:1972ri,Gribov:1972rt}. As a consequence, the $\ga^D_{ij}(x,\mu)$ coincide with the  anomalous dimensions of the parton distribution functions at one loop (where the roles of incoming and outgoing partons are interchanged). Therefore,
%%%
\begin{align} \label{eq:gaD}
  \ga^D_{qq}(x,\mu) & = \frac{\al_s(\mu) C_F}{\pi}\, \theta(x) P_{qq}(x)
  \,, \nn \\
  \ga^D_{qg}(x,\mu) &= \frac{\al_s(\mu) C_F}{\pi} \theta(x) P_{gq}(x)
  \,, \nn \\
  \ga^D_{gg}(x,\mu) &= \frac{\al_s(\mu)}{\pi}\,
  \theta(x)\, \Bigl[C_A P_{gg}(x) + \frac{1}{2} \beta_0\, \delta(1-x)\Bigr]
  \,, \nn \\
  \ga^D_{gq}(x,\mu) &= \frac{\al_s(\mu) T_F}{\pi}\,
  \theta(x)\, P_{qg}(x)
  \,,  
\end{align}
%%%
where $\beta_0 = (11 C_A - 4 n_f T_F)/3$, is the lowest order coefficient of the QCD $\beta$-function. The splitting functions are \cite{Altarelli:1977zs}
%%%
\begin{align} \label{eq:split}
  P_{qq}(x) &= \Big(\frac{1+x^2}{1-x}\Big)_+
  = (1+x^2)\, \cL_0(1-x) + \frac{3}{2}\, \de(1-x)
  \,, \nn \\
  P_{gq}(x) &= \theta(1-x) \,\frac{1+(1-x)^2}{x}
  \,, \nn \\
  P_{gg}(x)
  &= 2x\, \cL_0(1-x) + 2\,\theta(1-x)\Bigl[\frac{1-x}{x} +  x(1-x)\Bigr]
  \,,\nn\\
  P_{qg}(x) &= \theta(1-x)\, [x^2+(1-x)^2]
  \,,
\end{align}
%%%
in which we do not include the usual color factors for future convenience. The plus-distribution
%%%
\begin{equation}
\cL_0(1-x) = \bigg[\frac{\theta(1-x)}{1-x}\bigg]_+
\end{equation}
%%%
is defined in \eq{plusdef}. In section \ref{sec:gluondelta} we will show how our partonic calculation of $D_q^{j (1)}(x,\mu)$ also leads to \eq{gaD}.
The relation $\ga^D_{ij}(x,\mu) = \ga^f_{ji}(x,\mu)$ was shown no longer to hold beyond one loop in dimensional regularization and the $\overline{\rm MS}$ scheme in ref.~\cite{Curci:1980uw}, see also the discussion in ref.~\cite{Stratmann:1996hn}.

%===============================================================================
\subsection{Fragmenting Jet Functions}
\label{subsec:fragjet}
%===============================================================================

A high-energy single-inclusive hadron production process, where the invariant mass of the jet initiated by a light parton is measured, is described by a fragmenting jet function. We consider the case that the hadron mass is negligible (compared to the jet mass) and work in a frame where the perpendicular momentum of the jet is vanishing. 
For a fragmenting quark~\cite{Procura:2009vm}
%%%
\begin{align} \label{eq:Gqdef}
  \cG_{q,\bare}^h(s,z) & = 
  \int\! \df^4 y\, e^{\img k^+ y^-/2}\, \int\! \df p_h^+
\, \sum_X \,\frac{1}{4N_c}\,  \tr \,
  \Big[\frac{\bnslash}{2} \Mae{0}{[\de_{\w,\bnP}\, \de_{0,\cP_\perp} \chi_n(y)]} {X h}
  \Mae{X h}{\bar \chi_n(0)}{0} \Big]
  \nn \\
  & = 
  \frac{2(2\pi)^3}{ p_h^-} \int\! \frac{\df y^-}{4\pi}\, e^{\img k^+ y^-/2}
  \int\! \df^2 p_h^\perp\, \sum_X  \frac{1}{2 N_c} \tr  
  \Big[\frac{\bnslash}{2} \de(p_{X h, r}^-)\, \de^2(p_{X h, r}^\perp)\, 
  \nn \\ & \quad \times
  \Mae{0}{[\de_{\w,\bnP}\, \de_{0,\cP_\perp} \chi_n(y^-)]} {X h}
  \Mae{X h}{\bar \chi_n(0)}{0} \Big]~.
\end{align}
%%%
In the second equality we performed a translation of the collinear field $\chi_n(y)$, whose argument is associated with residual momenta, and carried out the integrals in $y^+$ and $y^\perp$. The integration over $y^-$ fixes the partonic jet invariant mass $s=k^+ \w$. Analogously, in the case of a gluon-initiated jet, in $d$ space-time dimensions,
%%%
\begin{align} \label{eq:Ggdef}
 \cG_{g,\bare}^h(s,z) & = 
  -\frac{2(2\pi)^3\,\w}{(d-2) (N_c^2-1)\, p_h^-} 
  \int\! \frac{\df y^-}{4\pi}\, e^{\img k^+ y^-/2}
  \int\! \df^2 p_{h}^\perp\, \sum_X
 \de(p_{X h, r}^-)\, \de^2(p_{X h, r}^\perp)\,
  \nn \\ & \quad \times
  \Mae{0}{ [\de_{\w, \bnP} \,\de_{0,\cP_\perp} \cB_{n\perp}^{\mu, a}(y^-)]} {X h}
 \Mae{X h}{\cB_{n\perp, \mu}^a(0)}{0} \,.
\end{align}
%%%
Evaluating the partonic fragmenting jet functions at tree-level, we find
%%%
\begin{align} \label{eq:Gtree}
  \cG_q^{q\zero} (s,z) 
  &= \frac{2 (2\pi)^3}{p^-}\, \de(k^+)\, \de(\w - p^-)\, \frac{1}{2}\, \tr  \Big[\frac{\bnslash}{2}  \sum_\text{spins} u_n(p) \bar u_n(p)\Big]
  =2 (2\pi)^3 \de(s) \de(1-z) 
\,,  \nn \\  
  \cG_g^{g\zero} (s,z)
  &= -\frac{2 (2 \pi)^3\, \w}{p^-} \, \de(k^+)\, \de(\w - p^-) \, \frac{1}{d-2} \sum_\text{pol} \ve_{n\perp}^* \sdt \ve_{n\perp} 
  = 2 (2\pi)^3 \de(s) \de(1-z) 
  \,.
\end{align}
%%%
Here we have used the fact that $p^+=0$, due to the on-shell condition $p^2=0$ and the choice of frame, which sets $p_\perp=0$.

A consequence of \eq{repl} is that the renormalization and RG evolution of these two functions are the same
%%%
\begin{align} \label{eq:ZGdef}
  \cG_{i,\text{bare}}^h (s,z) & = \int_0^s\! \df s'\, Z_\cG^i(s - s',\mu)\, \cG_i^h(s',z,\mu)
  \,, \qquad
  Z_\cG^i(s,\mu) =   Z_J^i(s,\mu)
  \,,
\end{align}
%%%
where the index $i$ is not summed over. In particular, the renormalization of $\cG_i^h$ does not affect its $z$-dependence and does not mix quark and gluon fragmenting jet functions, at any order in perturbation theory. We will see this explicitly in our one-loop calculation in sections \ref{sec:gluondelta} and \ref{sec:gluondimreg}. The corresponding RGE is given by
%%%
\begin{align} \label{eq:GRGE}
  &\mu \frac{\df}{\df \mu} \cG_i^h(s, x, \mu) = \int_0^s\! \df s'\, \ga_\cG^i(s-s',\mu)\, \cG_i^h(s', x, \mu)
\end{align}
%%%
where
%%%
\begin{align} \label{eq:gammaG}
   \ga_\cG^i(s,\mu) &= - \int_0^s\! \df s'\, \big(Z_\cG^i\big)^{-1}(s-s',\mu)\; \mu\frac{\df}{\df\mu} Z_\cG^i (s',\mu)
  \,,
\end{align}
%%%
and $\big(Z^i_\cG\big)^{-1}$ is defined as
\begin{equation}
  \int_0^s\! \df s'\, \big(Z^i_\cG\big)^{-1}(s-s',\mu)\, Z_\cG^i(s',\mu) = \de(s)
  \,.
\end{equation}
%%%
The structure of the anomalous dimension of the jet function $\ga_J^i(\al_s)$ implies
%%%
\begin{align} \label{eq:gaG}
  \ga_\cG^i(s,\mu) & = -2\Ga_\cusp^i(\al_s)\, \frac{1}{\mu^2} \cL_0\Big(\frac{s}{\mu^2}\Big) +
  \ga_\cG^i(\al_s)\, \de(s)
   \,, \qquad
  \ga_\cG^i(\al_s) = \ga_J^i(\al_s) 
   \,,
\end{align}
%%%
where the plus distribution $\cL_0$ is defined in \eq{plusdef}.
The cusp anomalous dimension $\Ga_\cusp^i(\al_s)$ \cite{Korchemsky:1987wg} and the non-cusp part of the anomalous dimension $\ga_\cG^i(\al_s)$ are collected in the \app{pert}, to make the paper self-contained in view of the numerical analysis in \sec{appl}. A cross-check of our partonic one-loop calculation of $\cG_i^{j \one}(s,z, \mu)$ will be provided by the one-loop evolution kernels $\ga_\cG^i(s,\mu)$ in \eq{gaG}. From the anomalous dimension we can obtain the one-loop renormalization factor $Z_\cG^{i\one}$ through \eq{gammaG}. This can be compared with our calculation by using \eq{GRGE} expanded to one loop:
%%%
\begin{align} \label{eq:GrenNLO}
  \cG_{i,\text{bare}}^{j \one}(s,z) 
  & = \int\! \df s'\, \Big[Z^{i\zero}_\cG (s-s',\mu) \, \cG_i^{j \one}(s',z,\mu) + 
  Z^{i\one}_\cG(s-s',\mu)\, \cG_i^{j \zero}(s',z,\mu)\Big]
  \nn \\
  & = \int\! \df s'\, \Big[\de(s-s')\, \cG_i^{j \one}(s',z,\mu) + 
  Z^{i\one}_\cG(s-s',\mu)\, \delta_{i j} \,\de(s')\, \de(1-z) \Big]
  \nn \\
  & = \cG_i^{j\one}(s,z,\mu) + 
  Z^{i\one}_\cG(s,\mu) \, \delta_{i j} \,\de(1-z) 
  \,.
\end{align}
%%%
Here we have noted that $\cG_i^j$ contributes at tree level only when $i=j$. Therefore for $i \ne j $,  $\cG_i^j$ are UV finite at one loop.

%===============================================================================
\subsection{Results for Matching onto Fragmentation Functions}
\label{subsec:OPE}
%===============================================================================

%%%
\begin{figure}[t]
\centering
\includegraphics[width=0.85\textwidth]{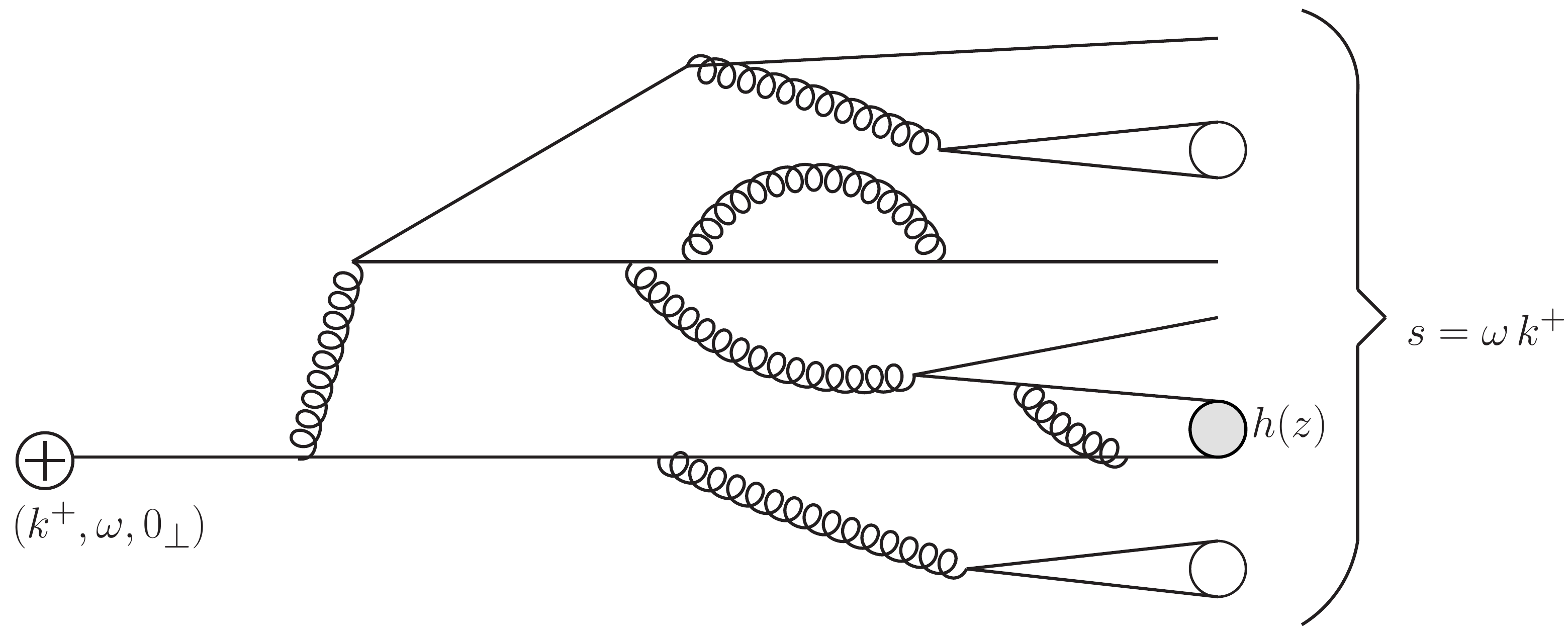}
\caption{The information encoded in $\cG_q^h(s, z,\mu)$ is exemplified here. The incoming quark creates a jet of invariant mass $s$, inside which a hadron $h$ with momentum fraction $z=p_h^-/\omega$ is produced. Initially, the large parton virtualities yield emissions at wider angles. This depends on $s$ and $z$, and can be described perturbatively by $\cJ_{ij}(s,z,\mu)$. At smaller parton virtualities, the emission is at smaller angles and essentially only affects $z$. Here the effect of hadronization also becomes important, and this is described by the standard fragmentation functions $D_j^h(z,\mu)$.}
\label{fig:fraginjet}
\end{figure}
%%%

By performing operator product expansions of the fragmenting jet functions in \eqs{Gqdef}{Ggdef}
about the $y^- \to 0$ limit, we can match onto the low-energy matrix elements in \eqs{Dqdef}{Dgdef} that correspond to the fragmentation functions. This amounts to the \SCETa onto \SCETb matching  (illustrated in \fig{fraginjet}) at the intermediate scale provided by the jet invariant mass $\mu_J \simeq \sqrt{s}$:
%%%
\begin{align} \label{eq:OPE}
  \cG_i^h(s,z,\mu_J) 
  & = \sum_{j=g,u, \bar u, d, \dots} \int_z^1 \frac{\df z'}{z'}
   \cJ_{ij}\Big(s,\frac{z}{z'},\mu_J\Big) D_j^h(z',\mu_J)
   \bigg[1+ \ORd{\frac{\lqcd^2}{s}}\bigg]
  \,.
\end{align}
%%%
The dependence on $z$ and $z'$ in $\cJ_{ij}$ is only through their ratio, since these coefficients can only depend on the perturbative variables associated with the partons $i$ and $j$, and not on the hadron $h$. Eq.~(\ref{eq:OPE}) is analogous to the matching of beam functions onto parton distribution functions performed in refs.~\cite{Fleming:2006cd,Stewart:2010qs}.

From the tree-level results in \eqs{Dtree}{Gtree}, we find by using \eq{OPE} that
%%%
\begin{align}
  \cJ_{ij}^{(0)}(s,z, \mu_J) = 2(2\pi)^3\, \de_{ij}\, \de(s)\,\de(1-z)
  \,.
\end{align}
%%%
This can simply be understood that at tree-level the hadron directly fragments from the parton $i$, without emitting radiation that would build up the jet.

The main purpose of this paper is to calculate $\cJ_{ij}(s,z,\mu_J)$ at one-loop. This completes the picture detailed in ref.~\cite{Procura:2009vm} with the information necessary to relate the factorization theorems for semi-inclusive processes, where the jet invariant mass is probed, to the standard $D_i^h(z,\mu)$ at NLO accuracy. We find that the one-loop matching coefficients $\cJ_{ij}^\one$ are given by
%%%
\begin{align} \label{eq:Jresult}
\frac{\cJ_{qq}^\one (s,z,\mu_J)}{2(2\pi)^3}
& = \frac{\al_s(\mu_J) C_F}{2\pi}\, \theta(z)\, \bigg\{
\frac{2}{\mu_J^2} \cL_1\Big(\frac{s}{\mu_J^2}\Big) \de(1\!-\!z) 
\!+\! \frac{1}{\mu_J^2} \cL_0\Big(\frac{s}{\mu_J^2}\Big) (1\!+\!z^2) \cL_0(1\!-\!z) 
\\ & \quad 
+ \de(s) \Big[(1+z^2)\cL_1(1-z) + P_{qq}(z) \ln z + \theta(1-z) (1-z) 
-\frac{\pi^2}{6}\de(1-z) \Big] \bigg\}
\,, \nn \\
\frac{\cJ_{qg}^\one (s,z,\mu_J)}{2(2\pi)^3}
&= \frac{\al_s(\mu_J) C_F}{2\pi}\, \theta(z)\, \bigg\{\Big[\frac{1}{\mu_J^2} \cL_0\Big(\frac{s}{\mu_J^2}\Big) +
\de(s) \ln{(z(1-z))} \Big] P_{gq}(z) + \de(s)\, \theta(1-z) z \bigg\} 
\,, \qquad \nn
\end{align}
%%%
%%%
\begin{align} \label{eq:Jgresult}
\frac{\cJ_{gg}^{\one}(s,z,\mu_J)}{2(2\pi)^3}
& = \frac{\al_s(\mu_J) C_A}{2\pi} \theta(z) \bigg\{ 
\frac{2}{\mu_J^2}\cL_1\Big(\frac{s}{\mu_J^2}\Big) \de(1-z) 
+ \frac{1}{\mu_J^2} \cL_0\Big(\frac{s}{\mu_J^2}\Big) P_{gg}(z)
\\ & \quad
 + \de(s) \Big[ \cL_1(1-z) \frac{2(1-z+z^2)^2}{z} + P_{gg}(z) \ln z - \frac{\pi^2}{6} \de(1-z) \Big] \bigg\}
 \,, \nn \\
 \frac{\cJ_{gq}^{\one}(s,z,\mu_J)}{2(2\pi)^3}
& =  \frac{\al_s(\mu_J) T_F}{2\pi}\, \theta(z) \bigg\{ \Big[\frac{1}{\mu_J^2} \cL_0\Big(\frac{s}{\mu_J^2}\Big) \!+\! \de(s) \ln[z(1\!-\!z)] \Big]  P_{qg}(z) \!+\! 2 \de(s) \theta(1\!-\!z) z(1\!-\!z)\bigg\}
\, , \nn \\
\cJ_{g\bar q}^{\one}(s,z,\mu_J) &= \cJ_{g q}^{\one}(s,z,\mu_J) \, , \nn
\end{align}
%%%
where the plus distributions $\cL_n$ are defined in \eq{plusdef}.
Furthermore $\cJ_{\bar q \bar q}= \cJ_{q q}$ and $\cJ_{\bar q g}=\cJ_{q g}$ by charge conjugation invariance in QCD.
The coefficients $\cJ_{q \bar q}$, $\cJ_{q q'}$ and $\cJ_{q \bar q'}$, where $q'$ denotes a quark of a different flavor than $q$, only start at two loops.

The $\cJ_{ij}^\one$ are extracted from our partonic calculations of $ \cG_i^{j \one} (s, z,\mu)$ and $D_i^{j\one} (z,\mu)$, by using \eq{OPE} expanded to one loop,
%%%
\begin{align} \label{eq:NLOmatch}
  \cG_i^{j \one} (s, z,\mu_J) & = \sum_k \int_z^1\! \frac{\df z'}{z'}\, 
  \Big[\cJ^\zero_{ik}\Big(s,\frac{z}{z'},\mu_J\Big) D_k^{j \one}(z',\mu_J) + 
  \cJ^\one_{ik}\Big(s,\frac{z}{z'},\mu_J\Big) D_k^{j \zero}(z',\mu_J)\Big]
  \nn \\
  & = \sum_k \! \int_z^1\! \frac{\df z'}{z'}
  \Big[2(2\pi)^3 \de_{ik} \,\de(s)\, \de\Big(1\!-\!\frac{z}{z'}\Big)\, D_k^{j\one}(z',\mu_J) \!+\! 
  \cJ^\one_{ik}\Big(s,\frac{z}{z'},\mu_J\Big)\, \de_{kj}\, \de(1\!-\!z' ) \Big]
  \nn \\
  & = 2(2\pi)^3\, \de(s) D_i^{j\one} (z,\mu_J) + \cJ^\one_{ij} (s,z,\mu_J)
  \,.
\end{align}
%%%
Therefore, the NLO matching coefficients can be obtained through subtractions of the one-loop renormalized partonic fragmenting jet functions and the one-loop fragmentation functions. The calculation of these two is the subject of sections \ref{sec:gluondelta}, \ref{sec:gluondimreg} and appendix \ref{app:offshell}.

%===============================================================================
\subsection{Relationship between $\cG_i^h(s,z,\mu)$ and the Jet Function $J_i(s, \mu)$}
\label{subsec:jet}
%===============================================================================

In light of the discussion in ref.~\cite{Procura:2009vm}, we re-derive the relationship between the fragmenting jet function and the jet function, exposing all subtleties. When we sum over all possible hadrons $h\in {\cal H}_i$ fragmenting from a parton $i$ and belonging to the jet, the fragmenting jet function can be related to the inclusive jet function $J_i(s, \mu)$, which is completely calculable in perturbation theory.
We use the completeness relation
%%%
\begin{equation} \label{eq:Xh}
\int_0^1 \df z\,z\,\sum_{h \in {\cal H}_i} \sum_X  |X h(z)\rangle \langle X h(z)| = \sum_{X_i} |X_i \rangle \langle X_i|= \dblone\,,
\end{equation}
%%%
where $\{ |X_i \rangle \}$ is a complete set of states in the jet-like kinematic region that we are interested in. The factor $z$ under the integral is needed to provide the correct symmetry factor for states with identical particles. This is easily seen from the following example: consider the case where $X$ consists of $n$ hadrons identical to $h$, {\it i.e.} $X=\{h(z_1) \dots h(z_n)\}$. The sum over $X$ in \eq{Xh} contains a phase-space integral over the momentum fractions $z_1, \dots, z_n$ that is subject to a momentum conserving delta function and has a symmetry factor of $1/n!$
%%%
\begin{align}
& \int_0^1 \df z\,z\, \frac{1}{n!} \int_0^1 \prod_{i=1}^n \df z_i\, |z_1,\dots z_n; z \rangle \langle z_1,\dots z_n; z|\, \de\bigg(1 - z - \sum_{i=1}^n z_i\bigg)
\nn \\ & \quad
= \frac{1}{(n+1)!} \int_0^1 \prod_{i=1}^{n+1} \df z_i\, |z_1,\dots z_n,z_{n+1} \rangle \langle z_1,\dots z_n, z_{n+1}|\, \de\bigg(1 - \sum_{i=1}^{n+1} z_i\bigg) \, (n+1) \,z_{n+1}
\nn \\ & \quad
= \frac{1}{(n+1)!} \int_0^1 \prod_{i=1}^{n+1} \df z_i\, |z_1,\dots z_n,z_{n+1} \rangle \langle z_1,\dots z_n, z_{n+1}|\, \de\bigg(1 - \sum_{i=1}^{n+1} z_i\bigg)
\,.
\end{align}
%%%
Since we integrate over $z$, the hadron $h$ is no longer distinguishable from $X$ and should therefore be grouped with the rest. In the first equality, we redefined $z_{n+1}=z$ and divided and multiplied by $n+1$ to get the correct symmetry factor in front. In the final step we replaced $(n+1) z_{n+1} \to \sum_{i=1}^{n+1} z_i$, which is justified because all particles are identical and all momentum fractions are integrated over. Using the momentum conserving delta function, $\sum_{i=1}^{n+1} z_i = 1$, which leads to the result.

Applying \eq{Xh} to the fragmentation function leads to
\begin{align} \label{eq:D_mom_cons}
  \sum_h \int_0^1 \df x\, x\, D_j^h(x,\mu) = 1
  \,,
\end{align}
which is consistent with momentum conservation and with the definition of $D_i^h(z,\mu)$ as the \emph{number density} of the hadron $h$ in the parton $i$~\cite{Collins:1981uw}. Similarly, in the case of the fragmenting jet function we obtain
%%%
\begin{equation}
  \sum_{h \in {\cal H}_i} \int_0^1 \df z\, z\; \cG_i^h(s,z,\mu) = 2(2\pi)^3 J_i(s,\mu)\,.
\end{equation}
%%%
Combining this with \eq{OPE}, leads to
%%%
\begin{align} \label{eq:jetrel}
    J_i(s,\mu) 
   & =\frac{1}{2(2\pi)^3}\, \sum_h \int_0^1 \df z\, z\, \sum_j \int_z^1 \frac{\df x}{x} 
   \cJ_{ij} \Big(s,\frac{z}{x},\mu\Big) D_j^h(x,\mu)
   \nn \\
   & =\frac{1}{2(2\pi)^3}\, \sum_j  \int_0^1 \df u\, u\, \cJ_{ij}(s,u,\mu)
   \,.
\end{align}
%%%
Here we introduced the variable $u=z/x$ to disentangle the integrations and used \eq{D_mom_cons}.
This relationship with the jet function does not constrain non-perturbative physics, but provides a cross-check of our perturbative calculation of $\cJ_{ij}$, since the quark and gluon jet functions are known.

%%%%%%%%%%%%%%%%%%%%%%%%%%%%%%%%%%%%%%%%%%%%%%%%%%%%%%%%%%%%%%%%%%%%%%%%%%%%%%%%
\section{Quark Matching Calculation with Gluon Mass and $\de$ Regulator}
\label{sec:gluondelta}
%%%%%%%%%%%%%%%%%%%%%%%%%%%%%%%%%%%%%%%%%%%%%%%%%%%%%%%%%%%%%%%%%%%%%%%%%%%%%%%%

In this section we present the calculation of the Wilson coefficients $\cJ_{qj}$ for matching the quark fragmenting jet function onto fragmentation functions at NLO.  These are extracted using \eq{NLOmatch}, for which we need to calculate the partonic fragmentation functions and fragmenting jet functions at NLO. 
To regulate UV divergences we employ dimensional regularization (DR) in $d=4-2\eps$ dimensions and renormalize according to the $\overline{\text{MS}}$-scheme. 
Concerning the choice of the IR regulator, we note that neither an offshellness nor a fictitious gluon mass takes care of the IR divergences for emissions from the collinear Wilson line $W_n$ in \eq{Wn}. According to ref.~\cite{Chiu:2009yz}, these can be regulated as follows:
%%%
\begin{equation} \label{eq:coll_Wilson_delta}
W_n(x) = \biggl[\sum_\text{perms} \exp\Bigl(-\frac{g}{\bn \sdt \cP_n - \de}\,\bn\sdt A_n(x)\Bigr)\biggr]
\, ,
\end{equation}
%%%
with $\de >0$.
In this section we adopt a gluon mass and $\de$ as IR regulators. A non-vanishing gluon mass violates gauge invariance when the calculation involves a triple gluon vertex, which is however not the case here. 

A naive calculation of the graphs includes the region when $\ell$ becomes soft, which was excluded in \eq{xi}; zero-bin contributions have to be subtracted. In these zero-bin subtractions we need to apply the $\de$-regulator prescription as it would have appeared in the soft Wilson line:
%%%
\begin{equation} 
Y_n(x) = \biggl[\sum_\text{perms} \exp\Bigl(-\frac{g}{n \sdt \hat p  - \de}\, n\sdt A_{us}(x)\Bigr)\biggr]
\,.
\end{equation}
%%%
In this framework DR does not regulate any IR singularity, and this enables us to show in a clean way how IR divergences get cancelled in the matching between $\cG_q$ and $D_q$.

In appendix \ref{app:offshell} we compute the same diagrams with a quark-offshellness regulator, where the IR divergences from eikonal propagators are regulated by DR. The resulting  $\cJ_{ij}$ turn out to agree with those computed in this section, as expected since these Wilson coefficients should be insensitive to the choice of IR regulators.
After having studied in detail the IR structure for the case of quark fragmentation here, we perform the gluon matching calculation using DR for both the UV and IR in section \ref{sec:gluondimreg}.

In our partonic calculation we replace the hadron $h$ in the intermediate state of \eqs{Dqdef}{Gqdef} by a quark or a gluon and the remainder $X$ by the vacuum or a gluon or a quark, as required at one-loop order. In this section we evaluate the graphs by integrating over the phase-space of the parton which replaces $X$. In appendix \ref{app:offshell} we compute the diagrams for the fragmenting jet function following an alternative approach  based on the optical theorem. 

%%%
\begin{figure}[t]
\centering
\includegraphics[width=0.9\textwidth]{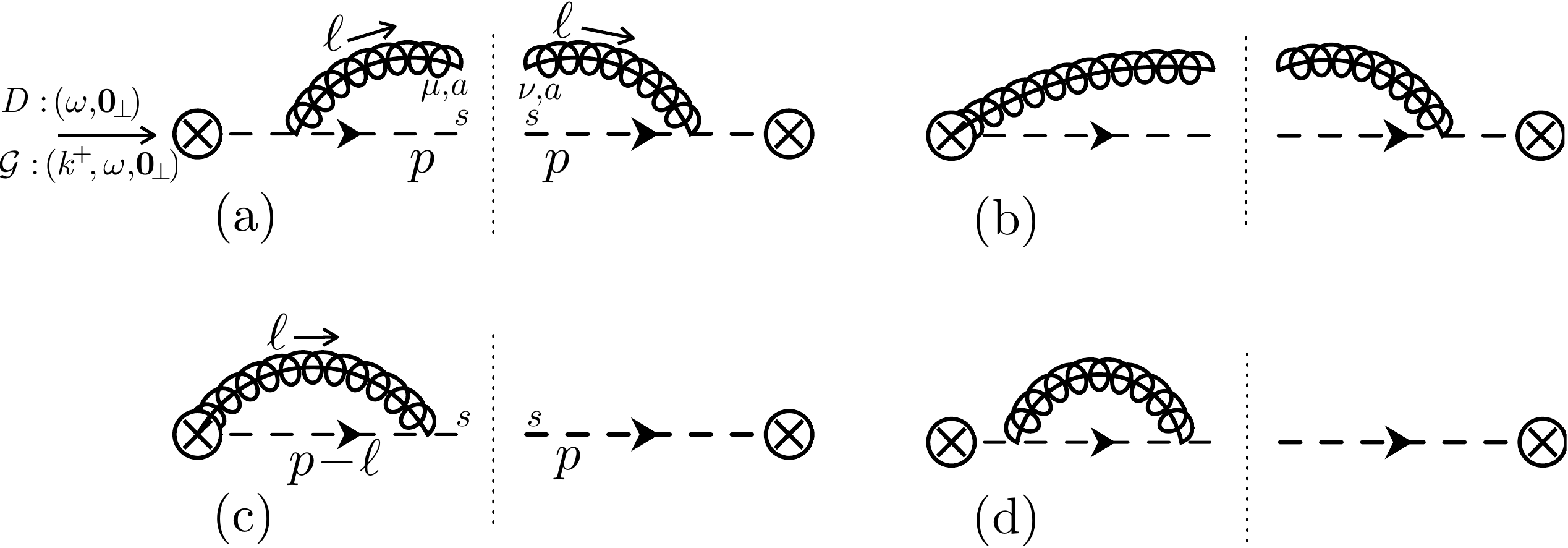}
\caption{Feynman graphs contributing to the partonic fragmentation function and fragmenting jet function that are non-zero at one-loop in Feynman gauge are shown here. Graphs (b) and (c) have a mirror image and (d) corresponds to the wave function renormalization.
For the partonic fragmentation function $D_q^i(z)$ the minus component $\w$ of the incoming momenta is fixed and the perpendicular components are zero by the choice of coordinates. For the fragmenting jet function the plus component $k^+$ is also fixed, determining the invariant mass of the jet.}
\label{fig:DGqgraphs}
\end{figure}
%%%

 The Feynman diagrams contributing to $D_q^i$ and $\cG_q^i$ at one loop are shown in figure \ref{fig:DGqgraphs}. By our choice of coordinates the incoming parton has no perpendicular momentum. Both in $D$ and in $\cG$ the ratio of the large components of the momentum of the incoming quark and the outgoing identified parton is measured.  For $\cG$, in addition, the virtuality of the incoming quark is specified
through $k^+$. The graphs in \fig{DGqgraphs}(a) and (b) correspond to a quark fragmenting into a quark while emitting a gluon, \fig{DGqgraphs}(c) corresponds to a virtual correction and \fig{DGqgraphs}(d) is the contribution from the quark wave-function renormalization. Figures 
\ref{fig:DGqgraphs}(b) and (c) have a mirror image which we do not draw separately, they also have nonvanishing zero-bin contributions. The graphs for a quark fragmenting into a gluon while emitting a quark are obtained by interchanging the momenta $p\lra \ell$ in \fig{DGqgraphs}(a) and (b).

We work in the Feynman gauge, without any loss of generality since the sum of the graphs is gauge-invariant. We use QCD Feynman rules to present the calculation here and have verified that the SCET Feynman rules give the same result, as expected.

%===============================================================================
\subsection{Quark Fragmentation Function at NLO}
\label{subsec:DNLO}
%===============================================================================

We start with graph (a) fig.~\ref{fig:DGqgraphs},
%%%
\begin{align} \label{eq:Da}
 D_{q,\bare}^{q(a)} & = \Big(\frac{e^{\ga_E} \mu^2}{4\pi}\Big)^\eps \frac{1}{2N_c\, x} 
  \int\!\df^{d-2} p_\perp\! \int\! \frac{\df^d \ell}{(2\pi)^{d-1}}\, \theta(\ell^0)\de(\ell^2-m^2) 
  \de(\w-\ell^- -p^-)\de^{d-2}(\ell_\perp + p_\perp)
  \nn \\ & \quad \times
 \tr\Big[\frac{\bnslash}{2} \frac{\img (\ellslash+\pslash)}{(\ell+p)^2}\, \img g\, \ga^\mu T^a \sum_{s, \pol} u_n^s(p) \ve_\mu(\ell)
  \ve^*_\nu(\ell) \bar u_n^s(p)\, \img g\, \ga^\nu T^a \frac{\img(\ellslash+\pslash)}{(\ell+p)^2}\Big] 
  \nn \\ &
  = \frac{\al_s(\mu) C_F}{2\pi}  
  (1-\eps)^2 \Ga(\eps) \Big(\frac{e^{\ga_E} \mu^2}{m^2}\Big)^\eps\, \theta(x)\theta(1-x)\, \frac{1-x}{x^\eps}
 \nn \\ &
  = \frac{\al_s(\mu) C_F}{2\pi}\, \theta(x)\theta(1-x)\, (1-x)
  \Big( \frac{1}{\eps} - 2 + \ln \frac{\mu^2}{x\,m^2} \Big) + \ord{\eps} 
 \,.
\end{align}
%%%
In the final step we expand in $\eps$ to extract the UV divergences.

Moving on to \fig{DGqgraphs}(b), 
%%%
\begin{align} \label{eq:Db}
 D_{q,\bare}^{q(b)} & = 2 \Big(\frac{e^{\ga_E} \mu^2}{4\pi}\Big)^\eps \frac{1}{2N_c\, x} 
  \int\!\df^{d-2} p_\perp\! \int\! \frac{\df^d \ell}{(2\pi)^{d-1}}\, \theta(\ell^0) \de(\ell^2-m^2) 
  \de(\w-\ell^- -p^-)\de^{d-2}(\ell_\perp + p_\perp)
  \nn \\ & \quad \times
 \tr\Big[\frac{\bnslash}{2} \frac{\img (\ellslash+\pslash)}{(\ell+p)^2}\, \img g\, \ga^\mu T^a \sum_{s,\pol} u_n^s(p) \ve_\mu(\ell)
  \ve^*_\nu(\ell) \bar u_n^s(p)\, \frac{g\, T^a\bn^\nu}{\ell^- + \de}\Big] 
  \nn \\ &
  = \frac{\al_s(\mu) C_F}{\pi}  
 \Ga(\eps) \Big(\frac{e^{\ga_E} \mu^2}{m^2}\Big)^\eps \, \theta(x)\theta(1-x)\, \frac{x^{1-\eps}}{1-x+\de/\w}
 \nn \\ &
  \!\!\stackrel{\de \to 0}{=} \frac{\al_s(\mu) C_F}{\pi} \,\theta(x)\theta(1-x)\,
  \Big(\frac{1}{\eps} + \ln \frac{\mu^2}{x\, m^2}\Big) \Big[x \cL_0(1-x) - \de(1-x) \ln \frac{\de}{\w}\Big]
 + \ord{\eps}
  \,,
\end{align}
%%%
where we include a factor of 2 for the mirror graph. 
At the end we take the limit $\de \to 0$ to isolate the IR divergences. There we use
%%%
\begin{align}
  \lim_{\de \to 0} \frac{\theta(1-x)}{1-x+\de/\w} 
  &= \lim_{\de \to 0} \frac{\theta(1 - \tilde x - \de/\w)}{1 - \tilde x}
  = \cL_0(1-x) - \de(1-x) \ln \frac{\de}{\w}
  \,,
\end{align}
%%%
which follows from the definition of $\cL_0$ in \eq{plusdef}, with $\tilde x = x - \de/\w$.

We need to subtract the zero-bin contribution, which comes from the region where the gluon becomes soft, $\ell^\mu \sim \la^2 p^-$ \cite{Manohar:2006nz}. Expanding \eq{Db} accordingly, and including the appropriate $\de$-regulator for the soft Wilson line that appears here,
%%%
\begin{align} \label{eq:Db0}
 D_{q,\bare}^{q(b)0} & = 2 \Big(\frac{e^{\ga_E} \mu^2}{4\pi}\Big)^\eps \frac{1}{2N_c\, x} 
  \int\!\df^{d-2} p_\perp\! \int\! \frac{\df^d \ell}{(2\pi)^{d-1}}\, \theta(\ell^0) \de(\ell^2-m^2)
  \de(\w -p^-)\de^{d-2}(p_\perp)
  \nn \\ & \quad \times
 \tr\Big[\frac{\bnslash}{2}\frac{\nslash}{2} \frac{\img}{\ell^+ + p^+ + \de} \, \img g\, \ga^\mu T^a 
 \sum_{s,\pol} u_n^s(p) \ve_\mu(\ell) \ve^*_\nu(\ell) \bar u_n^s(p)\, \frac{g\, T^a\bn^\nu}{\ell^- + \de}\Big] 
  \nn \\ &
 = \frac{\al_s(\mu) C_F}{\pi}\, \Ga(\eps)(e^{\ga_E} \mu^2)^\eps\, \de(1-x)
  \int_0^\infty \! \df \ell^-\, \frac{1}{(\ell^- + \de)( \ell^- \de + m^2)^\eps} \nn \\
& = \frac{\al_s(\mu) C_F}{\pi}\, \Ga(\eps)(e^{\ga_E} \mu^2)^\eps\, \de(1-x) \left [ \frac{\Gamma(\epsilon)\Gamma(1-\epsilon)}{(\de^2-m^2)^\epsilon} + \frac{m^{2-2\epsilon}}{\de^2 (\epsilon - 1)} \, _2F_1(1,1;2-\epsilon;m^2/\de^2) \right]
\end{align}
%%%
Note that $p^+=0$ from the on-shell condition $p^2=0$ and $p_\perp^\mu =0$. After expanding in $\epsilon$ and separating the IR divergences by taking $m^2 \to 0$, followed by $\de \to 0$, we get
%%%
\begin{align} \label{eq:Db0res}
 D_{q,\bare}^{q(b)0} & \stackrel{m^2\!,\,\de \to 0}{=}  \frac{\al_s(\mu) C_F}{\pi}\, \de(1-x)\, \Big[ \frac{1}{\eps^2} - \frac{2}{\eps} \ln \frac{\de}{\mu} + 2 \ln^2 \frac{\de}{\mu} + \frac{\pi^2}{4}\Big] + \ord{\eps}
 \,.
\end{align}
%%%
This result depends on the order of the limits $m^2 \to 0$, $\de \to 0$, but the sum of the diagrams does not. In \eq{Db0res}, $m^2$ is absent because it does not regulate anything for this diagram and can simply be set to zero.

For \fig{DGqgraphs}(c) we find
%%%
\begin{align} \label{eq:Dqqc}
 D_{q,\bare}^{q(c)} & = 2 \Big(\frac{e^{\ga_E} \mu^2}{4\pi}\Big)^\eps \frac{1}{2N_c\, x} 
  \int\!\df^{d-2} p_\perp\! \int\! \frac{\df^d \ell}{(2\pi)^d}\, \de(\w - p^-)\de^{d-2}(p_\perp)
 \nn \\ & \quad \times
 \tr\Big[\frac{\bnslash}{2} \sum_s u_n^s(p)  \bar u_n^s(p)\,  \img g\, \ga^\mu T^a\, 
 \frac{\img (\pslash-\ellslash)}{(p - \ell)^2 + \img 0}\, \frac{g\, T^a\bn_\mu}{\ell^- + \de}\Big] 
 \frac{-\img}{\ell^2 - m^2 + \img 0} 
  \nn \\ &
  = \Big(\frac{e^{\ga_E} \mu^2}{4\pi}\Big)^\eps 4\img\, g^2 C_F \,
  \de(1 - x) \! \int\! \frac{\df^d \ell}{(2\pi)^d}\, 
 \frac{\w - \ell^-}{(\ell^- \!+\! \de) [\ell^- \ell^+ \!+\! \ell_\perp^2 \!-\! m^2 \!+\! \img 0][(\ell^- \!-\! \w)\ell^+ \!+\! \ell_\perp^2 \!+\! \img 0]} 
  \nn \\ &
  = -\frac{\al_s(\mu) C_F}{\pi}  
 \Ga(\eps) \Big(\frac{e^{\ga_E} \mu^2}{m^2}\Big)^\eps \de(1-x) 
  \int_0^\w\! \df \ell^-\, \frac{(\w - \ell^-)^{1-\eps}}{\w^{1-\eps} (\ell^- + \de)} 
\nn \\ &
  \!\!\stackrel{\de \to 0}{=} \frac{\al_s(\mu) C_F}{\pi} \de(1-x) \Big(\frac{1}{\eps}+ \ln \frac{\mu^2}{m^2}\Big)
  \Big[1 + \ln \frac{\de}{\w} + \eps \Big(1 - \frac{\pi^2}{6}\Big) \Big] + \ord{\eps}
  \,.  
\end{align}
%%%
After integrating the delta functions, we perform the $\ell^+$ integral by contours. The poles are located at
%%%
\begin{equation}
  \ell^+ = \frac{m^2-\ell_\perp^2 -\img 0}{\ell^-}
  \,, \qquad
  \ell^+ = \frac{-\ell_\perp^2-\img 0}{\ell^- - \w}
  \, ,
\end{equation}
%%%
which are on opposite sides of the real axis for $0 < \ell^- < \w$. We pick up the first pole and perform the standard $\ell_\perp$-integral, yielding the second last line in \eq{Dqqc}. In the last step we perform the remaining $\ell^-$ integral, expand in $\eps$ and take the limit $\de \to 0$ to isolate the IR divergences.

The corresponding zero-bin contribution is obtained by expanding in the region $\ell^\mu \sim \la^2 p^-$,
%%%
\begin{align}
 D_{q,\bare}^{q(c)0} & = 2 \Big(\frac{e^{\ga_E} \mu^2}{4\pi}\Big)^\eps \frac{1}{2N_c\, x} 
  \int\!\df^{d-2} p_\perp\! \int\! \frac{\df^d \ell}{(2\pi)^d}\, \de(\w - p^-)\de^{d-2}(p_\perp) 
 \nn \\ & \quad \times
 \tr\Big[\frac{\bnslash}{2} \sum_s u_n^s(p)  \bar u_n^s(p)\,  \img g\, \ga^\mu T^a\, 
 \frac{\nslash}{2} \frac{-\img}{\ell^+ + \de - \img 0}\, \frac{g\, T^a\bn_\mu}{\ell^- + \de}\Big] 
 \frac{-\img}{\ell^2 - m^2 + \img 0} 
  \nn \\ &
  = -\Big(\frac{e^{\ga_E} \mu^2}{4\pi}\Big)^\eps 4\img\, g^2 C_F \,
  \de(1-x) \int\! \frac{\df^d \ell}{(2\pi)^d}\, 
 \frac{1}{(\ell^- + \de) (\ell^+ + \de - \img 0) (\ell^- \ell^+ + \ell_\perp^2 -m^2 + \img 0)} 
  \nn \\ &
 = \Big(\frac{e^{\ga_E} \mu^2}{4\pi}\Big)^\eps \frac{g^2 C_F}{\pi} \de(1-x)
 \int_{0}^\infty\! \df \ell^-\, \frac{1}{\ell^- + \de} 
 \int \frac{\df^{d-2} \ell_\perp}{(2\pi)^{d-2}}\, \frac{1}{\ell_\perp^2 - m^2 - \de \ell^-}
 \nn \\ &
  = -D_{q,\bare}^{q(b)0}
  \,.
\end{align}
%%%
The opposite sign of the $\img 0$ prescription in $\ell^+ + \de - \img 0$ comes from dividing out $-p^-<0$.
This time the $\ell^+$ poles are at
%%%
\begin{equation}
  \ell^+ = \frac{m^2-\ell_\perp^2 -\img 0}{\ell^-}
  \,, \qquad
  \ell^+ = -\de + \img 0
  \,,
\end{equation}
%%%
which are on opposite sides of the real axis for $\ell^- > 0$. Picking up the second pole bring us to the third line, which is equal to minus the second line of \eq{Db0}. Thus the zero bins from the real and virtual Wilson line emission cancel each other, which is no surprise because $D$ is insensitive to the scale associated to the soft radiation accompanying the final parton. This will no longer be true for the fragmenting jet functions.

Using on-shell wave-function renormalization for the massless quark,  
%%%
\begin{align}
 (1 - Z_\psi) \img \pslash + \ord{\eps}
 &=  \Big(\frac{e^{\ga_E} \mu^2}{4\pi}\Big)^\eps \int\! \frac{\df^d \ell}{(2\pi)^d}\,
  \img g\, \ga^\mu T^a\, \frac{\img (\ellslash+\pslash)}{(\ell+p)^2} \,
  \img g\, \ga_\mu T^a \frac{-\img}{\ell^2 - m^2}
  \nn \\ &
  = \img \pslash\, \frac{\al_s(\mu) C_F}{2\pi} \frac{1-\eps}{2-\eps}\, \Ga(\eps) \Big(\frac{e^{\eps \ga_E} \mu^2}{m^2}\Big)^\eps
  \,
\end{align}
%%%
from which we derive
%%%
\begin{equation}
  Z_\psi = 1 + \frac{\al_s(\mu) C_F}{2\pi} \Big( -\frac{1}{2\eps} + \frac{1}{4} - \frac{1}{2} \ln \frac{\mu^2}{m^2}\Big) \,.
\end{equation}
%%%
The contribution to $D$ from wave-function renormalization is therefore given by 
%%%
\begin{equation}
 D_{q,\bare}^{q(d)} = (Z_\psi-1) \de(1-x)
 \,.
\end{equation}
%%%

We will now calculate the diagrams for $D_q^g$, where the momentum fraction of the gluon is measured. We can obtain these from the corresponding expressions for the quark case in \eqs{Da}{Db} by taking $x \to 1-x$. To see how this comes about, we explicitly include the on-shell condition for $p$ and the definition of $x$ in \eq{Da} and \eq{Db} as follows:
%%%
\begin{align}
  \int\! \df^{d-2} p_\perp = \int\! \df^d p\, \de(p^2) \theta(p^0)\, \de\Big(x - \frac{p^-}{\w}\Big)
  \,.
\end{align}
%%%
As a consequence, the quark and the gluon in \fig{DGqgraphs}(a) and (b) are on completely equal footing, except that the momentum fraction of the quark was measured. Measuring the momentum fraction of the gluon therefore amounts to $x \to 1-x$. 

For the first diagram we obtain
%%%
\begin{align} \label{eq:Dap}
 D_{q,\bare}^{g(a)}(x) & = D_{q,\bare}^{q(a)}(1-x)
 \nn \\ &
  = \frac{\al_s(\mu) C_F}{2\pi}  
  (1-\eps)^2 \Ga(\eps) \Big(\frac{e^{\ga_E} \mu^2}{m^2}\Big)^\eps\, \theta(x)\theta(1-x)\, \frac{x}{(1-x)^\eps}
 \nn \\ &
  = \frac{\al_s(\mu) C_F}{2\pi}\, \theta(x)\theta(1-x)\, x
  \Big( \frac{1}{\eps} - 2 + \ln \frac{\mu^2}{(1-x)\,m^2} \Big) + \ord{\eps} 
 \,.
\end{align}
%%%
For the second graph we find
%%%
\begin{align} \label{eq:Dbp}
 D_{q,\bare}^{g(b)}(x) & = D_{q,\bare}^{q(b)}(1-x)
 \nn \\ &
  = \frac{\al_s(\mu) C_F}{\pi}  
 \Ga(\eps) \Big(\frac{e^{\ga_E} \mu^2}{m^2}\Big)^\eps \, \theta(x)\theta(1-x)\, \frac{(1-x)^{1-\eps}}{x+\de/\w}
 \nn \\ &
  \!\!\stackrel{\de \to 0}{=} \frac{\al_s(\mu) C_F}{\pi} \,\theta(x)\theta(1-x)\,
  \Big(\frac{1}{\eps} + \ln \frac{\mu^2}{(1-x)\, m^2}\Big) \Big[(1-x) \cL_0(x) - \de(x) \ln \frac{\de}{\w}\Big]
 + \ord{\eps}
 \nn \\ &
  = \frac{\al_s(\mu) C_F}{\pi} \,\theta(x)\theta(1-x)\,
  \Big(\frac{1}{\eps} + \ln \frac{\mu^2}{(1-x)\, m^2}\Big) \frac{1-x}{x}
 + \ord{\eps}
  \,,
\end{align}
%%%
where in the last step we dropped both the plus prescription and the $\de(x)$ terms 
since $x>0$. The zero bin only contributes at $x=0$ and therefore does not need to be taken into account.

Adding up our results yields
%%%
\begin{align} \label{eq:Dbare}
D_{q,\bare}^{q\one} 
&= D_{q,\bare}^{q(a)} + \Big(D_{q,\bare}^{q(b)} - D_{q,\bare}^{q(b)0}\Big) 
+ \Big(D_{q,\bare}^{q(c)} - D_{q,\bare}^{q(c)0}\Big) + D_{q,\bare}^{q(d)}
\nn \\
&= \frac{\al_s(\mu) C_F}{2\pi} \theta(x) \bigg[\Big( \frac{1}{\eps} + \ln \frac{\mu^2}{m^2} -\ln x \Big) P_{qq}(x)
- \Big(\frac{\pi^2 }{3}-\frac{9 }{4} \Big) \delta (1-x) - 2\theta(1-x) (1-x)  \bigg] 
\,, \nn \\
D_{q,\bare}^{g\one}  
&= D_{q,\bare}^{g(a)} + D_{q,\bare}^{g(b)} 
\nn \\
&=  \frac{\al_s(\mu) C_F}{2\pi} \theta(x) \bigg[\Big( \frac{1}{\eps} + \ln \frac{\mu^2}{m^2} -\ln(1-x) \Big) P_{gq}(x)
 - 2\theta(1-x) x  \bigg] 
\,,
\end{align}
%%%
where the splitting functions $P_{iq}$ are given in \eq{split}. Note that $1/\eps$-poles multiplying IR regulators cancel in the sum of the diagrams, as must be the case. In \eq{Dbare} $\de$ is absent since it regulates those IR divergences which cancel between the real and virtual emission diagrams for the fragmentation function.

%===============================================================================
\subsection{Quark Fragmenting Jet Function at NLO}
\label{subsec:GNLO}
%===============================================================================

Compared to the $D_i^j$ case, the partonic fragmenting jet function calculation involves an additional $\de[\w(k^+ - \ell^+ - p^+)]$, where $p^+ = {\vec p}_\perp^{\,2}/p^- = -p_\perp^2/p^-$. It would seem that this fixes all the momentum components in real radiation graphs \ref{fig:DGqgraphs}(a) and \ref{fig:DGqgraphs}(b), however, as we will see, this is not true for the zero-bin diagram associated with (b).
Starting with figure \ref{fig:DGqgraphs}(a),
%%%
\begin{align} \label{eq:Ga}
 \frac{\cG_{q,\bare}^{q(a)}}{2(2\pi)^3} & = \Big(\frac{e^{\ga_E} \mu^2}{4\pi}\Big)^\eps \frac{1}{2N_c\, z} 
  \int\!\df^{d-2} p_\perp\! \int\! \frac{\df^d \ell}{(2\pi)^{d-1}}\, \theta(\ell^0) \de(\ell^2-m^2)
  \de(\w-\ell^- -p^-)\de^{d-2}(\ell_\perp + p_\perp) 
  \nn \\ & \quad \times  
 \de[\w(k^+ \!-\! \ell^+ \!-\! p^+)]\,
 \tr\Big[\frac{\bnslash}{2} \frac{\img (\ellslash+\pslash)}{(\ell+p)^2}\, \img g\, \ga^\mu T^a \sum_{s,\pol} 
 u_n^s(p) \ve_\mu(\ell) \ve^*_\nu(\ell) \bar u_n^s(p)\, \img g\, \ga^\nu T^a \frac{\img(\ellslash+\pslash)}{(\ell+p)^2}\Big] 
  \nn \\ &
 = - \Big(\frac{e^{\ga_E} \mu^2}{4\pi}\Big)^\eps \frac{g^2 C_F (1-\eps)}{2\pi}\, \theta(z)\theta(1-z)\,z(1-z)^2
 \nn \\ & \quad \times
 \int\! \frac{\df^{d-2} \ell_\perp}{(2\pi)^{d-2}}\, \frac{\ell_\perp^2}{(\ell_\perp^2 - z\, m^2)^2}\,
 \de\Big[\ell_\perp^2 - z(1-z)\Big(\frac{m^2}{1-z} - s\Big)\Big]
 \nn \\ &
  = \frac{\al_s(\mu) C_F}{2\pi}  
  \frac{1-\eps}{\Ga(1-\eps)} (e^{\ga_E} \mu^2)^\eps\, \theta(z)\theta(1-z)\, z^{-\eps} (1-z)^{1-\eps}\,
  \theta\Big(s-\frac{m^2}{1-z}\Big) \frac{(s-\frac{m^2}{1-z})^{1-\eps}}{s^2}
 \nn \\ &
  \!\!\!\!\stackrel{m^2 \to 0}{=} \frac{\al_s(\mu) C_F}{2\pi}\, \theta(z)\theta(1-z)\, (1-z)
  \Big[ \frac{1}{\mu^2} \cL_0\Big(\frac{s}{\mu^2}\Big) + \de(s) \Big(\ln \frac{(1-z)\mu^2}{m^2} -  1\Big)\Big] + \ord{\eps}
 \,.
\end{align}
%%%
Here we rewrote
%%%
\begin{equation} \label{eq:newde}
  \de\Big[\w\Big(k^+ + \frac{\ell_\perp^2-m^2}{\w-p^-} + \frac{\ell_\perp^2}{p^-}\Big)\Big]
  = z(1-z) \de\Big[\ell_\perp^2 - z(1-z)\Big(\frac{m^2}{1-z} - s\Big)\Big]
  \,.
\end{equation}
%%%
Since all the dependence on $\ell_\perp$ is in terms of $\ell_\perp^2$, we used spherical coordinates to perform the $\ell_\perp$-integral
%%%
\begin{equation} \label{eq:lperp_sph}
  \int\! \frac{\df^{d-2} \ell_\perp}{(2\pi)^{d-2}} = \int_0^\infty \! \df( -\ell_\perp^2)\, \frac{(-\ell_\perp^2)^{-\eps}}{(4\pi)^{1-\eps} \Ga(1-\eps)}
  \,,
\end{equation}
%%%
which in combination with \eq{newde} demands that $s-m^2/(1-z) > 0$ for a non-vanishing contribution; hence the factor of $\theta(s-m^2/(1-z))$ in the  second to last line of \eq{Ga}. The graph is UV finite, so the expansion in $\eps$ is trivial. In the final step we take the limit $m^2\to 0$ to separate the IR divergences using \eqs{plusdef}{limits}. 

The calculation of \fig{DGqgraphs}(b) combines steps from the corresponding fragmentation function graph and the previous diagram,
%%%
\begin{align} \label{eq:Dbg}
 \frac{\cG_{q,\bare}^{q(b)}}{2(2\pi)^3} & = 2 \Big(\frac{e^{\ga_E} \mu^2}{4\pi}\Big)^\eps \frac{1}{2N_c\, z} 
  \int\!\df^{d-2} p_\perp\! \int\! \frac{\df^d \ell}{(2\pi)^{d-1}}\, \theta(\ell^0) \de(\ell^2-m^2)
  \de(\w-\ell^- -p^-)\de^{d-2}(\ell_\perp + p_\perp) 
  \nn \\ & \quad \times
 \de[\w(k^+ - \ell^+ - p^+)]\,
 \tr\Big[\frac{\bnslash}{2} \frac{\img (\ellslash+\pslash)}{(\ell+p)^2}\, \img g\, \ga^\mu T^a 
 \sum_{s,\pol} u_n^s(p) \ve_\mu(\ell) \ve^*_\nu(\ell) \bar u_n^s(p)\, \frac{g\, T^a\bn^\nu}{\ell^- + \de}\Big] 
  \nn \\ &
  = \frac{\al_s(\mu) C_F}{\pi}  
  \frac{1}{\Ga(1-\eps)} (e^{\ga_E} \mu^2)^\eps \theta(z)\theta(1-z)\, \frac{z^{1-\eps} (1-z)^{-\eps}}{1-z+\de/\w}\,   \theta\Big(s-\frac{m^2}{1-z}\Big)\, \frac{(s-\frac{m^2}{1-z})^{-\eps}}{s}
 \nn \\ &
  \!\!\!\!\!\stackrel{m^2\!,\,\de \to 0}{=} \frac{\al_s(\mu) C_F}{\pi}\, \theta(z)\theta(1-z)\, z\,
  \bigg\{  
  \frac{1}{\mu^2} \cL_0\Big(\frac{s}{\mu^2}\Big) \Big[\cL_0(1-z) - \ln \frac{\de}{\w}\, \de(1-z)\Big] 
  \nn \\ & \quad
  + \de(s) \Big[\cL_1(1\!-\!z)  + \ln \frac{\mu^2}{m^2}\, \cL_0(1\!-\!z) 
  - \Big(\frac{1}{2} \ln^2 \frac{\de}{\w} \!+\! \ln \frac{\mu^2}{m^2} \ln \frac{\de}{\w} \!+\! \frac{\pi^2}{6} \Big) \de(1\!-\!z)
  \Big] 
  \bigg\}  + \ord{\eps}
  \,.
\end{align}
%%%
We take the limit $m^2 \to 0$ followed by $\de \to 0$. Since the gluon mass occurs in the combination $m^2/(1-z)$, different order of limits give different results for individual graphs, but the sum of all graphs is independent of the order of limits.
We separately studied the cases $z<1$ and $s>0$, and determined the coefficient of the remaining $\de(s)\de(1-z)$ term by integration.

The zero bin corresponding to \eq{Dbg} is given by
%%%
\begin{align} 
 \frac{\cG_{q,\bare}^{q(b)0}}{2(2\pi)^3} & = 2 \Big(\frac{e^{\ga_E} \mu^2}{4\pi}\Big)^\eps \frac{1}{2N_c\, z} 
  \int\!\df^{d-2} p_\perp\! \int\! \frac{\df^d \ell}{(2\pi)^{d-1}}\,\theta(\ell^0) \de(\ell^2-m^2)
  \de(\w -p^-)\de^{d-2}(p_\perp) 
  \nn \\ & \quad \times
 \de[\w(k^+ \!-\! \ell^+ \!-\! p^+)]\,
 \tr\Big[\frac{\bnslash}{2}\frac{\nslash}{2} \frac{\img}{\ell^+ + p^+ + \de} \, \img g\, \ga^\mu T^a 
 \sum_{s,\pol} u_n^s(p) \ve_\mu(\ell) \ve^*_\nu(\ell) \bar u_n^s(p)\, \frac{g\, T^a\bn^\nu}{\ell^- + \de}\Big] 
  \nn \\ &
 = \frac{\al_s(\mu) C_F}{\pi}\, \Ga(\eps) e^{\eps \ga_E} \, \de(1-z) \,
 \frac{\theta(s)}{s+ \de\, \w} \Big(\frac{\mu^2}{m^2+ s\, \de/\w}\Big)^\eps
 \nn \\ &
 \!\!\!\!\stackrel{m^2 \to 0}{=} \frac{\al_s(\mu) C_F}{\pi}\, \Ga(\eps) e^{\eps \ga_E} \, \de(1-z) \,
 \frac{\theta(s) (s/\mu^2)^{-\eps}}{s+ \de\, \w} \Big(\frac{\de}{\w}\Big)^{-\eps}
 \nn \\ &
 \!\!\stackrel{\de \to 0}{=} \frac{\al_s(\mu) C_F}{\pi}\, \de(1-z)\, 
 \bigg\{\frac{1}{\eps} \Big[\frac{1}{\mu^2}\cL_0\Big(\frac{s}{\mu^2}\Big) - \de(s) \ln \frac{\de\,\w}{\mu^2}\Big]
 -\frac{1}{\mu^2} \cL_1\Big(\frac{s}{\mu^2}\Big)
  -\ln \frac{\de}{\w}\, \frac{1}{\mu^2} \cL_0 \Big(\frac{s}{\mu^2}\Big) 
  \nn \\ & \quad
  + \de(s)\Big[ \frac{1}{2} \ln^2 \frac{\de\, \w}{\mu^2} 
  + \ln \frac{\de\, \w}{\mu^2} \ln \frac{\de}{\w} + \frac{\pi^2}{6}\Big]\bigg\}
 +\ord{\eps}
 \,.
\end{align}
%%%
Since all divergences are regulated by $\de$, we can simply set $m^2=0$, which is consistent with the order of limits used previously. In taking $\de \to 0$ we used
%%%
\begin{align}
  \lim_{\de \to 0} \frac{\theta(s)}{s + \de \w} &= \lim_{\de \to 0} \frac{\theta(\tilde s - \de \w)}{\tilde s}
  = \frac{1}{\mu^2} \cL_0\Big(\frac{s}{\mu^2}\Big) - \de(s) \ln \frac{\de \, \w}{\mu^2}
  \,,\nn \\
  \lim_{\de \to 0} \frac{\theta(s) \ln (s/\mu^2)}{s + \de \w} &= \lim_{\de \to 0} \frac{\theta(\tilde s - \de \w) \ln[\theta(\tilde s - \de \w)/\mu^2]}{\tilde s}
  = \frac{1}{\mu^2} \cL_1\Big(\frac{s}{\mu^2}\Big) - \de(s) \Big(\frac{1}{2} \ln^2 \frac{\de \, \w}{\mu^2} + \frac{\pi^2}{6}\Big)
  \,,
\end{align}
%%%
which follows from \eqs{plusdef}{limits}, where $\tilde s = s + \de \w$.

The virtual emission graph and the wave-function renormalization contribution are unaffected by the restriction on the real radiation given by the $\de$-function involving $k^+$:
%%%
\begin{align}
   \cG_{q,\bare}^{q(r)}(s,z) =  2(2\pi)^3 \de(s)\, D_{q,\bare}^{q(r)}(z)
   \,, \qquad \text{with } (r)=(c), (c)0, (d)
   \,.
\end{align}
%%%

As for the fragmentation function, expressions for the diagrams where the momentum fraction of the gluon instead of the quark is measured can be readily obtained by $z \to 1-z$,
%%%
\begin{align} \label{eq:Gabp}
\frac{\cG_{q,\bare}^{g(a)}(z)}{2(2\pi)^3} & = 
\frac{\cG_{q,\bare}^{q(a)}(1-z)}{2(2\pi)^3}
 \nn \\ &
  = \frac{\al_s(\mu) C_F}{2\pi}  
  \frac{1-\eps}{\Ga(1-\eps)} (e^{\ga_E} \mu^2)^\eps\, \theta(z)\theta(1-z)\, z^{1-\eps} (1-z)^{-\eps}\,
  \theta\Big(s-\frac{m^2}{z}\Big) \frac{(s-\frac{m^2}{z})^{1-\eps}}{s^2}
 \nn \\ &
  \!\!\!\!\stackrel{m^2 \to 0}{=} \frac{\al_s(\mu) C_F}{2\pi}\, \theta(z)\theta(1-z)\, z
  \Big[ \frac{1}{\mu^2} \cL_0\Big(\frac{s}{\mu^2}\Big) + \de(s) \Big(\ln \frac{z \mu^2}{m^2} -  1\Big)\Big] + \ord{\eps}
 \,, \nn \\
  \frac{\cG_{q,\bare}^{g(b)}(z)}{2(2\pi)^3} & = \frac{\cG_{q,\bare}^{q(b)}(1-z)}{2(2\pi)^3}
  \nn \\ &
   = \frac{\al_s(\mu) C_F}{\pi}  
  \frac{1}{\Ga(1-\eps)} (e^{\ga_E} \mu^2)^\eps \theta(z)\theta(1-z)\, \frac{z^{-\eps} (1-z)^{1-\eps}}{z+\de/\w}\,   \theta\Big(s-\frac{m^2}{z}\Big) \frac{(s-\frac{m^2}{z})^{-\eps}}{s}
 \nn \\ &
  \!\!\!\!\stackrel{m^2 \to 0}{=} \frac{\al_s(\mu) C_F}{\pi} \theta(z)\theta(1\!-\!z)\, \frac{1\!-\!z}{z}\,
  \Big[ \frac{1}{\mu^2} \cL_0\Big(\frac{s}{\mu^2}\Big) + \de(s) \ln \frac{z \mu^2}{m^2}\Big] +\ord{\eps}
  \,.
\end{align}
%%%
Adding up all the diagrams,
%%%
\begin{align} \label{eq:Gbare}
\cG_{q,\bare}^{q\one} 
&= \cG_{q,\bare}^{q(a)} + \Big(\cG_{q,\bare}^{q(b)} - \cG_{q,\bare}^{q(b)0}\Big) 
+ \Big(\cG_{q,\bare}^{q(c)} - \cG_{q,\bare}^{q(c)0}\Big) + \cG_{q,\bare}^{q(d)}
\nn \\ &
= 2(2\pi)^3\, \frac{\al_s(\mu) C_F}{2\pi} \bigg\{ \frac{2}{\eps^2} \de(s) \de(1-z) + 
 \frac{2}{\eps} \Big[- \frac{1}{\mu^2} \cL_0\Big(\frac{s}{\mu^2}\Big) + \frac{3}{4} \de(s) \Big] \de(1-z) 
\nn \\ & \quad 
+ \frac{2}{\mu^2} \cL_1\Big(\frac{s}{\mu^2}\Big) \de(1-z) 
+ \frac{1}{\mu^2} \cL_0\Big(\frac{s}{\mu^2}\Big) (1+z^2) \cL_0(1-z) 
+ \de(s)\Big[P_{qq}(z) \ln \frac{\mu^2}{m^2} 
\nn \\ & \quad
+ (1+z^2)\cL_1(1-z) -\theta(1-z) (1-z) -  \Big( \frac{\pi^2}{2} - \frac{9}{4}\Big) \de(1-z)\Big] \bigg\}
\,, \nn \\
\cG_{q,\bare}^{g\one}  
&= \cG_{q,\bare}^{g(a)} + \cG_{q,\bare}^{g(b)} 
\nn \\ &
= 2(2\pi)^3\, \frac{\al_s(\mu) C_F}{2\pi}\, \theta(z)\, \bigg\{\Big[\frac{1}{\mu^2} \cL_0\Big(\frac{s}{\mu^2}\Big) +
\de(s) \ln \frac{z\mu^2}{m^2} \Big] P_{gq}(z) - \de(s)\, \theta(1-z) z \bigg\}
\,.
\end{align}
%%%
All $1/\eps$-poles here are of UV origin. The $\de$-regulator disappears from the final result. This is expected, since the fragmentation function in \eq{Dbare} contains no dependence on the $\de$-regulator either, and the IR divergences have to cancel in the matching between $\cG_i$ and $D_i$.

%===============================================================================
\subsection{Renormalization and Matching}
\label{subsec:JNLO}
%===============================================================================

According to \eq{DrenNLO} the relevant countertems are 
%%%
\begin{align} \label{eq:Zres}
 Z^D_{qq}(x,\mu) & = \de(1-x) + 
  \frac{\al_s(\mu) C_F}{2\pi} \frac{1}{\eps} \theta(x) P_{qq}(x)
 \,, \nn \\
 Z^D_{qg}(x,\mu) & = 
 \frac{\al_s(\mu) C_F}{2\pi} \frac{1}{\eps} \theta(x) P_{gq}(x)
\,,
\end{align}
%%%
which lead to the same anomalous dimensions as in \eq{gaD}. The renormalized one-loop quark fragmentation functions are given by
%%%
\begin{align} \label{eq:Dren}
D_q^{q\one} (x,\mu)
&= \frac{\al_s(\mu) C_F}{2\pi} \theta(x) \bigg[\Big( \ln \frac{\mu^2}{m^2} -\ln x \Big) P_{qq}(x)
- \Big( \frac{\pi^2 }{3}-\frac{9 }{4} \Big) \delta (1-x) - 2\theta(1-x) (1-x)  \bigg] 
\,, \nn \\
D_q^{g\one}  (x,\mu)
&=  \frac{\al_s(\mu) C_F}{2\pi} \theta(x) \bigg[\Big( \ln \frac{\mu^2}{m^2} -\ln(1-x) \Big) P_{gq}(x)
 - 2\theta(1-x) x  \bigg] 
\,.
\end{align}
%%%

Our one-loop result in \eq{Gbare} agrees with the fact that the $\mu$-dependence of $\cG_i(s,z,\mu)$ is the same as in the jet function $J_i(s,\mu)$, at any order in $\al_s$. In particular, from \eq{GrenNLO} we find
%%%
\begin{align} \label{eq:ZGres}
Z_\cG^q (s,\mu) = \de(s) + \frac{\al_s(\mu) C_F}{\pi} \bigg\{ \frac{1}{\eps^2} \de(s) + 
 \frac{1}{\eps} \Big[- \frac{1}{\mu^2} \cL_0\Big(\frac{s}{\mu^2}\Big) + \frac{3}{4} \de(s) \Big] \bigg\}
\,,
\end{align}
%%%
which yields the known one-loop anomalous dimension for $J_q$ in \eq{ga_jet}. Furthermore, $\cG_q^g$ is UV finite and therefore the renormalization does not mix quark and gluon fragmenting jet functions. Moreover, the UV divergences in \eq{Gbare} are multiplied by $\delta(1-z)$ implying that the renormalization does not affect the $z$-dependence.

For the renormalized one-loop quark fragmenting jet function we obtain
%%%
\begin{align} \label{eq:Gren}
\frac{\cG_{q}^{q\one} (s,z,\mu)}{2(2\pi)^3}
& = \frac{\al_s(\mu) C_F}{2\pi}\, \theta(z)\, \bigg\{
\frac{2}{\mu^2} \cL_1\Big(\frac{s}{\mu^2}\Big) \de(1-z) 
+ \frac{1}{\mu^2} \cL_0\Big(\frac{s}{\mu^2}\Big) (1+z^2) \cL_0(1-z) 
\nn \\ & \quad
+ \de(s)\Big[P_{qq}(z) \ln \frac{\mu^2}{m^2} \!+\! (1\!+\!z^2)\cL_1(1\!-\!z) \!-\!\theta(1\!-\!z) (1\!-\!z) \!-\!
\Big( \frac{\pi^2}{2} \!-\! \frac{9}{4}\Big) \de(1\!-\!z) \Big] \bigg\}
\,, \nn \\
\frac{\cG_{q}^{g\one} (s,z,\mu)}{2(2\pi)^3}
& = \frac{\al_s(\mu) C_F}{2\pi}\, \theta(z)\, \bigg\{\Big[\frac{1}{\mu^2} \cL_0\Big(\frac{s}{\mu^2}\Big) +
\de(s) \ln \frac{z\mu^2}{m^2} \Big] P_{gq}(z) - \de(s)\, \theta(1-z) z \bigg\}
\,.
\end{align}
%%%

Inserting our results from \eqs{Dren}{Gren} into \eq{NLOmatch}, we produce the matching coefficients in \eq{Jresult}. The IR divergences cancel in the matching, as they should. 

A final cross-check on our calculation is provided by the relationship in \eq{jetrel} to ${\cal O}(\alpha_s)$. We correctly find that
%%%
\begin{align}
  \int_0^1\! \df z\, z \Big[\frac{\cJ_{qq}^\one(s,z,\mu)}{2(2\pi)^3} + \frac{\cJ_{qg}^\one(s,z,\mu)}{2(2\pi)^3}\Big]
   &= \frac{\al_s(\mu) C_F}{2\pi} \Big[\frac{2}{\mu^2} \cL_1\Big(\frac{s}{\mu^2}\Big) \!-\! \frac{3}{2\mu^2} \cL_0\Big(\frac{s}{\mu^2}\Big) + \Big(\frac{7}{2} \!-\! \frac{\pi^2}{2} \Big) \de(s) \Big]
\nn \\& 
  = J_q^{(1)}(s,\mu)   
   \,.
\end{align}
%%%
where $J_q^{(1)}(s,\mu)$ is the one-loop piece of the renormalized quark jet function in \eq{jetf}.

%%%%%%%%%%%%%%%%%%%%%%%%%%%%%%%%%%%%%%%%%%%%%%%%%%%%%%%%%%%%%%%%%%%%%%%%%%%%%%%%
\section{Gluon Matching Calculation in Pure Dimensional Regularization}
\label{sec:gluondimreg}
%%%%%%%%%%%%%%%%%%%%%%%%%%%%%%%%%%%%%%%%%%%%%%%%%%%%%%%%%%%%%%%%%%%%%%%%%%%%%%%% 

In this section we calculate the matching coefficients $\cJ_{g j}$, using dimensional regularization for both UV and IR divergences. 
Since we cannot distinguish UV and IR divergences, we will not be able to check that $D_g^i$ and $\cG_g^i$ have the anomalous dimensions in \eqs{gaD}{gaG} \emph{and} that the IR divergences cancel in the matching \eq{NLOmatch}. We have verified both of these statements for the quark case in section \ref{sec:gluondelta}, where we used dimensional regularization for the UV and a gluon mass plus a $\delta$-regulator for the IR. Here, for simplicity, we will assume the anomalous dimensions in \eqs{gaD}{gaG} and verify that the IR divergences cancel in the matching. An equivalent procedure would be to assume that the IR divergences cancel in the matching and verify that we obtain the expected anomalous dimension for ${\cal G}_g^h$. An additional check on the UV-finite part of our calculation will come from the relation with the jet function in \eq{jetrel}.

%%%
\begin{figure}[t]
\centering
\includegraphics[width=0.9\textwidth]{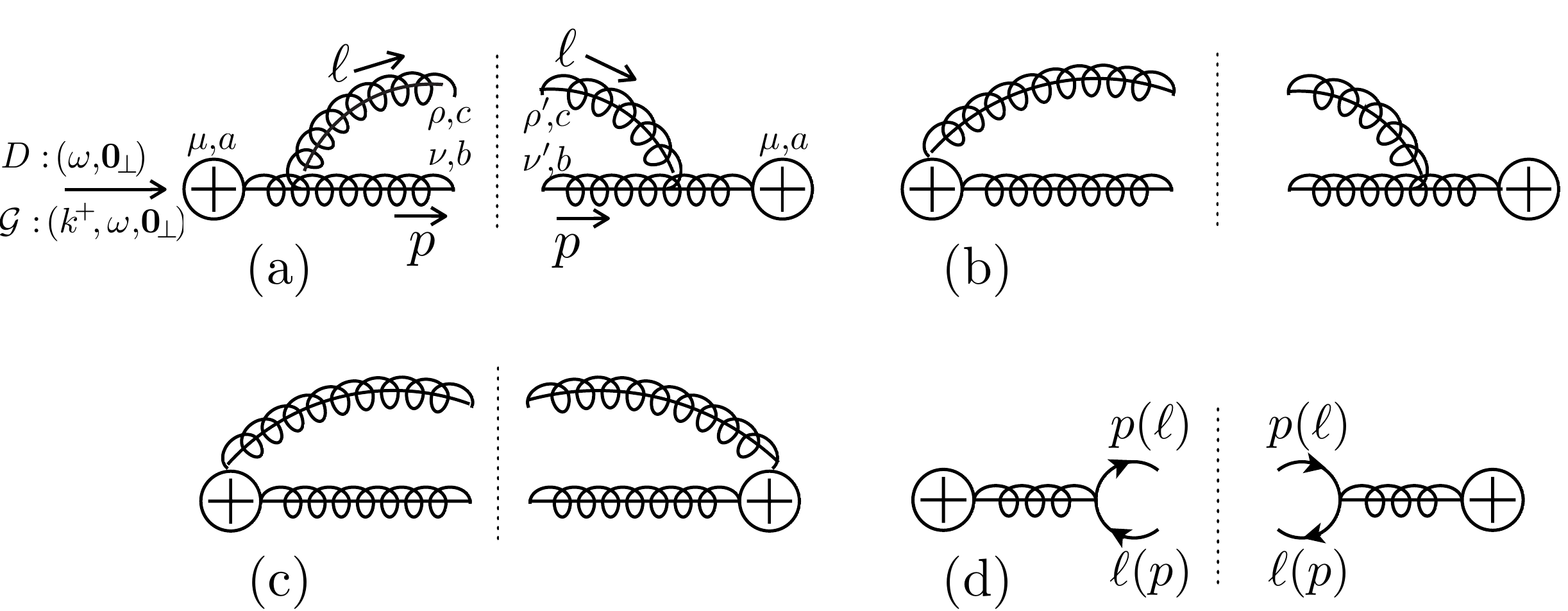}
\caption{Feynman graphs contributing to the gluon fragmentation function and the gluon fragmenting jet function at one-loop. We have not shown virtual diagrams here as they are scaleless and therefore trivially vanish in DR. Graphs (a) through (c) correspond to ${\cal G}_g^g$ and graph (d) to ${\cal G}_g^{q(\bar q)}$. Graphs (b) has a mirror image.}
\label{fig:DGggraphs}
\end{figure}
%%%

The partonic graphs for the gluon fragmentation function and gluon fragmenting jet function are shown in figure~\ref{fig:DGggraphs}. In pure dimensional regularization all integrals for the bare fragmentation function are scaleless, therefore they vanish. Inserting the known one-loop anomalous dimensions from \eq{gaD} into \eq{DrenNLO}, we find that the renormalized fragmentation functions up to one-loop are given by
%%%
\begin{align} \label{eq:DgNLO}
  D_g^{g}(x,\mu) & = \de(1-x) - \frac{1}{\eps} \frac{\al_s(\mu)}{2\pi}\, \theta(x)\, \Big[C_A P_{gg}(x) + \frac{1}{2} \bt_0 \de(1-x)\Big]
  \,, \nn \\
  D_g^{q}(x,\mu) & = -\frac{1}{\eps} \frac{\al_s(\mu) T_F}{2\pi}\, \theta(x)\, P_{qg}(x)
  \,,
\end{align}
%%%
where the $1/\eps$-poles are IR divergences.

For the fragmenting jet function the real emission graphs can give a non-zero contribution, because the measurement of $k^+$ now provides the Lorentz invariant quantity $s=\w k^+$ as a scale in the calculation. The virtual graphs are still scaleless because $k^+ = p^+ = 0$.  It is easiest to calculate using the sum over physical polarizations. Then the real emission graphs only contribute to physical degrees of freedom in the final state. The physical polarization sum in light-cone coordinates reads
%%%
\begin{equation} \label{eq:polsum}
  \sum_\text{pol} \ve^*_\mu(p) \ve_\nu(p) = 
  -g^\perp_{\mu\nu} + \frac{\bn_\mu p^\perp_\nu}{\bn \sdt p} + \frac{p^\perp_\mu \bn_\nu}{\bn \sdt p} - \frac{\bn_\mu \bn_\nu p_\perp^2}{(\bn \sdt p)^2}
\,,
\end{equation}
%%%
which gives a vanishing contribution for diagrams~\ref{fig:DGggraphs}(b) and \ref{fig:DGggraphs}(c) on contracting Lorentz indices with the 2-gluon vertex from the operator insertion; only graph~\ref{fig:DGggraphs}(a) contributes to $\cG_g^g$.
For figure~\ref{fig:DGggraphs}(a) we find 
%%%
\begin{align}
\frac{{\cal G}_{g,\rm{bare}}^{g(a)}}{2(2\pi)^3} & =  \Big(\frac{e^{\ga_E} \mu^2}{4\pi}\Big)^\eps \frac{-\theta(z)}{(d\!-\!2)(N_c^2 \!-\! 1) z} \int\! \df^{d-2} p_\perp \!\! \int\! \frac{\df^d \ell}{(2\pi)^{d-1}} \theta(\ell^0) \delta(\ell^2)  \delta(\omega \!-\! p^- \!-\! \ell^-) \delta^{d-2}(p_\perp \!+\! \ell_\perp) 
\nn \\ & \quad \times
 \delta(k^+ \!-\! \ell^+ \!-\! p^+) 
g f^{abc} \big[  g_\perp^{\mu\nu} (\ell + 2p)^\rho + g^{\nu\rho} (\ell-p)_\perp^\mu + g_\perp^{\rho\mu} (-2\ell - p)^\nu \big] \sum_\pol \ve^*_\rho(\ell) \ve_{\rho'}(\ell) 
\nn \\[-0.5ex] & \quad \times
\sum_\pol \ve^*_\nu(p) \ve_{\nu'}(p)\,
g f^{abc} \big[\de_{\perp\mu}^{\nu'} (-\ell - 2p)^{\rho'} \!+\! g^{\nu'\rho'} (-\ell+p)_{\perp\mu} \!+\! \de_{\perp\mu}^{\rho'} (2\ell + p)^{\nu'} \big]  \Big[ \frac{-\img}{(\ell\!+\!p)^2} \Big]^2 
\nn \\ &
=  \Big(\frac{e^{\ga_E} \mu^2}{4\pi}\Big)^\eps 2g^2 C_A
\frac{\theta(z)}{z s^2 }
\int\! \frac{\df^d \ell}{(2\pi)^{d-1}} 
\theta(\ell^0) \delta(\ell^2)  \delta(\omega - p^- - \ell^-)
 \delta(k^+ - \ell^+ - p^+) \nn \\ &
 \quad \times -\frac{\ell_\perp^2}{z(1-z)} \Bigl[\frac{2z}{1\!-\!z} \!+\! \frac{2(1\!-\!z)}{z} \!+\! 2 z(1\!-\!z)\Bigr]
\nn \\ & 
= \frac{\al_s(\mu) C_A}{2\pi} \frac{(e^{\ga_E} \mu^2)^\eps}{\Gamma(1-\eps)}
\frac{\theta(z)\theta(1-z)}{z^\eps(1-z)^\eps} \frac{\theta(s)}{s^{1+\eps}} \Bigl[\frac{2z}{1\!-\!z} \!+\! \frac{2(1\!-\!z)}{z} \!+\! 2 z(1\!-\!z)\Bigr]
\nn \\ & 
= \frac{\al_s(\mu) C_A}{2\pi} \theta(z) \bigg\{ \Big[\frac{2}{\eps^2}\, \de(s) - \frac{2}{\eps} \frac{1}{\mu^2} \cL_0\Big(\frac{s}{\mu^2}\Big) \Big] \de(1\!-\!z) - \frac{1}{\eps} \de(s) P_{gg}(z) +
\frac{2}{\mu^2}\cL_1\Big(\frac{s}{\mu^2}\Big) \de(1\!-\!z) 
\nn \\ & \quad
+ \frac{1}{\mu^2} \cL_0\Big(\frac{s}{\mu^2}\Big) P_{gg}(z) +
\de(s) \Big[ \cL_1(1-z) \frac{2(1-z+z^2)^2}{z} + P_{gg}(z) \ln z - \frac{\pi^2}{6} \de(1-z) \Big] \bigg\}
 \,,
\end{align}
%%%
where $s= (\ell+p)^2 = \w k^+$ and $z=p^-/\w$ as always. In the first line we already used $p_\perp+\ell_\perp = 0$ to simplify the Feynman rules for the ${\cal B}_\perp^{\mu a}$ operator. In the first step we work out the rather tedious contractions of Lorentz indices, using \eq{polsum} and set $\ell^- = (1-z) \w$ owing to $\delta(\w - p^- - \ell^-)$. In the second step we perform the straightforward $\delta$-function integrals that are much alike quark fragmenting jet function calculation.
In the final step we expand in $\eps$, dropping $\ord{\eps}$ terms.

Finally, we calculate the mixing graph in \fig{DGggraphs}(d) 
%%%
\begin{align}
\frac{{\cal G}_{g,\rm{bare}}^{q(d)}}{2(2\pi)^3} & =  \Big(\frac{e^{\ga_E} \mu^2}{4\pi}\Big)^\eps \frac{-\theta(z)}{(2\!-\!2\eps)(N_c^2 \!-\! 1)z} \int \df^{d-2} p_\perp \int \frac{\df^d \ell}{(2\pi)^{d-1}} \theta(\ell^0) \delta(\ell^2)  \delta(\omega \!-\! p^- \!-\! \ell^-) \delta^{d-2}(p_\perp \!+\! \ell_\perp)
\nn \\ & \quad \times
\delta(k^+ \!-\! \ell^+ \!-\! p^+)\, \sum_{s,s'}  \bar u_n^s(p)\, \img g \ga_\perp^\mu T^a\, v_n^{s'}(\ell)\, \bar v_n^{s'}(\ell)
\, \img g \ga^\perp_\mu \, u_n^s(p)   \Big[ \frac{-\img}{(\ell+p)^2}\Big]^2
\nn \\ &
= \frac{\al_s(\mu) T_F}{2\pi} \frac{(e^{\ga_E} \mu^2)^\eps}{\Ga(2-\eps)} \frac{\theta(z)\theta(1-z)}{z^\eps (1-z)^\eps} [(1-\eps)-2z(1-z)] \frac{\theta(s)}{s^{1+\eps}}
\nn \\ &
= \frac{\al_s(\mu) T_F}{2\pi}\, \theta(z) \bigg\{ \Big[-\frac{1}{\eps}\, \de(s) + \frac{1}{\mu^2} \cL_0\Big(\frac{s}{\mu^2}\Big) + \de(s) \ln[z(1-z)] \Big]  P_{qg}(z) 
\nn \\ & \quad
+ 2 \de(s) \theta(1-z) z(1-z)\bigg\}
\, ,
\end{align}
which we notice is invariant under the transformation $z \to 1-z$. This is expected because the diagram is symmetric under the exchange of quark and anti-quark lines. Therefore
\begin{align}
{\cal G}_{g,\rm{bare}}^{q(d)} & = {\cal G}_{g,\rm{bare}}^{\bar q(d)} \, .
\end{align}
%%%

Using the known anomalous dimension of the gluon jet function from \eq{gaG}, we obtain the renormalized one-loop fragmenting jet function with the aid of \eq{GrenNLO},
%%%
\begin{align}
\frac{\cG_g^{g\one}(s,z,\mu)}{2(2\pi)^3}
& \!=\! \frac{\al_s(\mu) C_A}{2\pi} \theta(z) \bigg\{ - \frac{1}{\eps} \de(s) \Big[P_{gg}(z) + \frac{\bt_0}{2C_A} \de(1-z)\Big] +
\frac{2}{\mu^2}\cL_1\Big(\frac{s}{\mu^2}\Big) \de(1-z) 
\nn \\ & \quad
\!+ \frac{1}{\mu^2} \cL_0\Big(\frac{s}{\mu^2}\Big) P_{gg}(z)
 + \de(s) \Big[ \cL_1(1\!-\!z) \frac{2(1\!-\!z\!+\!z^2)^2}{z} \!+\! P_{gg}(z) \ln z \!-\! \frac{\pi^2}{6} \de(1\!-\!z) \Big] \bigg\}
 \,, \nn \\
 \frac{\cG_g^{q\one}(s,z,\mu)}{2(2\pi)^3}
& =   \frac{\cG_g^{\bar q\one}(s,z,\mu)}{2(2\pi)^3} = \frac{\al_s(\mu) T_F}{2\pi}\, \theta(z) \bigg\{ \Big[\!-\!\frac{1}{\eps}\, \de(s) \!+\! \frac{1}{\mu^2} \cL_0\Big(\frac{s}{\mu^2}\Big) \!+\! \de(s) \ln[z(1\!-\!z)] \Big]  P_{qg}(z) 
\nn \\ & \quad \hspace{5.5cm}
+ 2 \de(s) \theta(1-z) z(1-z)\bigg\}
\, .
\end{align}
%%%
Here the $1/\eps$-poles are of IR origin.
By subtracting the one-loop fragmentation functions of \eq{DgNLO}, we arrive at the matching coefficients given in \eq{Jgresult}. Note that the IR divergences again cancel, as they should.

As a final check, we verified that these results satisfy the relationship in \eq{jetrel} with the jet function,
%%%
\begin{align} 
  \int_0^1\! \df z\, z \Big[\frac{\cJ_{gg}^\one(s,z,\mu)}{2(2\pi)^3} +  n_f \frac{\cJ_{gq}^\one(s,z,\mu)}{2(2\pi)^3} +  n_f \frac{\cJ_{g \bar q}^\one(s,z,\mu)}{2(2\pi)^3} \Big]
  = J_g^{(1)}(s,\mu)   
   \,.
\end{align}
%%%
Here, we included the factor $n_f$ to account for the number of light quark and antiquark flavors. The $J_g^{(1)}(s,\mu)$ denotes the one-loop terms of the renormalized gluon jet function in \eq{jetf}.

%%%%%%%%%%%%%%%%%%%%%%%%%%%%%%%%%%%%%%%%%%%%%%%%%%%%%%%%%%%%%%%%%%%%%%%%%%%%%%%%
\section{Numerical Analysis}
\label{sec:appl}
%%%%%%%%%%%%%%%%%%%%%%%%%%%%%%%%%%%%%%%%%%%%%%%%%%%%%%%%%%%%%%%%%%%%%%%%%%%%%%%%

Here we present plots of quark and gluon fragmenting jet functions for $\pi^+$-production (section~\ref{subsec:Gresults}), and a numerical study up to NNLL accuracy of single $\pi^+$-fragmentation in $e^+ e^-$ collisions where a cut on thrust is imposed (section~\ref{subsec:siresults}). 

As input we use the HKNS fragmentation functions~\cite{Hirai:2007cx}. For simplicity we will always utilize $D_i^{\pi^+}(z)$ at NLO, even though our formal counting (cf. table \ref{tab:counting} on page 31) would imply the use of LO fragmentation functions at LO, LL and NLL. Consistently with ref.~\cite{Hirai:2007cx}, we set $\alpha_s(m_Z) = 0.125$, used two-loop running, and matched $\alpha_s$ continuously across the $b$- and $c$-quark thresholds at $m_b = 4.3 \GeV$ and $m_c = 1.43 \GeV$. In our analysis we will not include the effects of the uncertainties associated with the fragmentation functions or $\al_s(m_Z)$.

%===============================================================================
\subsection{Fragmenting Jet Functions up to NLO}
\label{subsec:Gresults}
%===============================================================================

%%%
\begin{figure}[t]
\includegraphics[width=0.49\textwidth]{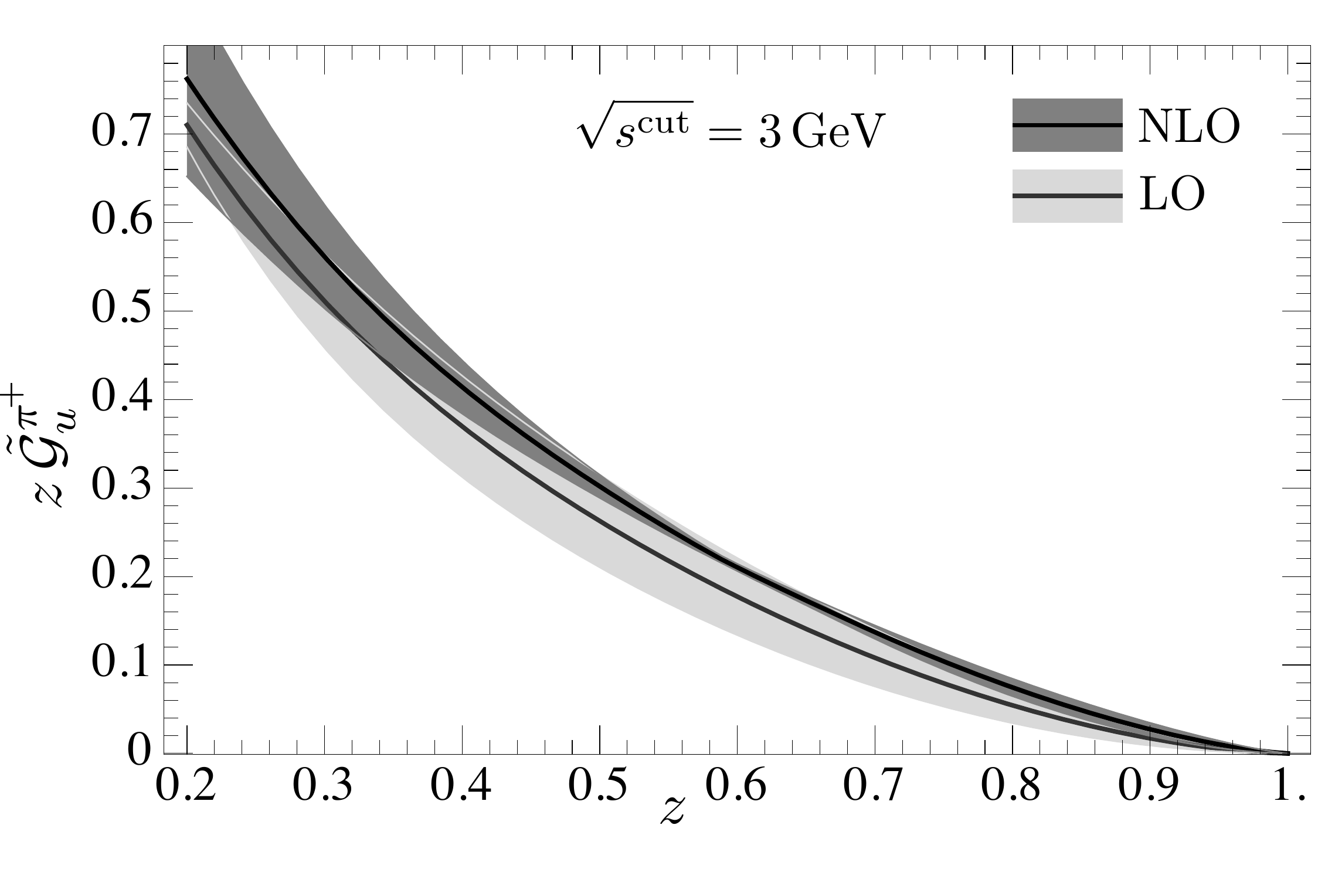} \hfill
\includegraphics[width=0.49\textwidth]{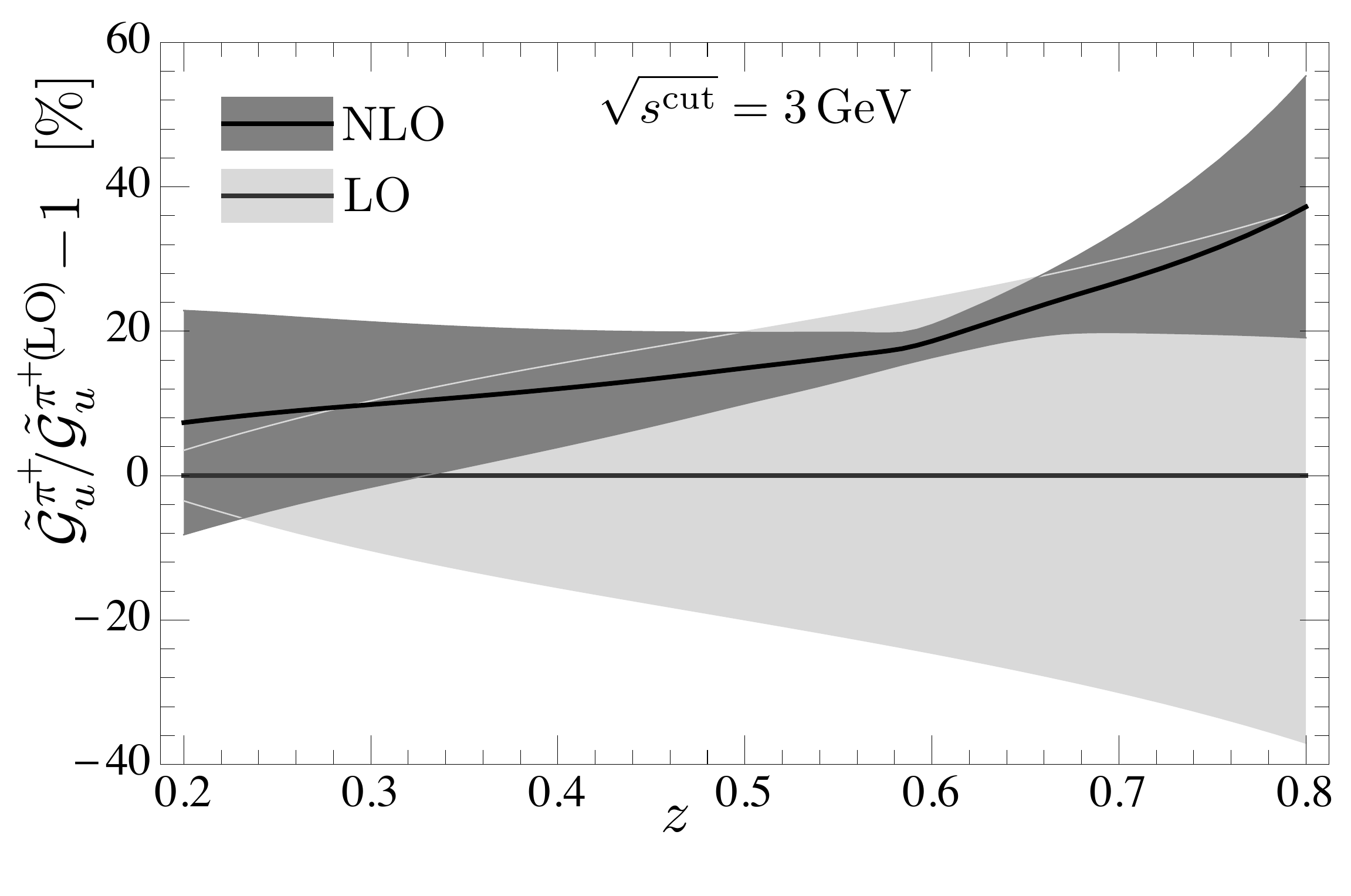} \\
\includegraphics[width=0.49\textwidth]{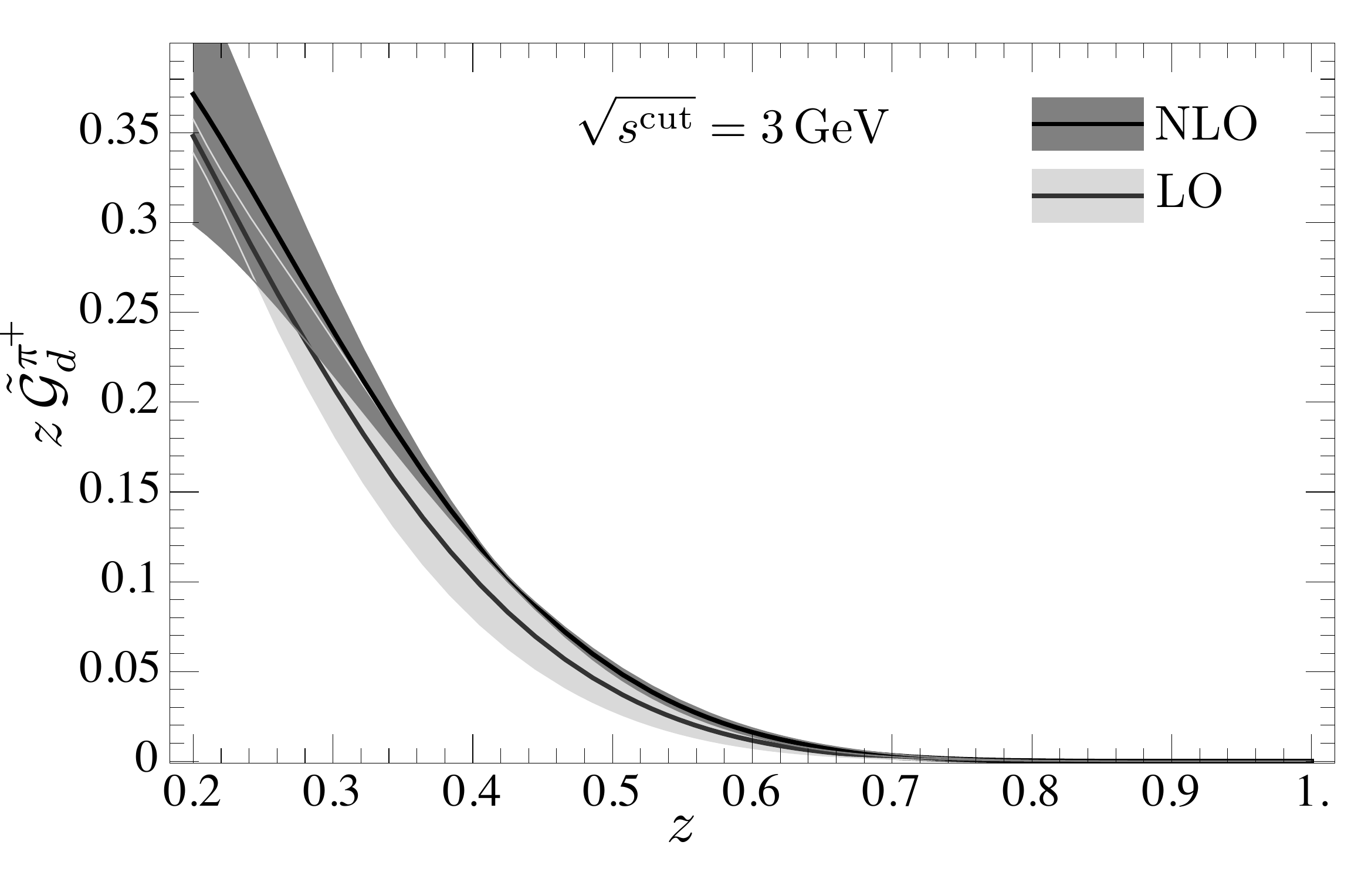} \hfill
\includegraphics[width=0.49\textwidth]{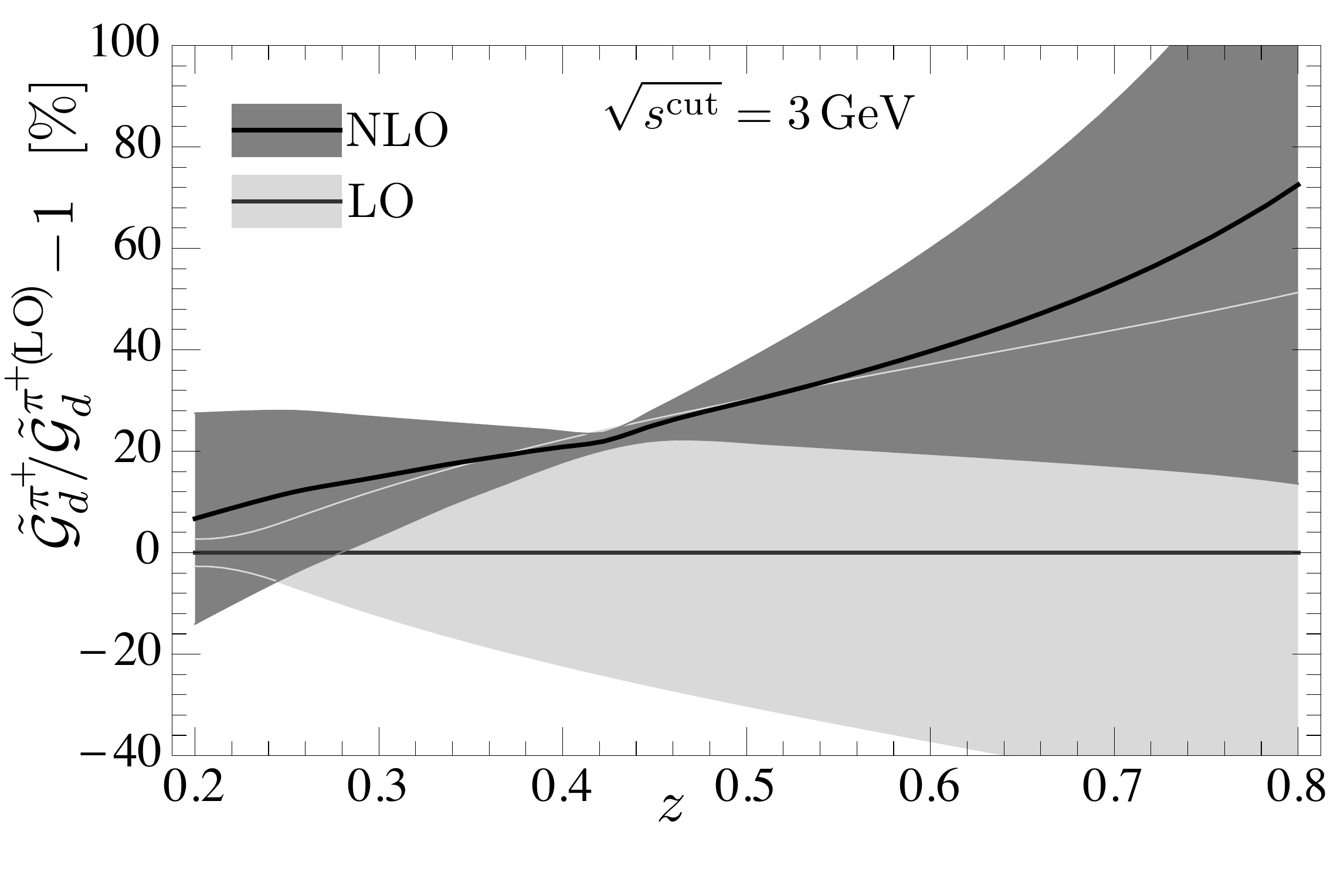}
\caption{Fragmenting jet functions for the $\pi^+$-fragmentation from a $u$-quark (top row) and a $d$-quark (bottom row). Shown are the LO and NLO results with the corresponding perturbative uncertainties, as explained in the text. The left panels display $z\, \tG_i^{\pi^+}\!(s^\cut,z, \mu)$ for $s^\cut=(3\GeV)^2$ as function of $z$, at the jet scale $\mu \simeq \sqrt{s^\cut}$. The right panels show the same curves relative to the LO.}
\label{fig:Gplots}
\end{figure}
%%%

%%%
\begin{figure}[t]
\includegraphics[width=0.49\textwidth]{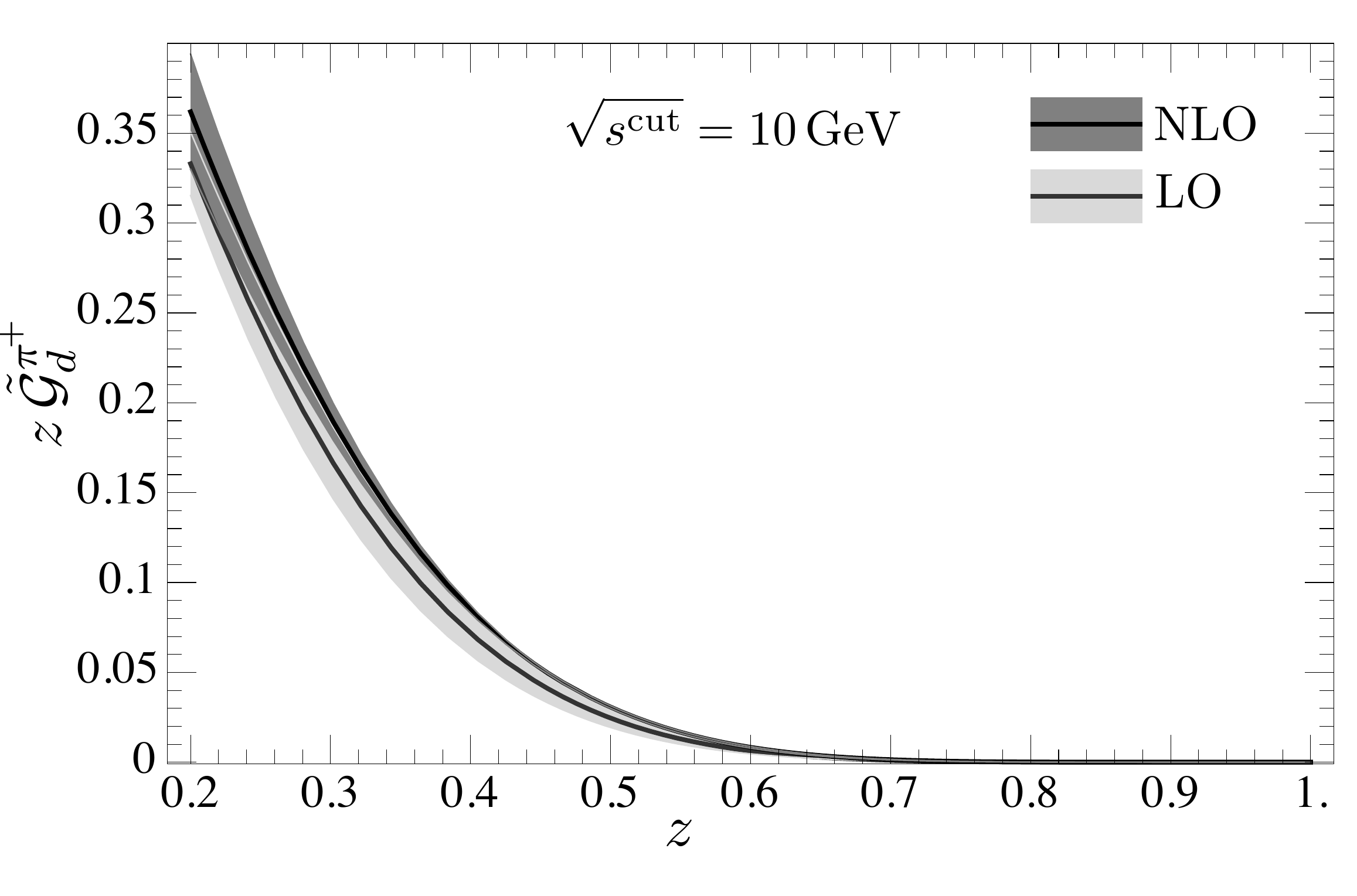} \hfill
\includegraphics[width=0.49\textwidth]{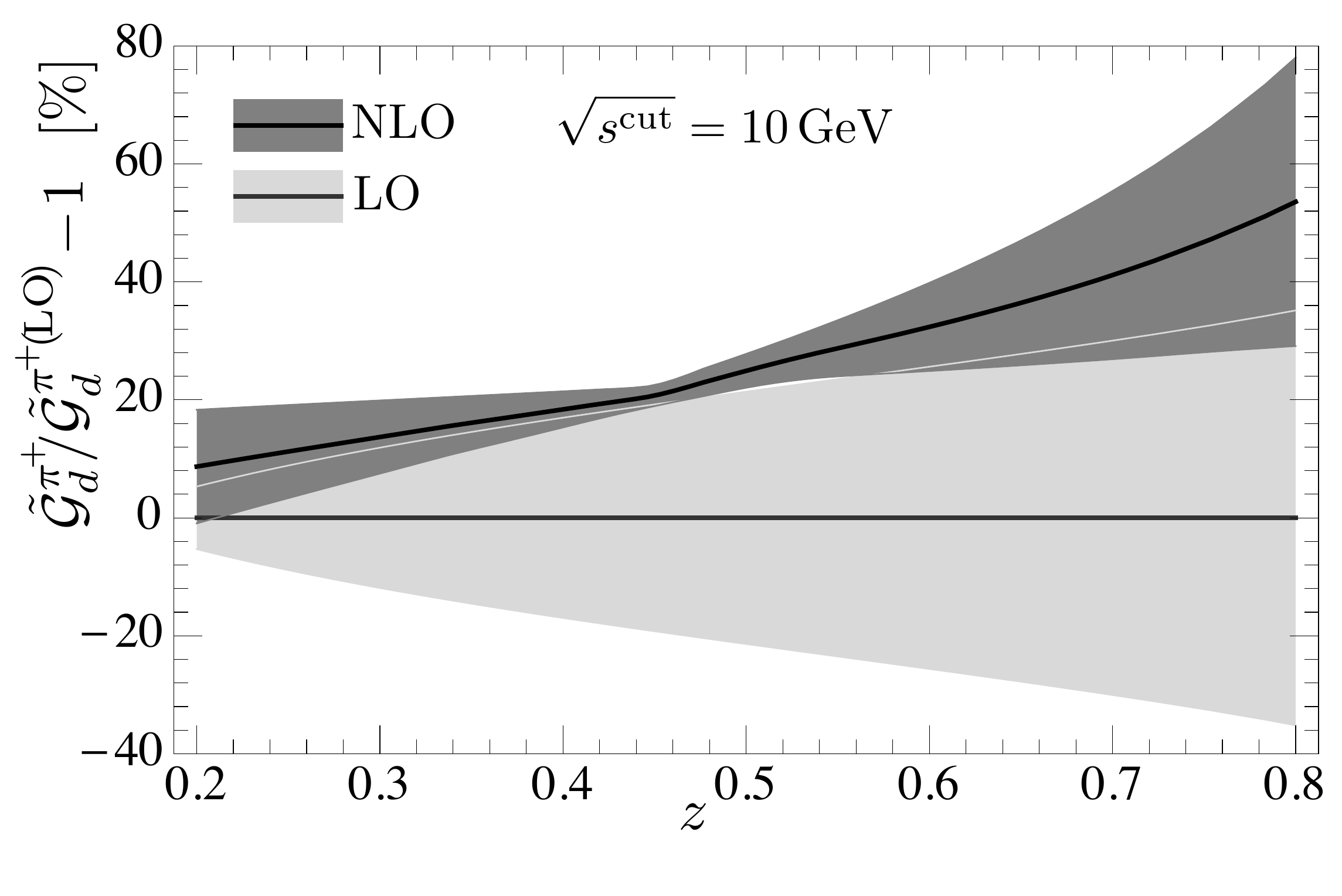}
\includegraphics[width=0.49\textwidth]{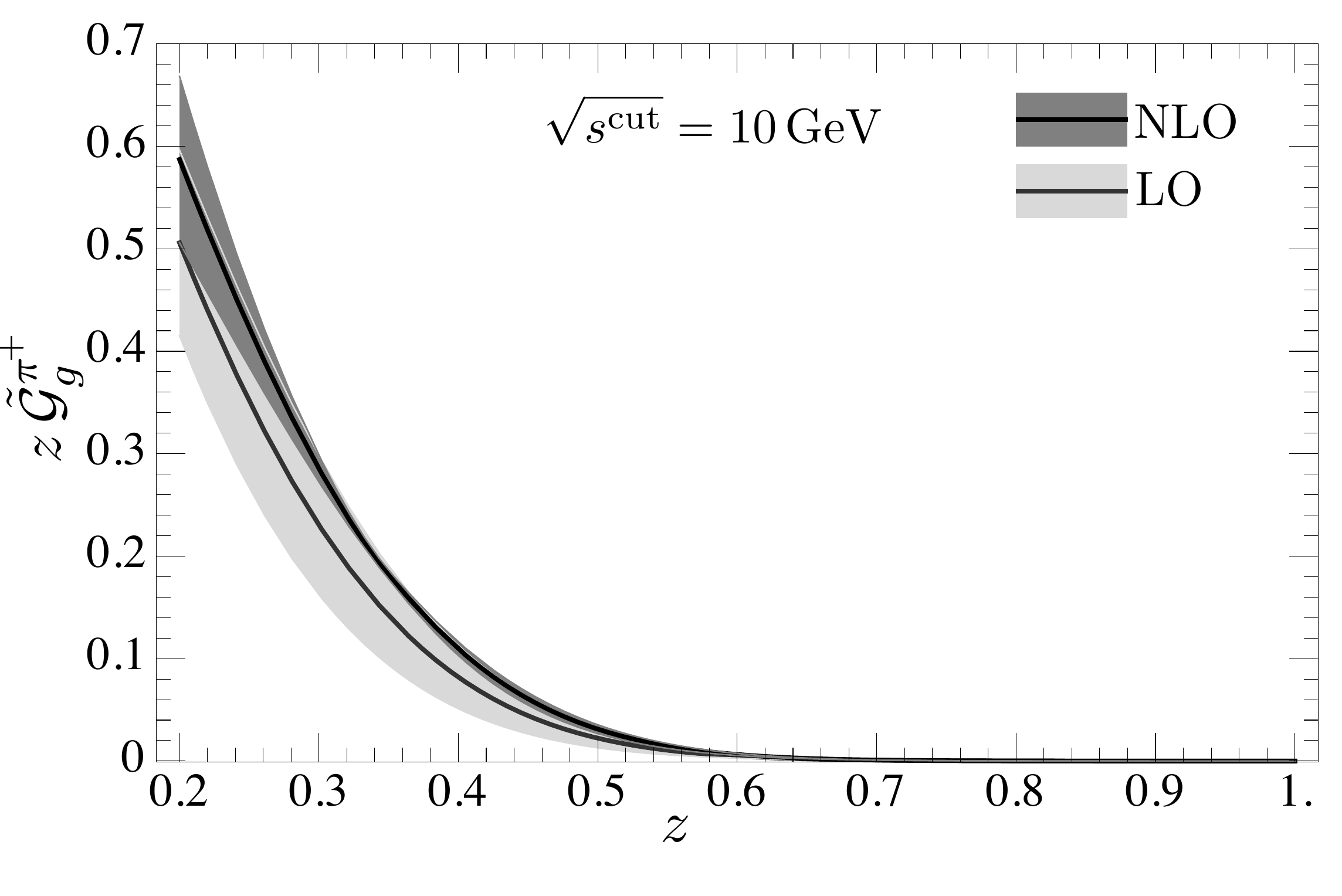} \hfill
\includegraphics[width=0.49\textwidth]{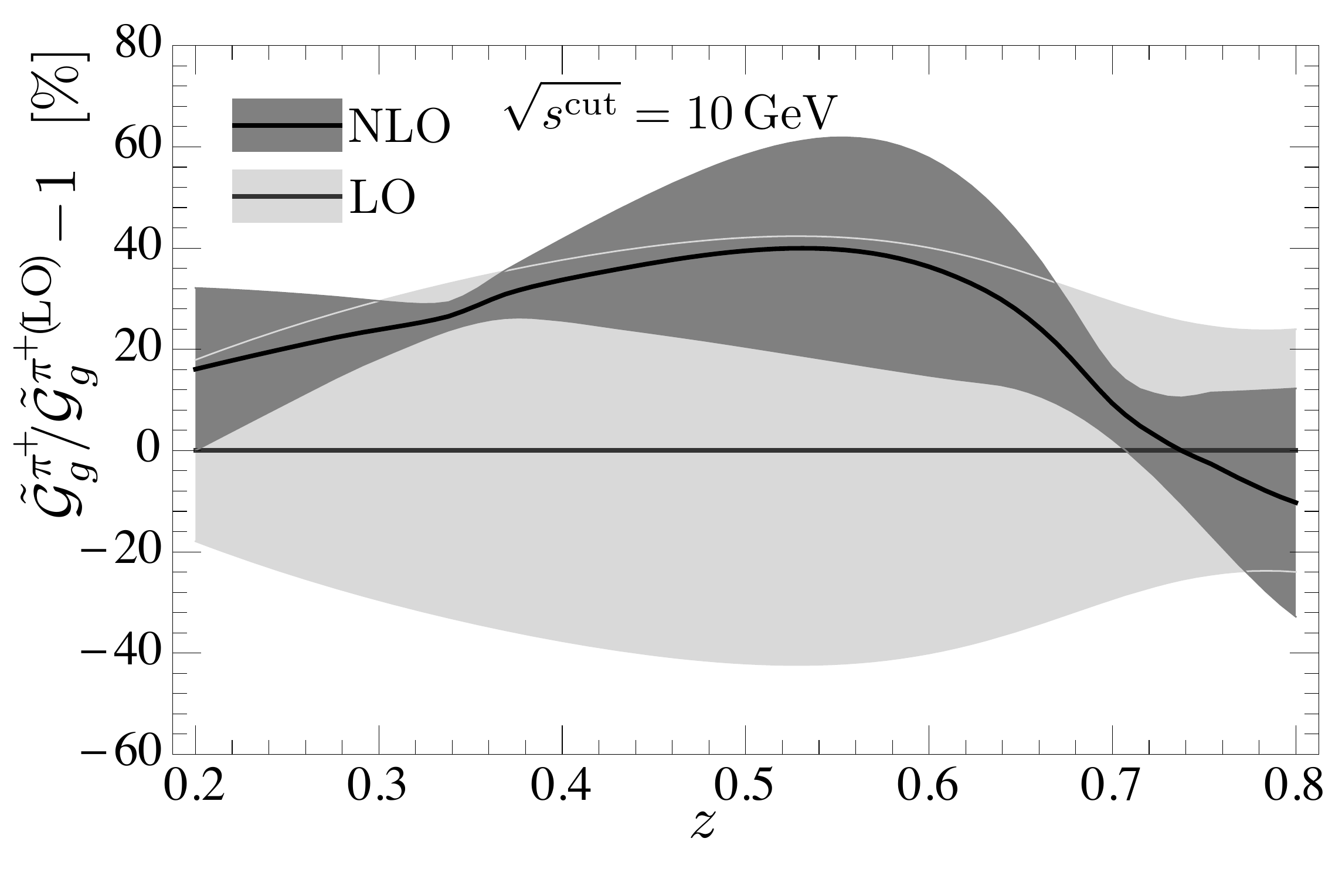}
\caption{Same as figure \ref{fig:Gplots} but for $d$-quark (top row) and gluon (bottom row) initiated fragmentation of $\pi^+$ with $\mu \simeq \sqrt{s^\cut} = 10 \GeV$.}
\label{fig:Gplotsb}
\end{figure}
%%%

In this section, we will show plots for the fragmenting jet functions, using \eq{GisD} and our results for $\cJ_{ij}$. We will study the (dimensionless) integral of the fragmenting jet functions over the jet invariant mass up to $s^\cut$
%%%
\begin{equation} \label{eq:cGintegral}
 \tG_i^h(s^\cut,z,\mu) = \frac{1}{2(2\pi)^3} \int_0^{s^\cut}\!\! \df s\, \cG_i^h(s,z,\mu)
\end{equation}
%%%
for scales $\mu \simeq \sqrt{s^{\rm cut}}$, where there are no large logarithms involving the variable $s$.
Our plots show the effects of  constraining the invariant mass of the jet both on the normalization and on the shape of the $z$-dependence.
In figure \ref{fig:Gplots} we show the $\pi^+$-fragmenting jet function $\tG_i^{\pi^+}\!(s^\cut,z,\mu)$ for a representative fixed value of $s^\cut = (3 \GeV)^2$ as a function of $z$. This choice of cut is motivated by the next section, where we consider $e^+e^-$ collisions at $Q=10.6\GeV$. The top row in figure \ref{fig:Gplots} corresponds to fragmentation from a $u$-quark, and the bottom row to fragmentation from a $d$-quark. The $\bar u$- and $\bar d$-quark results are identical to those for the $d$- and $u$-quark, as a consequence of the fit ansatz in the extraction of the HKNS fragmentation functions. The left panel shows $z\, \tG_i^{\pi^+}\!(s^\cut,z, \mu)$ and the right panel displays the corrections relative to the LO result $\tG_i^{\pi^+ ({\rm LO})}(s^\cut, z, \mu) = D_i^{\pi^+}(z, \mu)\, \theta(s^\cut)$.  To obtain the fragmenting jet function at a scale different from the jet scale, one can use the RGE in \eq{GRGE}, which does not affect the $z$-dependence. 

The plots are cut off at $z=0.2$, because our factorization formula is not valid for too small values of $z$. 
For $1-z \ll 1$, \eq{cGintegral} contains large logarithms of $1-z$ [see \eq{Jresult}], which should be summed. We avoid this issue by cutting off the relative plots at $z=0.8$. As is clear from the absolute plots, $z\, \tG_i^{\pi^+}\!(s^\cut,z, \mu)$ is very small in this region anyways.

To estimate the uncertainty from higher-order corrections, we vary $\mu$ between $\sqrt{s}/2$ and $2\sqrt{s}$, which is shown by the uncertainty bands. At LO the scale variation is simply that of the fragmentation function, whose maximum and minimum are obtained at $\mu = \sqrt{s}/2, 2\sqrt{s}$ with central value at $\mu = \sqrt{s}$. At NLO, the maximum and minimum for the $\mu$ variation do not occur at these values, due to the double logarithms in the $\cJ_{ij}$. We therefore sample over $\sqrt{s}/2 \leq \mu \leq 2\sqrt{s}$ to determine the uncertainty band, and take the central value to 
be the average of the maximum and minimum variation.

As can be seen in figure \ref{fig:Gplots}, for the $u$-quark the uncertainty at NLO is less than at LO and the uncertainty bands overlap, so perturbation theory is well-behaved. For the $d$-quark the uncertainty bands overlap as well, but they become rather large for $z \gtrsim 0.7$, both for the LO and NLO result. For these values of $z$, $D_d^{\pi^+}(z, \mu=1\,{\rm GeV})$ is tiny~\cite{Hirai:2007cx} and its increase is due to the running up to $\mu \simeq \sqrt{s^\cut}$, which leads to these large uncertainties. For larger scales this effect decreases: in the top row of figure \ref{fig:Gplotsb} we show the $d$-quark fragmenting jet function for $\mu \simeq \sqrt{s^\cut} = 10\GeV$, where the uncertainties are smaller and decrease from LO to NLO. The gluon fragmenting jet function exhibits the same feature, which is why we again choose $\mu \simeq \sqrt{s^\cut} = 10\GeV$ for the plots in the bottom row of figure \ref{fig:Gplotsb}. This scale choice is also relevant for hadron colliders, where one would expect to see gluon-initiated jets of higher invariant mass.

%===============================================================================
\subsection{Fragmentation at $e^+e^-$ Collisions, with a Cut on Thrust}
\label{subsec:siresults}
%===============================================================================

We will now show results for $e^+ e^- \to X\pi^+$ in the dijet limit, where the fragmentation variable $z$ is measured. We remind the reader that in our choice of frame the jet's perpendicular momentum vanishes. 

Including the RGE evolution kernels in \eq{factth}, the resummed cross section is given by
%%%
\begin{align} \label{eq:factth2}
\frac{\df^2\sigma}{\df \tau\, \df z}
&= 
H(Q^2, \mu_H)\, U_H(Q^2, \mu_H, \mu) \sum_{q=\{u, \bar u, d, \bar d, s, \bar s\}} \frac{\si_0^q}{2(2\pi)^3} 
\int\!\df s_a\,\df s_b
\nn\\ &\quad \times
\int\!\df s_a'\, \cG_q^h(s_a - s_a', z, \mu_J)\, U^q_\cG(s_a', \mu_J, \mu)
\int\!\df s_b'\, J_{\bar q}(s_b - s_b', \mu_J)\, U^q_J(s_b', \mu_J, \mu)
\nn\\ &\quad \times
\int\!\df k\, Q\, S_\tau\Bigl(Q\,\tau - \frac{s_a + s_b}{Q} - k, \mu_S \Bigr)\, U_S(k, \mu_S, \mu)
\,,\end{align}
%%%
where we sum over light quark flavors.
This formula only describes the singular contribution to the cross section, which goes like $\sim (\ln^k \tau)/\tau$ for small $\tau$. The nonsingular contribution is suppressed by $\ord{\tau}$ relative to the singular one and we therefore neglect it in our numerical analysis of the dijet limit $\tau \ll 1$.
The double logarithms of $\tau$ are summed by evaluating the hard, (fragmenting) jet, and soft function at their natural scales ($\mu_H \simeq Q$, $\mu_J \simeq \sqrt{\tau} Q$, $\mu_S \simeq \tau Q$, respectively) and then evolving them to an arbitrary common scale $\mu$. In terms of the Fourier conjugate variable of $\tau$, denoted by $y$, the cross section takes the following form
%%%
\begin{equation} \label{eq:logstruct}
\ln \frac{\df\sigma}{\df y} \sim \ln y (\alpha_s \ln y)^k + (\alpha_s \ln y)^k + \alpha_s (\alpha_s \ln y)^k + \dots
\,,
\end{equation}
%%%
where the $k$ runs over the positive integers. The terms on the right-hand side correspond to the LL, NLL and NNLL series. For $1-z \ll 1$ we cannot trust the convergence due to the large double logarithms of $1-z$, as discussed in the previous section. 

%%%
\begin{table}
  \centering
  \begin{tabular}{l | c c c c}
  \hline \hline
  & matching & $\gamma_x$ & $\Gamma_\cusp$ & $\beta$  \\ \hline
  LO & $0$-loop & - & - & $1$-loop \\
  NLO & $1$-loop & - & - & $2$-loop \\
  LL & $0$-loop & - & $1$-loop & $2$-loop\\  
  NLL & $0$-loop & $1$-loop & $2$-loop & $2$-loop\\
  NNLL & $1$-loop & $2$-loop & $3$-loop & $3$-loop\\
  \hline\hline
  \end{tabular}
\caption{Order counting in fixed-order and resummed perturbation theory.}
\label{tab:counting}
\end{table}
%%%

To calculate the cross section in \eq{factth2} at a specific order, we need the input summarized in table \ref{tab:counting}, where ``matching" refers to the fixed-order contribution, $\ga_x$ to the non-cusp anomalous dimension, $\Ga_\cusp$ to the cusp anomalous dimension and $\beta$ to the QCD $\beta$-function. The evolution factors and the one-loop hard, jet and soft function are all known and collected in appendix \ref{app:pert}. Our one-loop calculation of the matching coefficients $\cJ_{ij}$ is the remaining ingredient necessary to sum the logarithms of $\tau$ to NNLL order. The convolutions of plus distributions are carried out using the identities from appendix B of ref.~\cite{Ligeti:2008ac}.

We will now address our choice of scale for the central value of the cross section, as well as the scale variations used to estimate the perturbative uncertainties. We start by observing that the hard function for $e^+e^- \to $ dijet is the square of a time-like form factor and contains large $\pi^2$-terms for $\mu_H = Q$ from $\ln^2(-\img Q/\mu_H)$. To improve convergence we resum these $\pi^2$-terms by taking $\mu_H=-\img Q$~\cite{Parisi:1979xd, Sterman:1986aj, Magnea:1990zb, Eynck:2003fn}.

Following ref.~\cite{Abbate:2010xh}, we observe that there are three distinct kinematic regions where the resummation of the logarithms of $\tau$ must be handled differently:
%%%
\begin{align}
\text{1)}& \quad
\mu_H \simeq -\img Q\,,\qquad \mu_J \simeq \sqrt{\lqcd Q}\,, \qquad \mu_S = \lqcd
\,, \hspace{20ex}\nn\\
\text{2)}& \quad
\mu_H \simeq -\img Q\,, \qquad \mu_J \simeq \sqrt{\tau} Q\,, \qquad \mu_S \simeq \tau Q
\,,\nn\\
\text{3)}& \quad
\img \mu_H = \mu_J = \mu_S \simeq Q
\,.\nn\end{align}
%%%
Here we shall focus on region 2). However, our choice of scales and the scale uncertainties are affected because we get close to the regimes 1) and 3). In region 1) $\tau$ is small and the soft scale becomes of order $\lqcd$. The factorization theorem in~\eq{factth2} remains valid, but non-perturbative corrections to the soft function must be taken into account, which can be done using the methods of refs.~\cite{Hoang:2007vb, Ligeti:2008ac}. On the other hand, for large $\tau$ the resummation becomes irrelevant (except for the $\pi^2$ resummation) and nonsingular corrections should be taken into account. As was observed in ref.~\cite{Abbate:2010xh}, there is an important cancellation between the singular and nonsingular cross section in \eq{factth2} in the limit $\tau \to 0.5$, which requires the scales to merge in region 3).

Our choice of scales should smoothly connect these regions, which we achieve using profile functions. This approach has been previously used to analyze the $B\to X_s\gamma$ spectrum~\cite{Ligeti:2008ac}, thrust in $e^+e^-\to $ jets~\cite{Abbate:2010xh} and Higgs production through gluon fusion with a jet veto~\cite{Berger:2010xi}. Here we use the same profile functions as in eqs.~(2.52) and (2.53) of ref.~\cite{Berger:2010xi}, and estimate the perturbative uncertainty by taking the envelope of the three independent variations of profile parameters in eq.~(2.55) thereof (with the replacement $m_H \to Q$). Our central curve corresponds to the following choice of profile parameters: 
%%%
\begin{equation} \label{eq:profile}
\mu = Q
\,,\quad
e_B = e_S = 0
\,,\quad
\mu_0 = 2 \GeV
\,,\quad
\tau_1 = \frac{2 \GeV}{Q}
\,,\quad
\tau_2 = 0.25
\,,\quad
\tau_3 = 0.5
\,.\end{equation}
%%%

We will show here plots for the cumulant of the cross section in \eq{factth2}
%%%
\begin{equation} \label{eq:si_taucut}
  \frac{\df \si}{\df z}(\tau^\cut) = \int_0^{\tau^\cut}\!\! \df \tau\, \frac{\df^2\si}{\df \tau\, \df z}
  \, ,
\end{equation}
%%%
where the dijet limit is imposed by requiring $\tau \leq \tau^\cut \ll 1$. We only consider the contribution from the light quark flavors ($uds$). A strong cut on thrust almost entirely removes the $b$-quark events~\cite{Seidl:2008xc}. The $c$-quark contribution to $\pi^+$ fragmentation is small and therefore neglected. 

%%%
\begin{figure}[t]
\includegraphics[width=0.49\textwidth]{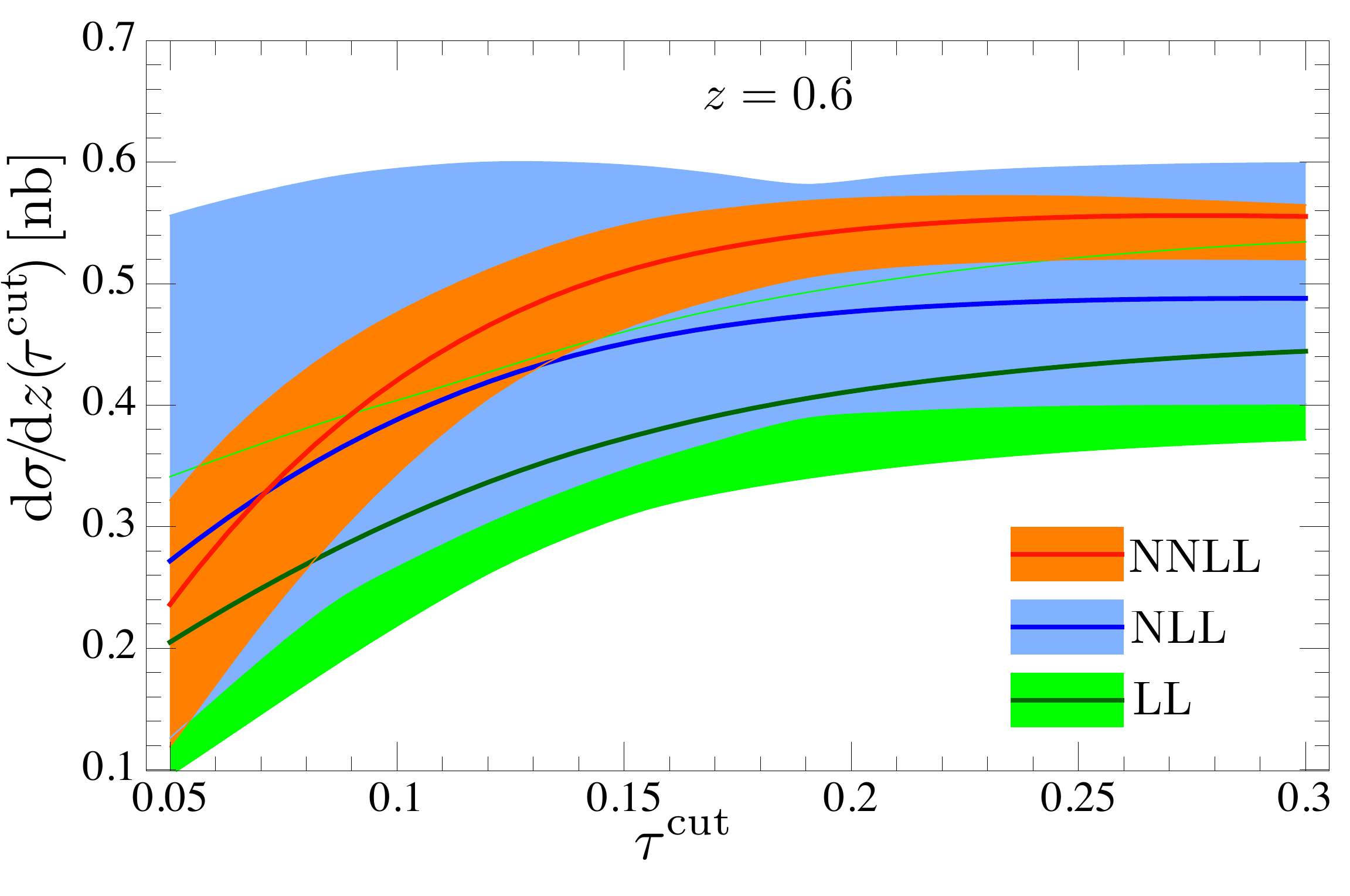} \hfill
\includegraphics[width=0.49\textwidth]{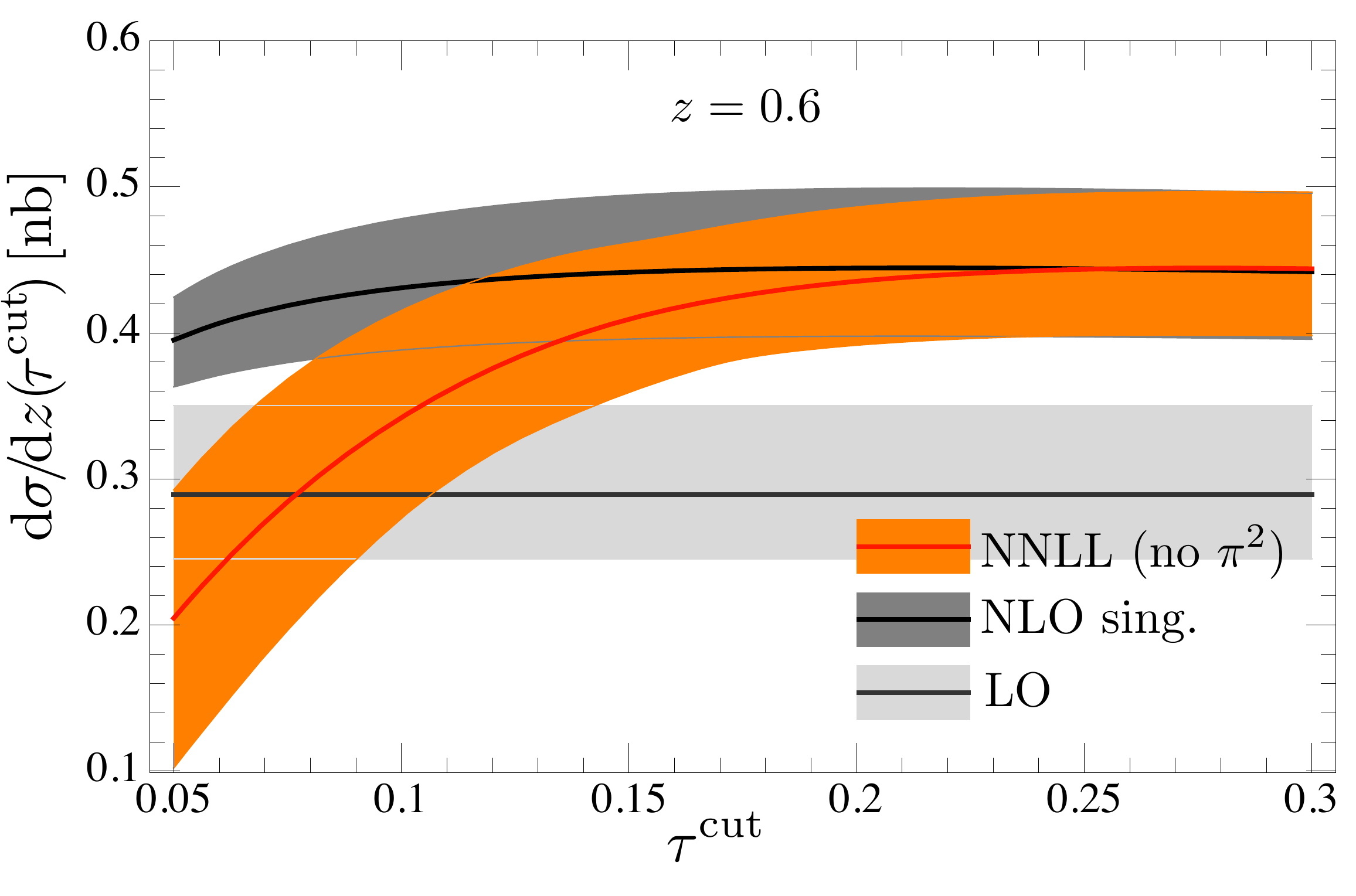}
\caption{The cross section for $e^+ e^- \to X\pi^+$, for $Q=10.6 \GeV$ and $z=0.6$, as function of the cut on thrust $\tau^\cut$. The left panel shows the resummed results at LL, NLL and NNLL order. The right panel shows the LO and singular NLO compared to the NNLL. In the right panel we switch off the $\pi^2$ resummation for the NNLL to show how it merges with the singular NLO. The bands correspond to the perturbative uncertainties as explained above \eq{profile}.
}
\label{fig:si_tau}
\end{figure}
%%%

%%%
\begin{figure}[t]
\includegraphics[width=0.49\textwidth]{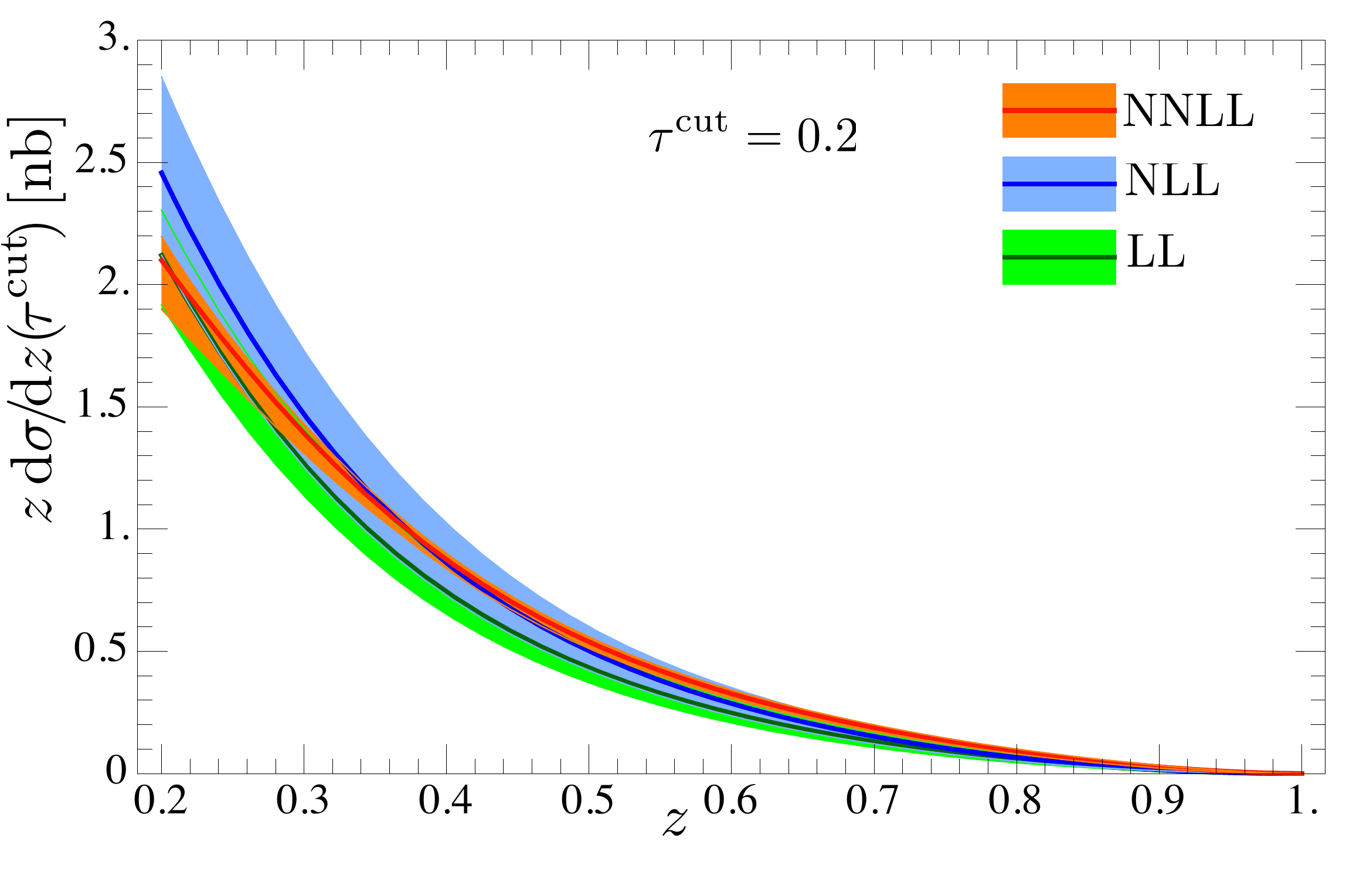} \hfill
\includegraphics[width=0.49\textwidth]{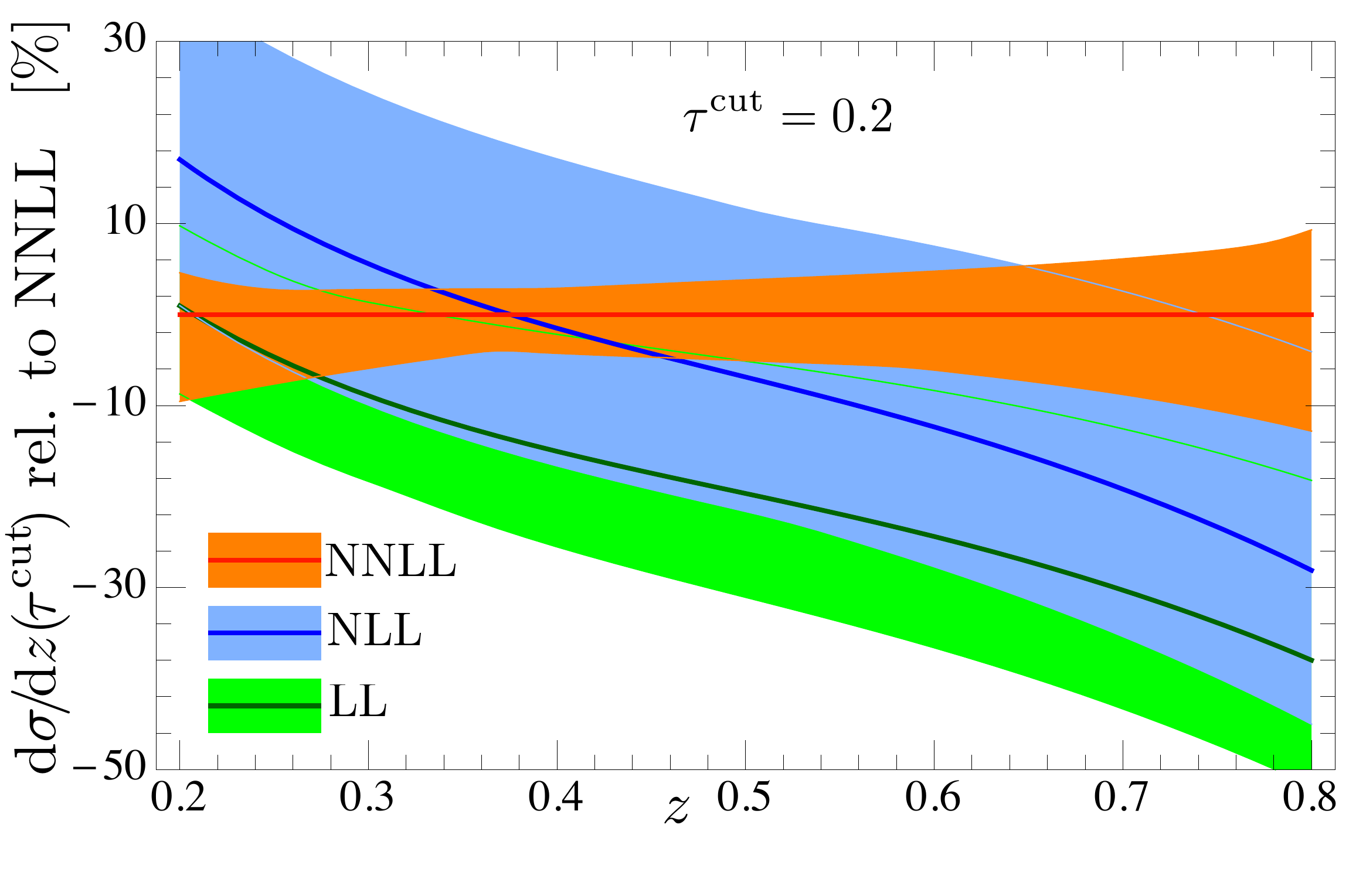} 
\caption{The cross section for $e^+ e^- \to X\pi^+$, for $Q=10.6 \GeV$ with a cut on thrust of $\tau\leq \tau^\cut = 0.2$, as a function of the momentum fraction $z$. In the left panel $z \df \si/\df z(\tau^\cut)$ is plotted at LL, NLL and NNLL. The right panel shows the same curves and bands as a percentage relative to the NNLL. The bands correspond to the perturbative uncertainties, see text above \eq{profile}.
}
\label{fig:si_z}
\end{figure}
%%%

In figure \ref{fig:si_tau}, we show the cross section for $e^+ e^- \to X\pi^+$ in \eq{si_taucut} for the Belle c.m. energy $Q=10.6 \GeV$ and a representative value of $z=0.6$ as a function of $\tau^\cut$. We checked that our numerical results are not sensitive to small variations of the scale $Q$. The left panel shows the result at LL, NLL and NNLL order in resummed perturbation theory. The uncertainties at LL and NLL are rather large, so only at NNLL we obtain a useful prediction. 
Furthermore the LL band is smaller than the NLL one and does not overlap the NNLL, which indicates that the LL is not reliable. The NLL and NNLL results are compatible within their uncertainties.

In the right panel of figure \ref{fig:si_tau}, the LO and singular NLO cross sections are plotted together with the NNLL without $\pi^2$-resummation. The singular NLO is obtained from \eq{factth2} by setting $\mu_H = \mu_J = \mu_S = \mu$. The remaining nonsingular terms that are present in the full NLO are suppressed by $\ord{\tau}$ relative to the singular ones. We take $\mu=Q$ for the central curve and vary $\mu$ between $Q/2$ and $2Q$ to estimate the uncertainties of the LO and NLO results which do not turn out to be compatible. For large $\tau$ the resummation is unimportant (except for the $\pi^2$ resummation, which we switched off in this plot) and so the NNLL merges with the singular NLO. However, below $\tau \sim 0.2$ the NNLL and singular NLO start to differ and below $\tau \sim 0.1$ this difference is no longer captured by the uncertainty bands, implying that resummation is necessary.

%%%
\begin{figure}[t]
\includegraphics[width=1.0\textwidth]{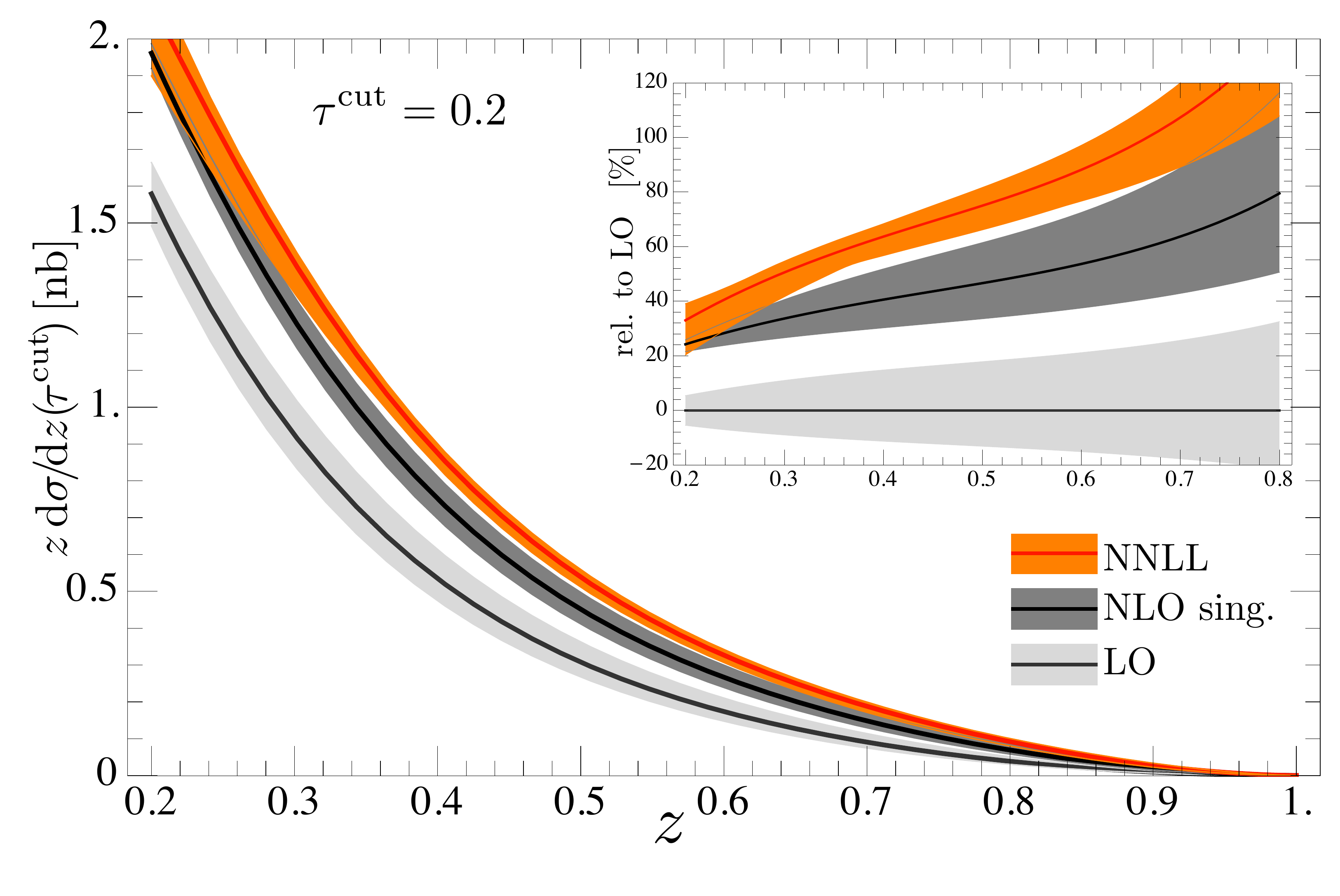}
\caption{The cross section for $e^+ e^- \to X\pi^+$, for $Q=10.6 \GeV$ with a cut on thrust of $\tau\leq \tau^\cut = 0.2$, as a function of the momentum fraction $z$. Here we compare the NNLL result with the corresponding LO and singular NLO. The inset shows the same curves and bands as a percentage relative to the LO. The bands correspond to the perturbative uncertainties.
}
\label{fig:si_z_main}
\end{figure}
%%%

In figure \ref{fig:si_z} we show results for the cross section for $Q=10.6 \GeV$ and $\tau^\cut=0.2$, as function of $z$. We plot the LL, NLL and NNLL results times $z$ and the right panel shows the same results relative to the NNLL (our best result). As in figure \ref{fig:si_tau}, the LL prediction is not reliable (it does not overlap with the NNLL) but the NLL is. 
Finally, in figure \ref{fig:si_z_main} we compared LO and singular NLO to our NNLL result. This illustrates the effect of the thrust cut on the dependence on the measured fragmentation variable $z$. As in figure \ref{fig:si_tau} the difference between the LO and singular NLO is not captured by their uncertainties. 

From the right panel of figure \ref{fig:si_tau}, one would expect resummation to be only marginally important for $\tau^\cut=0.2$. The difference between the singular NLO and NNLL in figure \ref{fig:si_z_main} is mainly due to the $\pi^2$ resummation, which is an overall factor. If one switches off the $\pi^2$ resummation, there is still a difference between the singular NLO and the NNLL, but for most of the plotted range this is within the uncertainties.

We also study the impact of our results on the determination of the fragmentation function parameters. For simplicity we only consider the contribution from the $u$-quark here.
The HKNS parametrization is given by \cite{Hirai:2007cx}
%%%
\begin{equation}
  D_u^{\pi^+} (z, \mu=1 \GeV) = \frac{M_u^{\pi^+}}{B(\al_u^{\pi^+}+2,\bt_u^{\pi^+}+1)}\, z^{\al_u^{\pi^+}} (1-z)^{\bt_u^{\pi^+}}
\,,
\end{equation}
%%%
where $M_u^{\pi^+}$ determines the normalization of $D_u^{\pi^+}$, $\al_u^{\pi^+} $ and $\bt_u^{\pi^+}$ describe its shape, and $B$ is the Euler beta function.
With $M_u^{\pi^+} = 0.401 \pm 0.052$, $\al_u^{\pi^+} = -0.963 \pm 0.177$ and $\bt_u^{\pi^+} = 1.370 \pm 0.144$ we reproduce our NNLL result that we will treat as ``data''. We then fit for the three parameters to these ``data'' using the LO formula for the cross section, $\df \si_\text{LO}^u/\df z = \si_0^u \,D_u^{\pi^+}(z,\mu=Q)$. We find that $\al_u^{\pi^+}$ and $\bt_u^{\pi^+}$ change by $\sim 30\%$ and $M_u^{\pi^+}$ changes by $\sim 70\%$. This clearly shows that if we use the LO result, rather than NNLL (or NLO), to extract the fragmentation function parameters in the presence of a cut on thrust, they may differ significantly from their true values.

%%%%%%%%%%%%%%%%%%%%%%%%%%%%%%%%%%%%%%%%%%%%%%%%%%%%%%%%%%%%%%%%%%%%%%%%%%%%%%%%
\section{Conclusions}
\label{sec:concl}
%%%%%%%%%%%%%%%%%%%%%%%%%%%%%%%%%%%%%%%%%%%%%%%%%%%%%%%%%%%%%%%%%%%%%%%%%%%%%%%%

In this paper we have calculated the matching coefficients $\cJ_{i j}(s,z/z', \mu)$ at one-loop, which are an important ingredient for factorization theorems that describe spin-averaged fragmentation of a light hadron $h$ fragmenting from a light quark or a gluon $i$ within a jet with constrained invariant mass. These matching coefficients contain the short-distance physics relating the fragmenting jet functions $\cG_i^h(s,z,\mu)$ -- that  depend both on the fragmentation variable $z$ and on the invariant mass $s$ of the jet -- to the standard fragmentation functions $D_j^h(z',\mu)$ via a convolution in $z^\prime$.
We have presented our calculation for $\cJ_{i j}$ in great detail, using various IR regulators for the partonic $\cG_i^j$ and $D_i^j$, exposing the structure of the zero-bin subtractions. A powerful cross-check on our results is provided by the relationship between $\cJ_{i j}$ and the leading jet function, which we have derived here in detail.

We have applied our results to study fragmentation of a $\pi^+$ in $e^+ e^-$ collisions, where we restrict to the dijet limit by a cut on thrust. Our calculation of $\cJ_{i j}$ enables us to resum the logarithms of $\tau^\cut = 1-T^\cut$ up to NNLL accuracy in the cross-section $\df \si/\df z (\tau^\cut)$. We analyzed this cross section for $\tau^\cut \lesssim 0.3$ and c.m.~energy equal to $10.6\,{\rm GeV}$, as in the study of light quark fragmentation in $B$-factories on the $\Upsilon(4S)$ resonance~\cite{Seidl:2008xc}. Here the convergence of resummed perturbation theory is better than that of fixed-order perturbation theory, and the perturbative uncertainties become reasonably small at NNLL accuracy. The NNLL cross section deviates from that at NLO for $\tau\lesssim0.2$, and below $\tau\lesssim 0.1$ resumming the logarithms of $\tau$ is necessary for a reliable prediction. 
Consistent with this observation, we have also shown that using cross sections at LO instead of NNLL (or NLO)  could have a sizeable impact on the extracted numerical values of the model parameters for $D_i^h$ from fits to experimental data for $e^+ e^- \to {\rm dijet} + h$.

We leave for future work the inclusion of nonsingular terms in the thrust distribution, and the effects of the uncertainties associated with the fragmentation functions and $\alpha_s(m_Z)$. This would provide a reliable theoretical framework to constrain fragmentation functions from $B$-factory data where cuts on thrust are applied.
However, the results presented have a more general applicability, and can be used to study fragmentation taking place inside any well separated jet. \\[2ex]

\noindent
{\bf Note added:} While we were completing this work, a paper \cite{Liu:2010ng} appeared which contains a calculation of the quark matching coefficient $\cJ_{q q}$ to one loop. We agree with this result.

%%%%%%%%%%%%%%%%%%%%%%%%%%%%%%%%%%%%%%%%%%%%%%%%%%%%%%%%%%%%%%%%%%%%%%%%%%%%%%%%

%%%%%%%%%%%%%%%%%%%%%%%%%%%%%%%%%%%%%%%%%%%%%%%%%%%%%%%%%%%%%%%%%%%%%%%%%%%%%%%%
\begin{acknowledgments}
We thank Iain Stewart for useful discussions and are grateful to him and Aneesh Manohar for comments on this manuscript.
We thank also Jui-Yu Chiu for discussions on IR regulators.  
M.P. acknowledges support by BMBF, by the DFG Cluster of Excellence ``Origin and Structure of the Universe'',  by the
``Innovations- und Kooperationsprojekt C-13'' of the Schweizerische Universit\"atskonferenz SUK/CRUS and by the Swiss National Science Foundation. A.J. is supported by DOE grant, 22645.1.1110173. A.J. would also like to thank the Max-Planck-Institut f\"ur Physik for hospitality while parts of this work were completed.

\end{acknowledgments}
%%%%%%%%%%%%%%%%%%%%%%%%%%%%%%%%%%%%%%%%%%%%%%%%%%%%%%%%%%%%%%%%%%%%%%%%%%%%%%%%

\appendix

%%%%%%%%%%%%%%%%%%%%%%%%%%%%%%%%%%%%%%%%%%%%%%%%%%%%%%%%%%%%%%%%%%%%%%%%%%%%%%%%
\section{Plus Distributions and Identities}
\label{app:plusdist}
%%%%%%%%%%%%%%%%%%%%%%%%%%%%%%%%%%%%%%%%%%%%%%%%%%%%%%%%%%%%%%%%%%%%%%%%%%%%%%%%

The standard plus distribution for some function $g(x)$ can be defined as
%%%
\begin{equation} \label{eq:plus_def}
\bigl[\theta(x) g(x)\bigr]_+
= \lim_{\beta \to 0} \frac{\df}{\df x} \bigl[\theta(x-\beta)\, G(x) \bigr]
\qquad\text{with}\qquad
G(x) = \int_1^x\!\df x'\, g(x')
\,,\end{equation}
%%%
satisfying the boundary condition $\int_0^1 \df x\, [\theta(x) g(x)]_+ = 0$. Two special cases we need are
%%%
\begin{align} \label{eq:plusdef}
\cL_n(x)
&\equiv \biggl[ \frac{\theta(x) \ln^n x}{x}\biggr]_+
 = \lim_{\beta \to 0} \biggl[
  \frac{\theta(x- \beta)\ln^n x}{x} +
  \delta(x- \beta) \, \frac{\ln^{n+1}\!\beta}{n+1} \biggr]
\,,\nn\\
\cL^\eta(x)
&\equiv \biggl[ \frac{\theta(x)}{x^{1-\eta}}\biggr]_+
 = \lim_{\beta \to 0} \biggl[
  \frac{\theta(x - \beta)}{x^{1-\eta}} +
  \delta(x- \beta) \, \frac{x^\eta - 1}{\eta} \biggr]
\,.\end{align}
%%%
In our calculations we will need the identity
%%%
\begin{equation} \label{eq:distr_id}
 \frac{\theta(x)}{x^{1+\eps}} = - \frac{1}{\eps}\,\delta(x) + \cL_0(x)
  - \eps \cL_1(x) + \ord{\eps^2}
\,,\end{equation}
%%%
and the two limits
%%%
\begin{align} \label{eq:limits}
\lim_{\bt\to 0}\biggl[\frac{\theta(x-\bt)\ln(x-\bt)}{x} + \delta(x-\bt)\,\frac{1}{2}\ln^2 \bt\biggr]
&= \cL_1(x) - \frac{\pi^2}{6}\,\delta(x)
\,,\nn\\
\lim_{\bt\to 0} \frac{\theta(x-\bt)\,\bt}{x^2}
&= \delta(x)
\,.\end{align}
%%%
Away from $x=0$ these identities are straightforward, while the behavior at $x = 0$ is obtained by taking the integral of both sides. General relations for the rescaling and convolutions of $\cL_n(x)$ and $\cL^\eta(x)$ can be found in App.~B of ref.~\cite{Ligeti:2008ac}.

%%%%%%%%%%%%%%%%%%%%%%%%%%%%%%%%%%%%%%%%%%%%%%%%%%%%%%%%%%%%%%%%%%%%%%%%%%%%%%%%
\section{Quark Matching Calculation with Offshellness IR Regulator}
\label{app:offshell}
%%%%%%%%%%%%%%%%%%%%%%%%%%%%%%%%%%%%%%%%%%%%%%%%%%%%%%%%%%%%%%%%%%%%%%%%%%%%%%%%

Here we present the one-loop calculation of $D_q^q(x,\mu)$, $D_q^g(x,\mu)$, $\cG_q^q(s,z,\mu)$ and $\cG_q^g(s,z,\mu)$ where the IR divergences are regulated through a small quark offshellness $p^2>0$. The real emission graphs for the fragmenting jet function are calculated using the Lehmann-Symanzik-Zimmermann (LSZ) reduction formulae combined with the optical theorem. We use dimensional regularization with $d=4-2\eps$ for the UV divergences and renormalize according to the $\overline{\text{MS}}$ scheme. The diagrams are computed in the Feynman gauge without any loss of generality since their sum is gauge-invariant, and we use the SCET Feynman rules. The zero-bin graphs vanish in this calculation.

%===============================================================================
\subsection{Quark Fragmentation Function}
%===============================================================================

For the virtual emission diagrams, the state $|X h \rangle$ in \eq{Dqdef} gets replaced by an off-shell quark. In the real graphs contributing to $D_q^q$, $X$ is an on-shell gluon, and in $D_q^g$ it is an off-shell quark. At the end of the calculation we expand in $\epsilon$ and neglect terms of $\ord{\eps}$.
Starting with $D_q^q$, the real emission graph in figure \ref{fig:DGqgraphs}a is given by 
%%%
\begin{align}
  D_{q, {\rm bare}}^{q (a)}(x) & = 
  \Big(\frac{e^{\ga_E} \mu^2}{4\pi}\Big)^\eps \frac{1}{2N_c\, x} \int\! \df^{d-2} p_\perp 
  \!\int\! \frac{\df^d \ell}{(2\pi)^{d-1}}\, \theta(\ell^0) \de(\ell^2)\, 
  \de(\w-\ell^- - p^-) \, \de^{d-2}(\ell_\perp + p_\perp)
  \nn \\ & \quad \times
 \tr \Big[\frac{\bnslash}{2}\,
  \img\, \frac{\nslash}{2}\,  \frac{\bar{n} \sdt (\ell+p)}{(\ell+p)^2+ \img 0}\,
  \img g T^a \Big(n^\mu + \frac{\pslash_\perp \ga_\perp^\mu}{\bn \sdt p}\Big) \frac{\bnslash}{2}\,
 \sum_{s, \rm{pol}} u_n^s(p)  \ve_\mu(\ell)\, \ve_\nu^*(\ell) \, {\bar u}_n^s(p)\, 
 \nn \\
 &\quad 
   \times \img g T^a \Big(n^\nu + \frac{\ga_\perp^\nu \pslash_\perp}{\bn \sdt p}\Big) \frac{\bnslash}{2}\,
  \img 
 \,\frac{\nslash}{2}\, \frac{\bar{n} \sdt (\ell+p)}{(\ell+p)^2+ \img 0}\Big]
  \nn \\
  &= \frac{\al_s(\mu) C_F}{2\pi}\, \theta(x)\, \theta(1-x)\, (1-x) \Big[\frac{1}{\eps} + \ln \frac{\mu^2}{p^2} -2 -\ln(1-x)\Big]
  \,,
\end{align}
%%%
with the momentum fraction $x = p^-/\omega$.
The contribution from the diagram in \fig{DGqgraphs}b plus its complex conjugate, ``mirror" graph, is
%%%
\begin{align}
   D_{q, {\rm bare}}^{q (b)}(x) & = 
   \Big(\frac{e^{\ga_E} \mu^2}{4\pi}\Big)^\eps \frac{1}{2N_c\, x} \int\! \df^{d-2} p_\perp 
  \!\int\! \frac{\df^d \ell}{(2\pi)^{d-1}}\, \theta(\ell^0) \de(\ell^2)\, 
    \de(\w-\ell^- - p^-) \,\de^{d-2}(\ell_\perp + p_\perp)
  \nn \\ & 
  %\quad 
  \times
\! \tr \Big[\frac{\bnslash}{2}\,
  \img\, \frac{\nslash}{2} \frac{\bar{n} \sdt (\ell+p)}{(\ell+p)^2+ \img 0}
  \img g T^a \! \Big(n^\mu + \frac{\pslash_\perp \ga_\perp^\mu}{\bn \sdt p}\Big) \frac{\bnslash}{2}
 \! \sum_{s, \rm{pol}} u_n^s(p) \ve_\mu(\ell)\, \ve_\nu^*(\ell)  {\bar u}_n^s(p) 
  \frac{g T^a \bn^\nu}{\bn  \sdt \ell} \Big] + {\rm c. c.}
  \nn \\
  &= \frac{\al_s(\mu) C_F}{2\pi}\, \theta(x)\theta(1-x) \,2 x \Big\{ \!- \frac{1}{\eps^2}\, \delta(1-x)+\frac{1}{\eps}\Big[ \cL_0(1-x) - \delta(1-x) \ln{\frac{\mu^2}{p^2}} \Big] 
  \nn \\ & \quad
   -\cL_1(1-x) + \cL_0(x) \ln{\frac{\mu^2}{p^2}} - \delta(1-x) \Big(\frac{1}{2}\, \ln{\frac{\mu^2}{p^2}}+ \frac{\pi^2}{12}\Big)  \Big\}\, .
\end{align}
%%%
For the virtual graphs in \fig{DGqgraphs}c and \fig{DGqgraphs}d and their mirror ones we find:
%%%
\begin{align} \label{eq:Dcd_os}
  D_{q, {\rm bare}}^{q (c)}(x) & = 
    \Big(\frac{e^{\ga_E} \mu^2}{4\pi}\Big)^\eps \frac{1}{2N_c\, x} \int\! \df^{d-2} p_\perp 
  \!\int\! \frac{\df^d \ell}{(2\pi)^{d}}\, 
    \de(\w - p^-) \,\de^{d-2}(p_\perp)\, \tr \Big[ \frac{\bnslash}{2}\, \sum_s u_n^s(p)\,{\bar u}_n^s(p)
  \nn \\ & 
  \quad 
  \times
  \img g T^a \Big(n^\mu + \frac{\ga_\perp^\mu (\pslash_\perp-\ellslash_\perp)}{\bn \sdt (p-\ell)}\Big) \frac{\bnslash}{2}\,
   \frac{- \img}{\ell^2+\img 0}\, \frac{g T^a \bn_\mu}{\bn  \sdt \ell}\,\img \frac{\nslash}{2} \frac{\bar{n} \sdt (p-\ell)}{(p-\ell)^2+ \img 0} \Big] + {\rm c.c.}
  \nn \\
  & = \frac{\al_s(\mu) C_F}{2\pi} \de(1-x)\, 2 \Big[\frac{1}{\eps^2}+\frac{1}{\eps} \Big(1+ \ln{\frac{\mu^2}{p^2}} \Big) + \frac{1}{2} \ln^2 \frac{\mu^2}{p^2} + \ln{\frac{\mu^2}{p^2}+2 -\frac{7}{12} \pi^2}  \Big]
 \,, \nn \\
  D_{q, {\rm bare}}^{q (d)}(x)  & = (Z_q^{1/2}-1) \, \de(1-x) + {\rm c.c.}
  = \frac{\al_s(\mu) C_F}{4\pi} \de(1-x) \Big(-\frac{1}{\eps} - \ln {\frac{\mu^2}{p^2}} - 1 \Big)
 \,,
\end{align}
%%%
using the one-loop on-shell wave function renormalization with an offshellness IR regulator. 
As we noted in section \ref{subsec:DNLO}, the zero bin does not contribute, because the fragmentation function is insensitive to the soft region.
Adding up all these graphs, we obtain the same renormalization factor $Z_{q q}^D(x,\mu)$ as in \eq{Zres}. For the renormalized quark fragmentation function we find
%%%
\begin{align}
D_q^{q\one}(x,\mu) 
& = \frac{\al_s(\mu) C_F}{2\pi}\,\theta(x)\, \theta(1-x)\, \Big[ P_{q q}(x) \ln \frac{\mu^2}{p^2}
-2x \cL_1(1-x) + \Big(\frac{7}{2} - \frac{4 \pi^2}{3}\Big) \de(1-x)
\nn \\ & \quad
- (1-x) (2 + \ln(1-x))\Big]
\,.
\end{align}
%%%

The one-loop $D_q^g$ is given by the real emission graphs in \fig{DGqgraphs}a and b, with the role of the quark and gluon interchanged. Therefore, from the results above,
%%%
\begin{align}
   D_{q, {\rm bare}}^{g (a)}(x) & 
   = \frac{x}{1-x} \! \times \! D_{q, {\rm bare}}^{q (a)}(x)
  = \frac{\al_s(\mu) C_F}{2\pi}\, \theta(x)\, \theta(1-x)\, x \Big[\frac{1}{\eps} +
  \ln \frac{\mu^2}{p^2} -2 -\ln(1-x)\Big]
  \,, \nn \\
  D_{q, {\rm bare}}^{g (b)}(x) & 
  =\Big(\frac{1-x}{x}\Big)^2 \!\times \! D_{q, {\rm bare}}^{q (b)}(x)  
  = \frac{\al_s(\mu) C_F}{2\pi}\, \theta(x)\,\theta(1-x)\, \frac{2(1-x)}{x} 
  \Big[\frac{1}{\eps} + \ln \frac{\mu^2}{p^2} - \ln (1-x)\Big]
  \,.
\end{align}
%%%
This leads to $Z_{q g}^D(x,\mu)$ as in \eq{Zres} and
%%%
\begin{align}
D_q^{g\one}(x,\mu) 
& = \frac{\al_s(\mu) C_F}{2\pi}\, \theta(x)\, \theta(1-x)
\Big[ P_{g q}(x) \Big(\ln\frac{\mu^2}{p^2} - \ln(1-x)\Big) -2x\Big]
\,.
\end{align}
%%%

%===============================================================================
\subsection{Quark Fragmenting Jet Function Via The Optical Theorem}
%===============================================================================

We now move on to the computation of the one-loop real emission graphs corresponding to $\cG_q^q(s,z,\mu)$ and $\cG_q^g(s,z,\mu)$. We use the LSZ reduction together with the optical theorem, along the lines of the calculations in the so-called cut vertex formalism in refs.~\cite{Mueller:1978xu,Baulieu:1979mr}. This procedure leads to the same result as directly integrating over the parton phase space, which has been employed everywhere else in this paper.

Applying the LSZ formalism to the collinear matrix elements in the definition of $\cG_q^q$ yields
%%%
\begin{align}
  \cG_q^q(s,z) & = 
  \int\! \df^4 y\, e^{\img k^+ y^-/2} \int\! \df^2 p_\perp
  \,\frac{1}{4N_c \pi p^-}\,  \tr \sum_X
  \Big[\frac{\bnslash}{2} \Mae{0}{[\de_{\w,\bnP}\, \de_{0,\cP_\perp} \chi_n(y)]} {X q(p_\ell, p_r)}
  \nn \\
  & \quad \times 
  \Mae{X q(p_\ell, p_r)}{\bar \chi_n(0)}{0} \Big] 
  \nn \\
  &= -\frac{R_q^{-1}}{4N_c \pi p^-}\int\! \df^2 p_\perp \int\! \df^4 y \int\! \df^4 x \int\! \df^4 x' \, e^{\img k^+ y^-/2}\, e^{\img p_r \cdot (x'-x)}
  \nn \\
  & \quad \times 
 \sum_X  \tr   \Big[\frac{\bnslash}{2} \Big\langle 0 \Big|  \bar{T} \Big\{ \bar{\xi}_{n,p_\ell}^{\rm amp}(x)\, [\de_{\w,\bnP}\, \de_{0,\cP_\perp} \chi_n(y)] \Big \} \Big| X \Big\rangle \Big\langle X \Big| T \Big\{ \xi_{n, p_\ell}^{\rm amp}(x')\, \bar{\chi}_n(0) \Big \} \Big| 0 \Big\rangle  \Big]\,,
\end{align}
%%%
where we distinguished label and residual momenta in the collinear quark states and fields. 
The superscript ``amp" indicates that the lines corresponding to the collinear quark fields $ \xi_n^{\rm amp}$ and $ \bar{\xi}_n^{\rm amp}$ should be amputated and replaced by the associated spinors. $R_q$ is the residue of the quark two-point function. In our calculations $R_q=1$ because we chose to absorb the finite terms of the self-energy diagram into the wave function renormalization [see \eq{Dcd_os}]. Following the optical theorem, we obtain
%%%
\begin{align} \label{eq:Gopt}
  \cG_q^q(s,z) & = 
   2\, {\rm Im} \frac{ -\img}{4N_c \pi p^-}\int\! \df^2 p_\perp \int\! \df^4 y \int\! \df^4 x \int\! \df^4 x' \, e^{\img k^+ y^-/2}\, e^{\img p_r \cdot (x'-x)}
  \nn \\
  & \quad \times 
  \tr \left[\frac{\bnslash}{2} \Big\langle 0 \Big|  T \Big\{ [\de_{\w,\bnP}\, \de_{0,\cP_\perp} \chi_n(y)] \, \bar{\xi}_{n,p_\ell}^{\rm amp}(x)\,  \xi_{n,p_\ell}^{\rm amp}(x')\, \bar{\chi}_n(0) \Big \} \Big| 0 \Big\rangle  \right]\, . 
\end{align}
%%%
Applying \eq{Gopt} to the contribution to the fragmenting jet function coming from figure \ref{fig:DGqgraphs}a, leads to
%%%
\begin{align}
  \frac{\cG_{q, {\rm bare}}^{q (a)}(s,z)}{2 (2\pi)^3} & = 
  2\, {\rm Im} \Big(\frac{e^{\ga_E} \mu^2}{4\pi}\Big)^\eps \frac{-\img}{2N_c\, x} \int\! \df^{d-2} p_\perp 
  \!\int\! \frac{\df^d \ell}{(2\pi)^{d}}\,  
  \de(\w-\ell^- - p^-) \, \de^{d-2}(\ell_\perp + p_\perp)\, 
   \\ & \quad \times
  \de(k^+ - l^+ - p^+)\,
  \tr \Big[\frac{\bnslash}{2}\,
  \img\, \frac{\nslash}{2}\, \frac{\bar{n} \sdt (\ell+p)}{(\ell+p)^2+ \img 0} \,
   \img g T^a \Big(n^\mu + \frac{\pslash_\perp \ga_\perp^\mu}{\bn \sdt p}\Big) \frac{\bnslash}{2}\, 
  \nn \\
  & \quad \times
  \sum_s u_n^s(p)\,\frac{- \img g_{\mu \nu}}{\ell^2 + i0} \, {\bar u}_n^s(p)\,
  \img g T^a \Big(n^\nu + \frac{\ga_\perp^\nu \pslash_\perp}{\bn \sdt p}\Big) \frac{\bnslash}{2}\,
  \img 
 \,\frac{\nslash}{2}\, \frac{\bar{n} \sdt (\ell+p)}{(\ell+p)^2+ \img 0} \Big]
  \nn \\
  &=2 {\rm Im}  \frac{\al_s(\mu) C_F}{(2\pi)^2} (e^{\ga_E} \mu^2)^{\eps} (1-\eps)  \Ga(\eps)\theta(z)  \theta(1-z) z^{- \eps} (1-z)^{1-\eps} 
  \frac{(p^2/z-s-\img 0)^{1-\eps}}{(s + \img 0)^2} \nn
\end{align}
%%%
where $s > p^2 > 0$.
Expanding in $\eps$ and evaluating the imaginary part yields
%%%
\begin{align}
  \frac{\cG_{q, {\rm bare}}^{q (a)}(s,z)}{2 (2\pi)^3} & = 
   \frac{\al_s(\mu) C_F}{2\pi}\, \theta(z)\theta(1-z) (1-z)\, 
  \frac{(s-p^2/z) \theta(s-p^2/z)}{s^2}
  \nn \\ 
  & \!\!\! \stackrel{p^2 \to 0}{=} \frac{\al_s(\mu) C_F}{2\pi}\, \theta(z)\theta(1-z) (1-z)\, 
  \Big[\frac{1}{\mu^2}\cL_0\Big(\frac{s}{\mu^2}\Big)
  + \de(s) \Big(\ln \frac{z\mu^2}{p^2} - 1\Big) \Big]
  \,.
\end{align}
%%%
In the last step we made use of  \eqs{plusdef}{limits} as $p^2 \to 0$ to isolate the IR divergences.

For \fig{DGqgraphs}b plus its mirror graph we find:
%%%
\begin{align}
   \frac{\cG_{q, {\rm bare}}^{q (b)}(s,z)}{2(2\pi)^3} \!& = 
 \! 2{\rm Im} \Big(\frac{e^{\ga_E} \mu^2}{4\pi}\Big)^\eps \!\frac{- \img}{2N_c\, x} \!\int\! \df^{d-2} p_\perp 
  \!\!\int\!\! \frac{\df^d \ell}{(2\pi)^{d}}
    \de(\w-\ell^- - p^-) \de^{d-2}(\ell_\perp + p_\perp) \de(k^+ - l^+ - p^+)
  \nn \\ & \quad \!\times
 \!\tr \Big[\frac{\bnslash}{2}\,
  \img\, \frac{\nslash}{2}\, \frac{\bar{n} \sdt (\ell+p)}{(\ell+p)^2+ \img 0} \,
  \img g T^a \Big(n^\mu + \frac{\pslash_\perp \ga_\perp^\mu}{\bn \sdt p}\Big) \frac{\bnslash}{2}\,
 \! \sum_s u_n^s(p)\frac{- \img g_{\mu \nu}}{\ell^2 + i0} \,  {\bar u}_n^s(p)
  \frac{g T^a \bn^\nu}{\bn  \sdt \ell} \Big] \!+ {\rm c. c.}
  \nn \\
  &\!\!\! \stackrel{p^2 \to 0}{=}  \frac{\al_s(\mu) C_F}{2\pi} \theta(z)\theta(1-z) \Big\{ \!-\frac{2}{\eps} \de(1-z) 
  \Big[\frac{1}{\mu^2}\cL_0\Big(\frac{s}{\mu^2}\Big)
  + \de(s) \ln \frac{\mu^2}{p^2}\Big] 
   \nn \\ & \quad
   + \frac{2}{\mu^2}\cL_1\Big(\frac{s}{\mu^2}\Big) \de(1-z)
  + \frac{1}{\mu^2}\cL_0\Big(\frac{s}{\mu^2}\Big)\, 2z \cL_0(1-z) +
  \de(s) \Big[\ln \frac{z\mu^2}{p^2}\, 2z \cL_0(1-z)
   \nn \\ & \quad
  - \ln^2 \frac{\mu^2}{p^2}\, \de(1-z) - \frac{\pi^2}{3}\, \de(1-z) \Big] \Big \}
  \, ,
\end{align}
%%%
again using \eqs{plusdef}{limits} to take the $p^2 \to 0$ limit.

The contribution of the virtual diagrams can be derived directly from the previous calculation of the partonic fragmentation function since
%%%
\begin{equation}
\frac{\cG_{q, {\rm bare}}^{q (r)}(s,z)}{2(2\pi)^3} = \de(s) \, D_{q, {\rm bare}}^{q (r)}(z)\, , \qquad  {\rm with} \; r=c,d~. 
 \end{equation}
%%%
The zero bin vanishes for this choice of IR regulator ($1/\eps_\text{UV} -1/\eps_\text{IR} = 0$) but still contributes to the fragmenting jet function turning the IR divergences into UV divergences.

Furthermore, as for $D_q^g$ at one loop, we obtain:
%%%
\begin{align}
   \frac{\cG_{q, {\rm bare}}^{g (a)}(s,z)}{2 (2\pi)^3} & = \frac{z}{1-z} \! \times \! \frac{\cG_{q, {\rm bare}}^{q (a)}(s,z)}{2 (2\pi)^3}
  =\frac{\al_s(\mu) C_F}{2\pi}\, \theta(z)\theta(1-z) z 
  \Big[\frac{1}{\mu^2}\cL_0\Big(\frac{s}{\mu^2}\Big)
  + \de(s) \Big(\ln \frac{z\mu^2}{p^2} - 1\Big) \Big]
   \nn \\
   \frac{\cG_{q, {\rm bare}}^{g (b)}(s,z)}{2 (2\pi)^3} & = \Big(\frac{1-z}{z}\Big)^2\! \times \! \frac{\cG_{q, {\rm bare}}^{q (b)}(s,z)}{2 (2\pi)^3}
   \nn \\ &
  =\frac{\al_s(\mu) C_F}{2\pi}\,\theta(z) \theta(1-z) \,\frac{2(1-z)}{z} \,
  \Big[\frac{1}{\mu^2}\cL_0\Big(\frac{s}{\mu^2}\Big)
  + \de(s) \ln \frac{z\mu^2}{p^2}  \Big]
  \,.
\end{align}
%%%
The UV divergences only occur in $ \cG_{q, {\rm bare}}^{q}$, as was already pointed out in \eq{ZGdef}, and we find that the corresponding renormalization factor $Z_{\cG}^q$ coincides with the one derived in \eq{ZGres}. The renormalized partonic  fragmenting jet functions are then given by
%%%
\begin{align}
  \frac{\cG_q^{q\one}(s,z,\mu)}{2 (2\pi)^3}& =  \frac{\al_s(\mu) C_F}{2\pi}\, \theta(z)\, \Big\{\frac{2}{\mu^2} \cL_1\Big(\frac{s}{\mu^2}\Big) \de(1-z) +
  \frac{1}{\mu^2} \cL_0\Big(\frac{s}{\mu^2}\Big) (1+z^2) \cL_0(1-z)
  \nn \\ & \quad
  + \de(s) \Big[P_{q q}(z) \ln \frac{z\mu^2}{p^2} 
   + \Big(\frac{7}{2} - \frac{3 \pi^2}{2}\Big) \de(1-z) - \theta(1-z)(1-z)\Big]\Big\}
   \nn \\
   \frac{\cG_q^{g\one}(s,z,\mu)}{2 (2\pi)^3}& =
  \frac{\al_s(\mu) C_F}{2\pi}\, \theta(z) 
  \Big\{
  \Big[\frac{1}{\mu^2} \cL_0\Big(\frac{s}{\mu^2}\Big)
  + \de(s) \ln \frac{z\mu^2}{p^2} \Big] P_{g q}(z)
  -\de(s) \theta(1-z) z \Big\}
  \,.
\end{align}
%%%
By applying \eq{NLOmatch}, we find that the matching coefficients $\cJ_{qq}^\one (s,z,\mu)$ and $\cJ_{qg}^\one (s,z,\mu)$ obtained from this calculation agree with the ones given in \eq{Jresult}. This had to be the case since the $\cJ_{i j}$ are insensitive to the choice of IR regulators.

%%%%%%%%%%%%%%%%%%%%%%%%%%%%%%%%%%%%%%%%%%%%%%%%%%%%%%%%%%%%%%%%%%%%%%%%%%%%%%%%
\section{Perturbative Results}
\label{app:pert}
%%%%%%%%%%%%%%%%%%%%%%%%%%%%%%%%%%%%%%%%%%%%%%%%%%%%%%%%%%%%%%%%%%%%%%%%%%%%%%%%

%===============================================================================
\subsection{Fixed-Order Results}
%===============================================================================

The Born cross section $\si_0^q$ in \eq{factth} is given by (see e.g. appendix A of ref.~\cite{Abbate:2010xh})
%%%
\begin{equation} \label{eq:si0}
  \si_0^q = \frac{4\pi \al_\text{em}^2 N_c}{3Q^2} \biggl[ Q_q^2 + \frac{(v_q^2 + a_q^2) (v_e^2+a_e^2) - 2 Q_q v_q v_e (1-m_Z^2/Q^2)}
{(1-m_Z^2/Q^2)^2 + \Gamma_Z^2/m_Z^2} \biggr]
\,,
\end{equation}
%%%
where $q$ denotes the (anti)quark flavor, $Q_q$ is the quark charge in units of $\abs{e}$, $v_{q,e}$ and $a_{q,e}$ are the vector and axial couplings of the (anti)quark $q$ and the electron to the $Z$  as e.g. in eq.(A3) of ref.~\cite{Abbate:2010xh}. Here $m_Z$ and $\Gamma_Z$ denote the mass and the width of the $Z$ boson.

The hard function for thrust, at leading order in the electroweak interactions, is the square of the Wilson coefficient in the matching of the quark current from QCD onto SCET,
%%%
\begin{equation} \label{eq:H_match}
H(Q^2,\mu_H)
=  \Abs{C(Q^2, \mu_H)}^2
\,.\end{equation}
%%%
The SCET matching was computed at one-loop in refs.~\cite{Manohar:2003vb, Bauer:2003di}, yielding
%%%
\begin{equation} \label{eq:C_hard}
C(Q^2, \mu_H) = 1 + \frac{\alpha_s(\mu_H)\,C_F}{4\pi} \biggl[-\ln^2 \Bigl(\frac{-Q^2-\img 0}{\mu_H^2}\Bigr) + 3 \ln \Bigl(\frac{-Q^2-\img 0}{\mu_H^2}\Bigr) - 8 + \frac{\pi^2}{6} \biggr]
\,.\end{equation}
%%%

The one-loop quark jet function~\cite{Bauer:2003pi} and one-loop gluon jet function~\cite{Fleming:2003gt,Becher:2009th} are given by
%%%
\begin{align} \label{eq:jetf}
   J_q(s,\mu) &= \de(s) + \frac{\al_s(\mu) C_F}{2\pi} \Big[ \frac{2}{\mu^2} \cL_1\Big(\frac{s}{\mu^2}\Big) - \frac{3}{2 \mu^2} \cL_0\Big(\frac{s}{\mu^2}\Big) - \Big(\frac{\pi^2}{2} - \frac{7}{2} \Big) \de(s) \Big]
 \,, \nn \\
   J_g(s,\mu) &= \de(s) + \frac{\al_s(\mu)}{2\pi} \Big\{\frac{2 C_A}{\mu^2} \cL_1\Big(\frac{s}{\mu^2}\Big) - \frac{\bt_0}{2\mu^2} \cL_0\Big(\frac{s}{\mu^2}\Big) + \Big[\Big(\frac{2}{3} - \frac{\pi^2}{2} \Big)C_A + \frac{5}{6} \bt_0 \Big] \de(s)  \Big\}
  \,.
\end{align}
%%%

The one-loop perturbative soft function for thrust can be obtained from refs.~\cite{Schwartz:2007ib, Fleming:2007xt}:
%%%
\begin{equation}
S_\tau(k,\mu_S) = \delta(k) + \frac{\alpha_s(\mu_S)\,C_F}{2\pi} \biggl[
-\frac{8}{\mu_S} \cL_1\Bigl(\frac{k}{\mu_S}\Bigr) + \frac{\pi^2}{6}\, \delta(k) \biggr]
\,.\end{equation}
%%%

The one-loop Wilson coefficients $\cJ_{ij}(s,z,\mu)$, for matching the fragmenting jet functions onto fragmentation functions in \eq{OPE}, are given in \eq{Jresult} for $i=q$ and \eq{Jgresult} for $i=g$.

%===============================================================================
\subsection{Renormalization Group Evolution}
%===============================================================================

The RGE and anomalous dimension for the hard Wilson coefficient in \eq{C_hard} are~\cite{Manohar:2003vb, Bauer:2003di}
%%%
\begin{equation} \label{eq:C_RGE}
\mu \frac{\df}{\df\mu} C(Q^2, \mu) = \gamma_H(Q^2, \mu)\, C(Q^2, \mu)
\,,\quad
\gamma_H(Q^2, \mu) =
\Gamma_\cusp^q[\alpha_s(\mu)] \ln\frac{- Q^2 - \img 0}{\mu^2} + \gamma_H^q[\alpha_s(\mu)]
\,. 
\end{equation}
%%%
The coefficients of the $\al_s$-expansion of $\Gamma_\cusp^q(\alpha_s)$ and $\gamma_H^q(\alpha_s)$ are given below in \eqs{Gacuspexp}{gaHexp}. By solving the RGE in \eq{C_RGE} we obtain the evolution of the hard function:
%%%
\begin{align} \label{eq:Hrun}
H(Q^2, \mu) &= H(Q^2, \mu_0)\, U_H(Q^2, \mu_0, \mu)
\,,\qquad
U_H(Q^2, \mu_0, \mu)
= \Bigl\lvert e^{K_H(\mu_0, \mu)} \Bigl(\frac{- Q^2 - \img 0}{\mu_0^2}\Bigr)^{\eta_H(\mu_0, \mu)} \Bigr\rvert^2
\,,\nn \\
K_H(\mu_0,\mu) &= -2 K^q_\Gamma(\mu_0,\mu) + K_{\gamma_H^q}(\mu_0,\mu)
\,, \qquad
\eta_H(\mu_0,\mu) = \eta_\Gamma^q(\mu_0,\mu)
\,,\end{align}
%%%
where $K_\Gamma^q(\mu_0, \mu)$, $\eta_\Gamma^q(\mu_0, \mu)$ and $K_\gamma$ are given below in \eq{Keta_def}.

The jet function RGE and anomalous dimension are
%%%
\begin{align} \label{eq:ga_jet}
\mu \frac{\df}{\df \mu} J_i(s, \mu) &= \int_0^s\! \df s'\, \gamma_J^i(s-s',\mu)\, J_i(s', \mu)
\,,\nn\\
\gamma_J^i(s, \mu)
&= -2 \Gamma^i_{\cusp}[\alpha_s(\mu)] \,\frac{1}{\mu^2}\cL_0\Bigl(\frac{s}{\mu^2}\Bigr) + \gamma_J^i[\alpha_s(\mu)]\,\delta(s)
\,,\end{align}
%%%
where the index $i=\{q,g\}$ is not summed over. Its solution is given by~\cite{Balzereit:1998yf, Neubert:2004dd, Fleming:2007xt, Ligeti:2008ac}
%%%
\begin{align} \label{eq:Grun_full}
J_i(s,\mu) & =  \int_0^s\! \df s'\, U_J^i(s - s',\mu_0, \mu)\, J_i(s',\mu_0)
\,, \nn \\
U_J^i(s, \mu_0, \mu) & = \frac{e^{K_J^i -\gamma_E\, \eta_J^i}}{\Gamma(1+\eta_J^i)}\,
\biggl[\frac{\eta_J^i}{\mu_0^2} \cL^{\eta_J^i} \Bigl( \frac{s}{\mu_0^2} \Bigr) + \delta(s) \biggr]
\,, \nn \\
K_J^i(\mu_0,\mu) &= 4 K^i_\Gamma(\mu_0,\mu) + K_{\gamma_J^i}(\mu_0,\mu)
\,, \quad
\eta_J^i(\mu_0,\mu) = -2\eta^i_{\Gamma}(\mu_0,\mu)
\,.\end{align}
%%%
According to  \eqs{GRGE}{gaG}, we obtain the RGE of the fragmenting jet function as well as its solution by simply replacing $J_i(s,\mu) \to \cG_i^h(s,z,\mu)$ in the expressions above.

The RGE of the thrust soft function is given by
%%%
\begin{align} \label{eq:S_RGE}
\mu\frac{\df}{\df\mu} S_\tau(k, \mu)
&= \int_0^k\! \df k'\, \gamma_S(k - k', \mu)\, S_\tau(k', \mu)
\,, \\
\gamma_S(k, \mu)
&= 4\Gamma_\cusp^q[\alpha_s(\mu)]\, \frac{1}{\mu} \cL_0\Big(\frac{k}{\mu}\Big) +
\gamma_S[\alpha_s(\mu)]\, \de(k)
\,, \nn \end{align}
%%%
whose solution is completely analogous to \eq{Grun_full}:
%%%
\begin{align} \label{eq:Srun}
S_\tau(k,\mu) & =  \int_0^k\! \df k'\, U_S(k - k',\mu_0,\mu)\, S_\tau(k',\mu_0)
\,, \nn \\
U_S(k, \mu_0, \mu) & = \frac{e^{K_S -\gamma_E\, \eta_S}}{\Gamma(1+\eta_S)}\,
\biggl[\frac{\eta_S}{\mu_0} \cL^{\eta_S} \Big( \frac{k}{\mu_0} \Big) + \delta(k) \biggr]
\,, \nn \\
K_S(\mu_0,\mu) &= - 4 K_\Gamma^q(\mu_0,\mu) + K_{\gamma_S}(\mu_0,\mu)
\,, \quad
\eta_S(\mu_0,\mu) = 4\eta_\Gamma^q(\mu_0,\mu)
\,.\end{align}
%%%

The functions $K_\Gamma^i(\mu_0, \mu)$, $\eta_\Gamma^i(\mu_0, \mu)$, $K_\gamma(\mu_0, \mu)$ in the above RGE solutions are defined as
%%%
\begin{align} \label{eq:Keta_def}
K_\Gamma^i(\mu_0, \mu)
& = \int_{\alpha_s(\mu_0)}^{\alpha_s(\mu)}\!\frac{\df\alpha_s}{\beta(\alpha_s)}\,
\Gamma_\cusp^i(\alpha_s) \int_{\alpha_s(\mu_0)}^{\alpha_s} \frac{\df \alpha_s'}{\beta(\alpha_s')}
\,,\quad
\eta_\Gamma^i(\mu_0, \mu)
= \int_{\alpha_s(\mu_0)}^{\alpha_s(\mu)}\!\frac{\df\alpha_s}{\beta(\alpha_s)}\, \Gamma_\cusp^i(\alpha_s)
\,,\nn \\
K_\gamma(\mu_0, \mu)
& = \int_{\alpha_s(\mu_0)}^{\alpha_s(\mu)}\!\frac{\df\alpha_s}{\beta(\alpha_s)}\, \gamma(\alpha_s)
\,.\end{align}
%%%
Expanding the $\beta$-function and the anomalous dimensions in powers of $\alpha_s$,
%%%
\begin{align}
\beta(\alpha_s) =
- 2 \alpha_s \sum_{n=0}^\infty \beta_n\Bigl(\frac{\alpha_s}{4\pi}\Bigr)^{n+1}
\,, \ \
\Gamma^i_\cusp(\alpha_s) = 
\sum_{n=0}^\infty \Gamma^i_n \Bigl(\frac{\alpha_s}{4\pi}\Bigr)^{n+1}
\,, \ \
\gamma(\alpha_s) = \sum_{n=0}^\infty \gamma_n \Bigl(\frac{\alpha_s}{4\pi}\Bigr)^{n+1}
\,,\end{align}
%%%
their explicit expressions at NNLL are
%%%
\begin{align} \label{eq:Keta}
K_\Gamma(\mu_0, \mu) &= -\frac{\Gamma_0}{4\beta_0^2}\,
\biggl\{ \frac{4\pi}{\alpha_s(\mu_0)}\, \Bigl(1 - \frac{1}{r} - \ln r\Bigr)
   + \biggl(\frac{\Gamma_1 }{\Gamma_0 } - \frac{\beta_1}{\beta_0}\biggr) (1-r+\ln r)
   + \frac{\beta_1}{2\beta_0} \ln^2 r
\nn\\ & \hspace{10ex}
+ \frac{\alpha_s(\mu_0)}{4\pi}\, \biggl[
  \biggl(\frac{\beta_1^2}{\beta_0^2} - \frac{\beta_2}{\beta_0} \biggr) \Bigl(\frac{1 - r^2}{2} + \ln r\Bigr)
  + \biggl(\frac{\beta_1\Gamma_1 }{\beta_0 \Gamma_0 } - \frac{\beta_1^2}{\beta_0^2} \biggr) (1- r+ r\ln r)
\nn\\ & \hspace{10ex}
  - \biggl(\frac{\Gamma_2 }{\Gamma_0} - \frac{\beta_1\Gamma_1}{\beta_0\Gamma_0} \biggr) \frac{(1- r)^2}{2}
     \biggr] \biggr\}
\,, \nn\\
\eta_\Gamma(\mu_0, \mu) &=
 - \frac{\Gamma_0}{2\beta_0}\, \biggl[ \ln r
 + \frac{\alpha_s(\mu_0)}{4\pi}\, \biggl(\frac{\Gamma_1 }{\Gamma_0 }
 - \frac{\beta_1}{\beta_0}\biggr)(r-1)
\nn \\ & \hspace{10ex}
 + \frac{\alpha_s^2(\mu_0)}{16\pi^2} \biggl(
    \frac{\Gamma_2 }{\Gamma_0 } - \frac{\beta_1\Gamma_1 }{\beta_0 \Gamma_0 }
      + \frac{\beta_1^2}{\beta_0^2} -\frac{\beta_2}{\beta_0} \biggr) \frac{r^2-1}{2}
    \biggr]
\,, \nn\\
K_\gamma(\mu_0, \mu) &=
 - \frac{\gamma_0}{2\beta_0}\, \biggl[ \ln r
 + \frac{\alpha_s(\mu_0)}{4\pi}\, \biggl(\frac{\gamma_1 }{\gamma_0 }
 - \frac{\beta_1}{\beta_0}\biggr)(r-1) \biggr]
\,.\end{align}
%%%
Here $r = \alpha_s(\mu)/\alpha_s(\mu_0)$ and we have suppressed the superscript $i$ on $K_\Ga^i$, $\eta_\Ga^i$ and $\Ga^i_n$. Note that the expressions in \eq{Keta} cannot be used across quark thresholds, where $n_f$ changes.

Up to three loops, the coefficients of the $\beta$-function~\cite{Tarasov:1980au, Larin:1993tp} and cusp anomalous dimension~\cite{Korchemsky:1987wg, Moch:2004pa} in $\overline{\mathrm{MS}}$ are
%%%
\begin{align} \label{eq:Gacuspexp}
\beta_0 &= \frac{11}{3}\,C_A -\frac{4}{3}\,T_F\,n_f
\,,\nn\\
\beta_1 &= \frac{34}{3}\,C_A^2  - \Bigl(\frac{20}{3}\,C_A\, + 4 C_F\Bigr)\, T_F\,n_f
\,, \nn\\
\beta_2 &=
\frac{2857}{54}\,C_A^3 + \Bigl(C_F^2 - \frac{205}{18}\,C_F C_A
 - \frac{1415}{54}\,C_A^2 \Bigr)\, 2T_F\,n_f
 + \Bigl(\frac{11}{9}\, C_F + \frac{79}{54}\, C_A \Bigr)\, 4T_F^2\,n_f^2
\,,\nn\\[2ex]
\Gamma^q_0 &= 4C_F
\,,\nn\\
\Gamma^q_1 &= 4C_F \Bigl[\Bigl( \frac{67}{9} -\frac{\pi^2}{3} \Bigr)\,C_A  -
   \frac{20}{9}\,T_F\, n_f \Bigr]
\,,\nn\\
\Gamma^q_2 &= 4C_F \Bigl[
\Bigl(\frac{245}{6} -\frac{134 \pi^2}{27} + \frac{11 \pi ^4}{45}
  + \frac{22 \zeta_3}{3}\Bigr)C_A^2
  + \Bigl(- \frac{418}{27} + \frac{40 \pi^2}{27}  - \frac{56 \zeta_3}{3} \Bigr)C_A\, T_F\,n_f
\nn\\* & \hspace{8ex}
  + \Bigl(- \frac{55}{3} + 16 \zeta_3 \Bigr) C_F\, T_F\,n_f
  - \frac{16}{27}\,T_F^2\, n_f^2 \Bigr]
\,,\\[2ex]
\Gamma^g_n &= \frac{C_A}{C_F}\, \Gamma^q_n \quad (\text{known to hold for }n\leq 2).
\end{align}
%%%
The $\overline{\mathrm{MS}}$ anomalous dimension for the hard function can be obtained~\cite{Idilbi:2006dg, Becher:2006mr} from the IR divergences of the on-shell massless quark form factor, which is known to three loops~\cite{Moch:2005id}. At the order we are working we only need the two-loop result,
%%%
\begin{align} \label{eq:gaHexp}
\gamma^q_{H\,0} &= -6 C_F
\,,\nn\\
\gamma^q_{H\,1}
&= - C_F \Bigl[
  \Bigl(\frac{82}{9} - 52 \zeta_3\Bigr) C_A
+ (3 - 4 \pi^2 + 48 \zeta_3) C_F
+ \Bigl(\frac{65}{9} + \pi^2 \Bigr) \beta_0 \Bigr]
\,.\end{align}
%%%
The anomalous dimension of the fragmenting jet function $\cG_i$ and jet function $J_i$ are equal, so in particular $\ga_\cG^i(\al_s) = \ga_J^i(\al_s)$. These anomalous dimensions were extracted for the quark jet function in ref.~\cite{Becher:2006mr} from ref.~\cite{Moch:2004pa}, and for the gluon jet function in ref.~\cite{Becher:2009th} from ref.~\cite{Vogt:2004mw}, at three loop order. We only need the quark and gluon jet function anomalous dimension up to two loops, which are given by
%%%
\begin{align}
\gamma_{\cG\,0}^q &= 6 C_F
\,,\nn\\
\gamma_{\cG\,1}^q
&= C_F \Bigl[
  \Bigl(\frac{146}{9} - 80 \zeta_3\Bigr) C_A
+ (3 - 4 \pi^2 + 48 \zeta_3) C_F
+ \Bigl(\frac{121}{9} + \frac{2\pi^2}{3} \Bigr) \beta_0 \Bigr]
\,,\nn\\
\ga_{\cG\,0}^g &= 2 \bt_0
\,,\nn\\
\ga_{\cG\,1}^g
&= \Big(\frac{182}{9} - 32\zeta_3\Big)C_A^2 +
\Big(\frac{94}{9}-\frac{2\pi^2}{3}\Big) C_A\, \bt_0 + 2\bt_1
\,. \end{align}
%%%
The consistency of the RGE for the factorization theorem in \eq{factth2} implies that $\ga_S(\al_s) = - 2\ga_H(\al_s) - 2\ga_\cG^q(\al_s)$, fixing the non-cusp anomalous dimension of the soft function.

\bibliographystyle{jhep}
\bibliography{../Fragmentation}

%%%%%%%%%%%%%%%%%%%%%%%%%%%%%%%%%%%%%%%%%%%%%%%%%%%%%%%%%%%%%%%%%%%%%%%%%%%%%%%%

\end{document}